\documentclass[preprint, review, 12pt, 3p]{elsarticle}
\usepackage{lscape}
\usepackage{amsmath}
\usepackage[colorlinks=true]{hyperref}
\usepackage{boldline}
\usepackage{dirtytalk}
\usepackage{graphicx}
\usepackage{subfig}
\usepackage{multirow}
\usepackage{lipsum}
\usepackage{booktabs}
\usepackage{multicol}
\usepackage{longtable}
\usepackage{caption}
\usepackage{amsfonts}
\usepackage{makecell}
\usepackage{adjustbox}
\usepackage{tabularx}
\usepackage{floatrow} 
\usepackage[vlines]{tabularht}
\usepackage{pbox}
\usepackage{pdflscape}
\usepackage{pifont}

\newcommand*\colourcheck[1]{%
  \expandafter\newcommand\csname #1check\endcsname{\textcolor{#1}{\ding{52}}}%
}
\colourcheck{blue}
\colourcheck{green}
\colourcheck{red}

\usepackage{nomencl}
\makenomenclature

\usepackage{float}
\floatstyle{plaintop}
\restylefloat{table}

\makeatletter
\newcommand{\thickhline}{%
    \noalign {\ifnum 0=`}\fi \hrule height 1pt
    \futurelet \reserved@a \@xhline
}
\newcolumntype{"}{@{\vrule width 1pt}}
\makeatother

\newlength{\Oldarrayrulewidth}

\usepackage[linesnumbered,ruled,vlined]{algorithm2e}
\SetKwInput{KwInput}{Input}                
\SetKwInput{KwOutput}{Output}              

\usepackage[table,xcdraw]{xcolor}
\newcommand{\MKH}[1]{\textcolor{blue}{#1}}

\newsavebox\MBox

\biboptions{square,sort&compress}

\journal{Computers in Biology and Medicine}

\begin{document}

\begin{frontmatter}

\title{A survey, review, and future trends of skin lesion segmentation and classification}

\author{Md. Kamrul Hasan\fnref{label2,label1}\corref{cor1}}
\ead{k.hasan22@imperial.ac.uk}

\author[label1]{Md. Asif Ahamad}
\ead{asif.fx@live.com}

\author[label2]{Choon Hwai Yap\corref{cor2}}
\ead{c.yap@imperial.ac.uk}

\author{Guang Yang\fnref{label3,label4}\corref{cor2}}
\ead{g.yang@imperial.ac.uk}

\address[label2]{Department of Bioengineering, Imperial College London, UK}

\address[label1]{Department of Electrical and Electronic Engineering (EEE), Khulna University of Engineering \& Technology (KUET), Khulna-9203, Bangladesh}

\address[label3]{National Heart and Lung Institute, Imperial College London, UK}

\address[label4]{Cardiovascular Research Centre, Royal Brompton Hospital, UK}

\cortext[cor1]{Corresponding author}
\cortext[cor2]{Co-senior last author}

\begin{abstract}
The Computer-aided Diagnosis or Detection (CAD) approach for skin lesion analysis is an emerging field of research that has the potential to alleviate the burden and cost of skin cancer screening. Researchers have recently indicated increasing interest in developing such CAD systems, with the intention of providing a user-friendly tool to dermatologists to reduce the challenges encountered or associated with manual inspection. \MKH{This article aims to provide a comprehensive literature survey and review of a total of 594 publications (356 for skin lesion segmentation and 238 for skin lesion classification) published between 2011 and 2022.} These articles are analyzed and summarized in a number of different ways to contribute vital information regarding the methods for the development of CAD systems. These ways include: relevant and essential definitions and theories, input data (dataset utilization, preprocessing, augmentations, and fixing imbalance problems), method configuration (techniques, architectures, module frameworks, and losses), training tactics (hyperparameter settings), and evaluation criteria. We intend to investigate a variety of performance-enhancing approaches, including ensemble and post-processing. We also discuss these dimensions to reveal their current trends based on utilization frequencies. In addition, we highlight the primary difficulties associated with evaluating skin lesion segmentation and classification systems using minimal datasets, as well as the potential solutions to these difficulties. Findings, recommendations, and trends are disclosed to inform future research on developing an automated and robust CAD system for skin lesion analysis.
\end{abstract}


\end{frontmatter}

\section*{Nomenclature}
{\footnotesize
\begin{longtable}{ll}
\textbf{\textbf{Acronym}} & \textbf{\textbf{Full Form}} \\ 
\endhead

\endfoot
\endlastfoot
ACC &	Accuracy\\
AdB &	AdaBoost\\
AI &	Artificial Intelligence\\
AK &	Actinic Keratosis\\
ANN &	Artificial Neural Network\\
AUC &	Area Under Curve \\
BACC &	Balanced Accuracy\\
BCC &	Basal Cell Carcinoma \\
BEMD &	Bi-dimensional Empirical Mode Decomposition\\
CAD &	Computer-aided Diagnosis\\
CAM &   Class Activation Map \\
CC &	Correlation Coefficient\\
DF &	Dermatofibroma\\
DL &	Deep Learning \\
DSC &	Dice Similarity Coefficient \\
DT &	Decision Tree\\
EBC &	Ensemble Binary Classifier\\
ENN &	Elman Neural Network \\
F1S &	F1-score \\
FFBPNN &	Feed Forward Back Propagation Neural Network\\
FNR &	False Negative Rate \\
FOM &	Figure of Merit \\
FPR &	False Positive Rate \\
FCN & Fully Convolutional Network \\
GAN &	Generative Adversarial Network\\
GLCM &	Gray Level Co-occurrence Matrix\\
GLDM &	Gray Level Difference Method\\
HcCNN &	Hyper-Connected CNN\\
HD &	Hausdorff  Distance \\
HMD &	Hammoude Distance \\
IAD &	Interactive Atlas of Dermoscopy\\
IRMA &	Image Retrieval in Medical Applications \\
ISIC &	International Skin Imaging Collaboration\\
JI &	Jaccard Index \\
KMC &	K-Means Clustering\\
KNN &	K-Nearest Neighbors \\
LBP &	Local Binary Patterns\\
LDA &	Linear Discriminant Analysis\\
LIME & Local Interpretable Model-agnostic Explanation \\
LR &	Learning Rate \\
MCC &	Matthew Correlation Coefficient \\
Mel &	Melanoma \\
ML &	Machine Learning \\
MPNN &	Multilayer Perceptron Neural Network\\
MSM-CNN &	Multi-Scale Multi-CNN \\
NB &	Naive Bayes \\
Nev &	Nevus\\
NPV &	Negative Predictive Value \\
PNN &	Probabilistic Neural Network\\
PRE &	Precision \\
PRISMA &	Preferred Reporting Items for Systematic Reviews and Meta-Analyses\\
PSNR &	Peak to Signal Ratio \\
QDA &	Quadratic Discriminant Analysis\\
RF &	Random Forest\\
SCC &	Squamous Cell Carcinoma\\
SE &	Segmentation Error \\
SEN &	Sensitivity \\
SENet &	Squeeze-and-Excitation Networks\\
SGD &	Stochastic Gradient Descent \\
SHAP & Shapley Additive Explanations \\
SIFT &	Scale-Invariant Feature Transform\\
SK &	Seborrheic Keratosis\\
SLA &	Skin Lesion Analysis \\
SLC &	Skin Lesion Classification\\
SLS &	Skin Lesion Segmentation \\
SPE &	Specificity\\
SSIM &	Structural Similarity Indices \\
SVM &	Support Vector Machine\\
TDS &	Total Dermatoscopy Score\\
VL &	Vascular Lesion\\
XAI &   Explainable Artificial Intelligence \\
\end{longtable}}

\section{Introduction}
\label{Introduction}
\begin{figure}[!ht]
  \centering
\subfloat[\MKH{Per year articles for lesion segmentation}]{\includegraphics[width=8.3cm, height= 2.7cm]{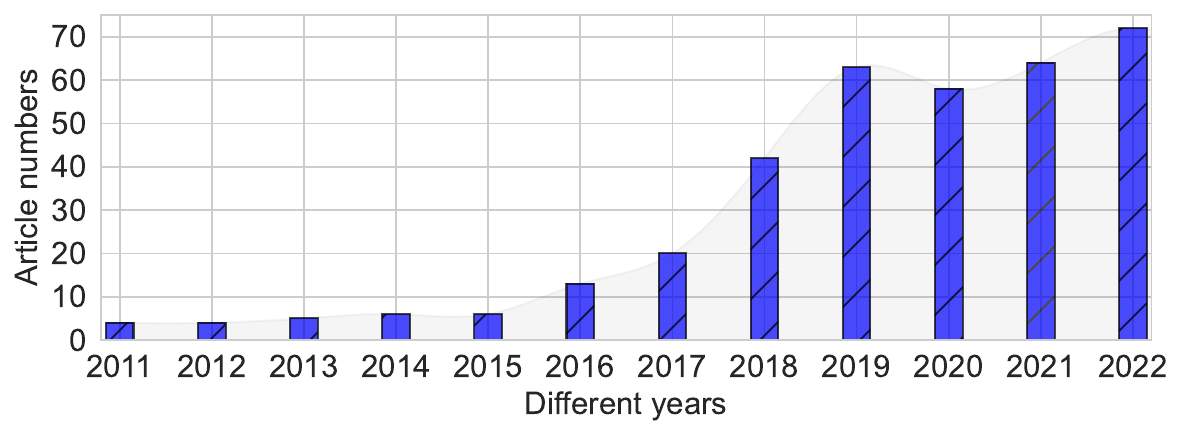}}
\subfloat[\MKH{Per year articles for lesion classification}]{\includegraphics[width=8.3cm, height= 2.7cm]{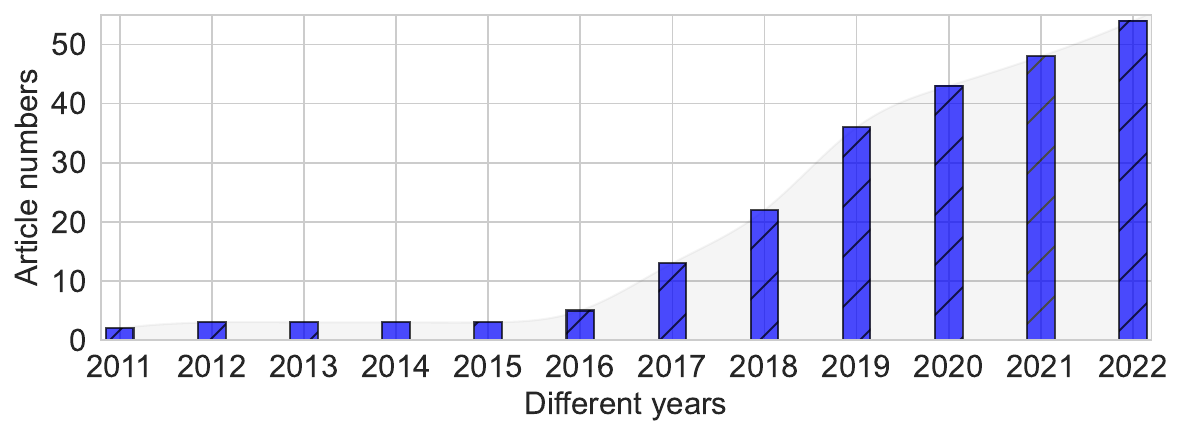}}
\caption{\MKH{The number of publications (in blue bars) in the past from 2011 to 2022 on lesion segmentation (left) and lesion classification (right), collected from Google Scholar (see details in section~\ref{Articles_selection}) by searching for ``skin lesion segmentation" and ``skin lesion classification". The publication numbers in 2020 (left figure) have probably decreased due to the COVID-19 pandemic. This figure shows the growing interest in this field and the complexity of the science involved due to the increasing number of contributions, requiring that these two factors necessitate a review such as ours to help readers navigate the complexity.}}
\label{fig:year_wise_paper}
\end{figure}

Malignant melanoma is the deadliest skin cancer \citep{chatterjee2019integration}, yet early diagnosis can cure it. Doctors and radiologists utilize gold-standard dermatoscopy to diagnose pigmented skin lesions using hand-held instruments and computer vision algorithms. Medical image processing has recently grown with more effective techniques to help dermatologists recognize and classify skin lesions \citep{pereira2020dermoscopic}. Therefore, a Computer-aided Diagnosis (CAD) is an inescapable tool for physicians and/or dermatologists in decision-making, especially when dealing with a large number of patients in a short time \citep{hasan2021dermo, tschandl2020human, tschandl2019expert}. CAD computational techniques comprise image acquisition, preprocessing, segmentation, feature extraction, and classification \citep{oliveira2018computational}. It is noteworthy that some academics consider segmentation a preprocessing step for feature extraction and classification. Segmented lesion masks and classification results can be used for contemporaneous lesion detection and recognition \citep{hasan2021dermo, hasan2021dermoexpert} (see example on YouTube\footnote{\url{https://youtu.be/nlfr_NCPy4U} [Access date: 12-Mar-2022]}). From 2011 to 2022, various studies were conducted on Skin Lesion Analysis (SLA) (see Fig.~\ref{fig:year_wise_paper}), especially lesion segmentation and classification, applying different techniques: computer vision algorithms, manual feature engineering, and automated Artificial Intelligence (AI).
Fig.~\ref{fig:year_wise_paper} remarks that SLA is a fast-growing study area, and the publication numbers have increased yearly for both tasks, especially after 2016. Henceforth, the extensive number of articles necessitates a systematic survey and in-depth discussions of datasets, preprocessing methodologies, segmentation strategies, manual feature engineering procedures, classification techniques, and evaluation benchmarks for automated SLA procedure(s) to help researchers strategize future research.

This article provides a systematic survey, assessment, and analysis of SLA approaches from 2011 through 2022, focusing on Skin Lesion Segmentation (SLS) and Skin Lesion Classification (SLC). SLS schemes employ image analysis (and/or computer vision) algorithms \citep{korotkov2012computerized, oliveira2018computational} or end-to-end Deep Learning (DL) methodologies \citep{jafari2016skin}, while SLC uses Machine Learning (ML) \citep{kassem2021machine} or end-to-end DL algorithms \citep{dildar2021skin, kassem2021machine}. It is noteworthy that the employment of ML-based SLC requires comprehensive feature engineering \citep{oliveira2018computational}. Therefore, we discuss SLA's feature extraction steps in the SLC task. This article summarizes input data (dataset utilization, preprocessing, augmentations, and solving imbalance problems), method configuration (techniques, architectures, module frameworks, and losses), training tactics (hyperparameter settings), and evaluation criteria (metrics) from past publications. It systematically presents their simple technological fundamentals and usage trends with recommendations to help researchers. It can make contributions in the following contexts:
\begin{itemize}
    \item \MKH{Provide essential knowledge regarding SLA tasks by compiling articles published over the past twelve years (2011–2022), covering SLS and SLC mechanisms with a variety of integral strategies and insightful back-and-forth exchanges.}
    
    \item Investigate the insight scenario of a variety of datasets with potential future trends and their existing requirements in the field of survey analysis that is being addressed.
    
    \item Analyze the many different preprocessing and augmentation procedures that are used in SLA in order to reveal their effectiveness, trends, and necessity to construct a robust supervised SLA model and mitigate class imbalance issues. 
    
    \item \MKH{Summarize the many SLS and SLC strategies, such as automated DL algorithms, manual feature engineering, ML models, ensemble models, and SLS post-processing, employed over the past twelve years.}
    
    \item Examine the trends in employment over a range of training conditions, as well as the hyperparameters such as batch size, learning rate, loss function, optimizer, and epoch. 
    
    \item Review a variety of assessment benchmarks along with adequate explanation details to disclose the tendencies in the SLS and/or SLC tasks.
    
    \item \MKH{In the end, categorize all the necessary SLA actions according to the number of times they have been employed in the past 594 articles. These categories are \textbf{\textit{High-frequency}} (mostly applied), \textbf{\textit{Medium-frequency}} (moderately applied), and \textbf{\textit{Low-frequency}} (less applied). Additionally, uncover the prospective essentials in the SLA tasks that will be an open challenge for researchers interested in this field.}
    
\end{itemize}

This paper is organized as follows: Section~\ref{Articles_selection} clarifies the inclusion and exclusion criteria for SLA article selection. Section~\ref{Datasets} describes various available skin lesion datasets and scrutinizes their utilization trends. The employment of various preprocessing and augmentations for the automated SLA task is explored in Section~\ref{Preprocessing_Augmentations} with the class imbalance problem solution analysis. Section~\ref{Segmentation_Techniques} reports different SLS techniques along with past trends. This section also highlights different post-processing schemes in the SLS task. Various SLC elements, like lesion features and ML- and DL-based classifiers, are investigated in Section~\ref{Classification_Techniques}. \MKH{The training schemes with different hyperparameter settings and evaluation benchmarks are described in Section~\ref{Training_and_Evaluation}, while Section~\ref{Explainability_of_SLA_Methods} explores different explainability schemes in the SLA techniques.} Lastly, Section~\ref{Observations_and_Recommendations} provides informative observations, recommendations, and trends for future research directions in related fields of interest, with the article's conclusion in Section~\ref{Conclusion}.

\section{Article Selection}
\label{Articles_selection}
\begin{figure*}[!ht]
  \centering
\includegraphics[width=12cm, height= 6.5cm]{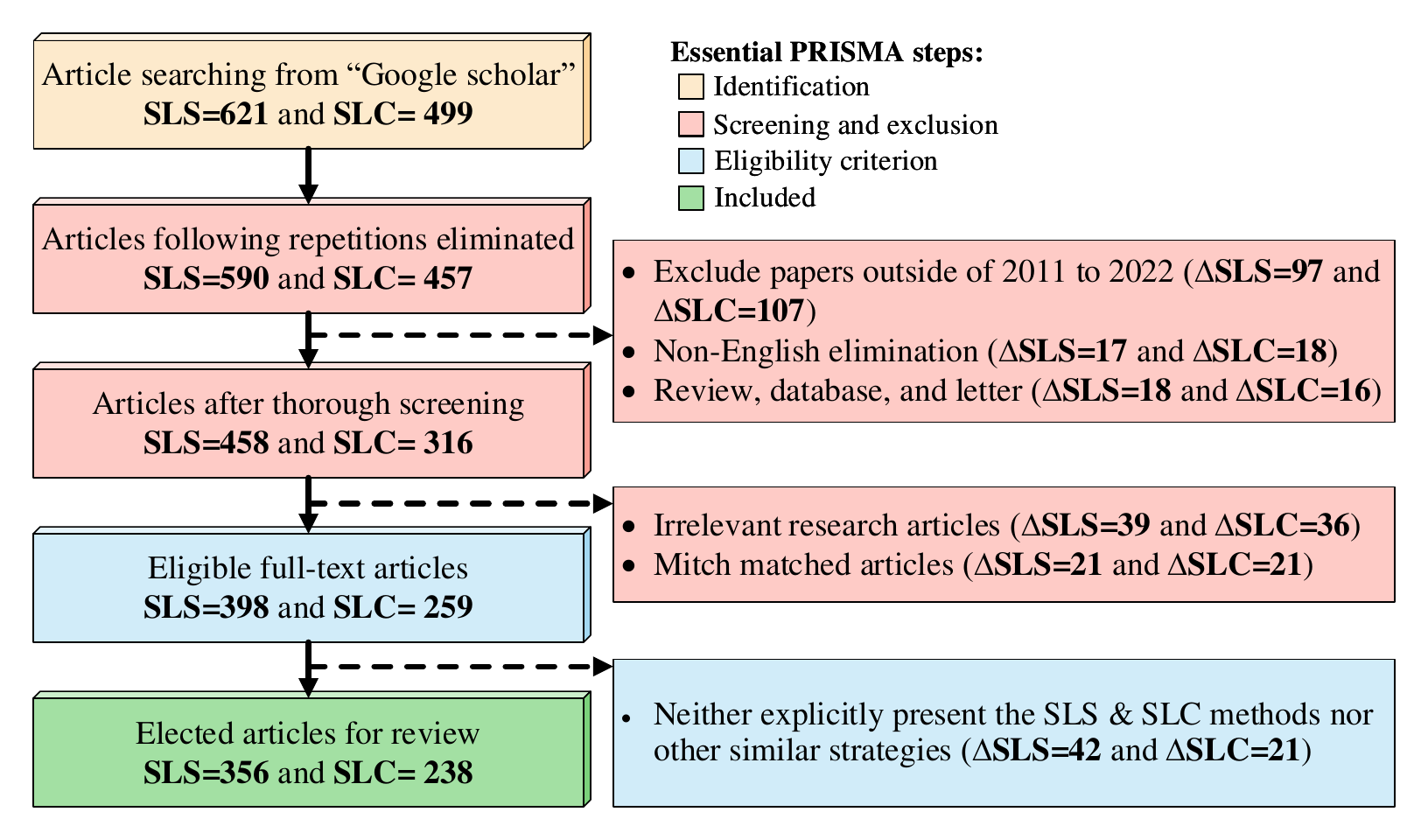}
\caption{\MKH{The employed PRISMA process for articles exploring approach, where we describe the evidence for the article's inclusion and exclusion standards.}}
\label{fig:PRISMA}
\end{figure*}

This review's article-searching approach adopts the Preferred Reporting Items for Systematic Reviews and Meta-Analyses (PRISMA) strategy \citep{islam2021emotion} (see Fig.~\ref{fig:PRISMA}), along with inclusion and exclusion benchmarks \citep{hasan2021missing} for paper selection.
\MKH{A google scholar search yields 621 SLS and 499 SLC publications with the keywords \say{skin lesion segmentation} and \say{skin lesion classification}, respectively. After removing duplicates, there are 590 SLS and 457 SLC articles. Then, in the first screening round (first-level exclusion), non-English articles ($\Delta SLS=17$ and $\Delta SLC=18$) and review, database, or letter-type articles ($\Delta SLS=18$ and $\Delta SLC=16$) are discarded. Only publications from 2011-2022 are kept in this cycle, returning 458 SLS and 316 SLC articles (see Fig.~\ref{fig:PRISMA}). 39 SLS and 36 SLC articles with research objectives that are not relevant to the current review, and 21 SLS and 21 SLC articles with schemes that are not suitable to the current review are excluded. This second exclusion produces 398 SLS and 259 SLC articles (see Fig.~\ref{fig:PRISMA}). Again, $\Delta SLS=42$ and $\Delta SLC=21$ articles are deleted since they do not explicitly illustrate the SLS and SLC systems or other similar strategies. Finally, 356 SLS and 238 SLC articles are included in this review. Fig.~\ref{fig:year_wise_paper} depicts the per-year distribution of those included papers from 2011 to 2022 on the SLS and SLC, revealing that publications have substantially increased, especially after 2016.}

\section{Image Acquisition}
\label{Datasets}
A computerized SLA system acquires images using non-invasive techniques such as dermoscopy, photography, confocal scanning laser microscopy, optical coherence tomography, ultrasound imaging, magnetic resonance imaging, and spectroscopic imaging. However, macroscopic and dermoscopic images are widely utilized for SLA \citep{oliveira2018computational}. Accumulating acquired images in a dataset is key to any automated SLS and/or SLC in the SLA system. Long-used public datasets for SLA are International Skin Imaging Collaboration (ISIC) \citep{gutman2016skin, codella2018skin, codella2019skin, tschandl2018ham10000, combalia2019bcn20000, rotemberg2021patient}, PH2 \citep{mendoncca2013ph}, Interactive Atlas of Dermoscopy (IAD) \citep{argenziano2002dermoscopy}, Image Retrieval in Medical Applications (IRMA) \citep{codella2015deep}, Dermaquest \citep{amelard2014high}, etc. The ISIC datasets include five distinct open versions with various image numbers and classes for five consecutive years: ISIC-16 \citep{gutman2016skin}, ISIC-17 \citep{codella2018skin}, ISIC-18 \citep{codella2019skin, tschandl2018ham10000}, ISIC-19 \citep{combalia2019bcn20000}, and ISIC-20 \citep{rotemberg2021patient} respectively for the years from 2016 to 2020. The next section~\ref{Datasets_explain} briefly summarizes those datasets.

\subsection{Datasets}
\label{Datasets_explain}
The \textbf{ISIC-16}\footnote{\url{https://challenge.isic-archive.com/landing/2016} [Access date: 21-Jun-2022]} is a dataset for binary SLS and SLC containing 900 training and 379 testing images but no validation images (see Table~\ref{tab:data_distribution}). It distinguishes Nevus (Nev) from Melanoma (Mel) using images with resolutions of $556\times 679$ to $2848\times 4828$ pixels. The \textbf{ISIC-17}\footnote{\url{https://challenge.isic-archive.com/landing/2017} [Access date: 21-Jun-2022]} is a binary SLS and multi-class SLC challenge. The latter SLC has Nev, Seborrheic Keratosis (SK), and Mel (see Table~\ref{tab:data_distribution}) classes. This dataset comprises 2750 images for training, validation, and testing with resolutions ranging from $453\times 679$ to $4499\times 6748$ pixels. 
The \textbf{ISIC-18}\footnote{\url{https://challenge.isic-archive.com/landing/2018} [Access date: 21-Jun-2022]} images are derived from the HAM10000 dataset \citep{tschandl2018ham10000}, including 10015 images for multi-class SLC and 12500 images for binary SLS. This dataset does not publish validation or test images, and all images have resolutions of $450\times 600$ pixels. It comprises seven classes: Nev, SK, Basal Cell Carcinoma (BCC), Actinic Keratosis (AK), Dermatofibroma (DF), Vascular Lesion (VL), and Mel (see Table~\ref{tab:data_distribution}).
The \textbf{ISIC-19}\footnote{\url{https://challenge.isic-archive.com/landing/2019} [Access date: 21-Jun-2022]} has an extension of one more class, i.e., Squamous Cell Carcinoma (SCC), in ISIC-18 with almost double sample images. It has 25331 images from multiple sites (HAM10000, BCN20000 \citep{combalia2019bcn20000}, and MSK) applying different preprocessing methods. The images of ISIC-19 have resolutions of $600\times 450$ to $1024\times 1024$ pixels, without explicit validation and test images like ISIC-18 (see Table~\ref{tab:data_distribution}). 
The \textbf{ISIC-20}\footnote{\url{https://www.kaggle.com/c/siim-isic-melanoma-classification/overview} [Access date: 21-Jun-2022]} contains 33126 dermoscopic images with the resolutions of $1024\times 1024$ pixels (see Table~\ref{tab:data_distribution}). Similar to ISIC-16, this dataset also contains binary SLC images from over 2000 patients, aiming to be classified as Nev and Mel. 
\begin{table}[!ht]
\scriptsize
\caption{Publicly available SLS and SLC datasets with variable amounts of classes and image samples.}
\label{tab:data_distribution}
\begin{tabular}{ll|cccccccccc}
\Xhline{1pt}
 & &
 \multicolumn{10}{c}{\textbf{Number of images for various tasks and classes}} \\\cline{3-12}
 
 & & \multicolumn{2}{c|}{\textbf{SLS task}} &
  \multicolumn{8}{c}{\textbf{Class-wise images for SLC task}} \\ \cline{3-12}
\multirow{-3}{*}{\textbf{Datasets}} &
  \multirow{-3}{*}{} &
  \textbf{Images} &
  \multicolumn{1}{c|}{\textbf{Masks}} &
  \textbf{Nev} &
  \textbf{SK} &
  \textbf{BCC} &
  \textbf{AK} &
  \textbf{DF} &
  \textbf{VL} &
  \textbf{Mel} &
  \textbf{SCC}  \\ \Xhline{1pt}
                                                     & Training                    & 900  &  \multicolumn{1}{c|}{900}  & 727  & $-$  & $-$  & $-$  & $-$  & $-$  & 173  &  $-$   \\ 

\multirow{-2}{*}{ISIC-16}      & Testing                     & 379  & \multicolumn{1}{c|}{379}  & 304  & $-$  & $-$  &   &$-$   &$-$   & 75  &  $-$   \\ \hline

                                                    & Training                    & 2000  &  \multicolumn{1}{c|}{2000}  & 1372  & 254  &   $-$&$-$   &$-$   & $-$  & 374  & $-$    \\ 
                                                    & Validation               & 150  & \multicolumn{1}{c|}{150}  &  78 & 42  & $-$  &  $-$ & $-$  &  $-$ & 30  & $-$     \\ 
\multirow{-3}{*}{ISIC-17} &      Testing                     & 600  & \multicolumn{1}{c|}{600}  & 393  &  90 &$-$   &   & $-$  & $-$  & 117  &  $-$   \\ \hline

\multirow{-1}{*}{ISIC-18} & Training & 12500  & \multicolumn{1}{c|}{12500}  & 6705  & 1099  & 514  &  327 & 115  & 142  & 1113  &  $-$   \\ \hline

\multirow{-1}{*}{ISIC-19}        & Training                     & $-$  & \multicolumn{1}{c|}{$-$}  & 12875  &  2624 & 3323  & 867  & 239  & 253  & 4522  & 624    \\ \hline

\multirow{-1}{*}{ISIC-20}        & Training                     & $-$  & \multicolumn{1}{c|}{$-$}  &  32542 & $-$  &  $-$ & $-$  & $-$  & $-$  & 548  & $-$    \\ \hline

\multirow{-1}{*}{PH2}     & $-$                     & 200  & \multicolumn{1}{c|}{200}  & 160  & $-$  & $-$  & $-$  & $-$  & $-$  & 40  & $-$    \\ \hline

\multirow{-1}{*}{IAD}      & $-$                     & 100  & \multicolumn{1}{c|}{100}  & 70  & $-$  & $-$  & $-$  & $-$  & $-$  & 30  & $-$    \\ \hline

\multirow{-1}{*}{Dermaquest}        & $-$                     & 137  & \multicolumn{1}{c|}{137}  & 61  & $-$  & $-$  & $-$  & $-$  & $-$  & 76  & $-$    \\ \hline

\multirow{-1}{*}{IRMA}     & $-$                     & $-$  & \multicolumn{1}{c|}{$-$}  & 560  & $-$  & $-$  & $-$  & $-$  & $-$  & 187  & $-$    \\ \Xhline{1pt}

\end{tabular}
\end{table}
The \textbf{PH2}\footnote{\url{https://www.fc.up.pt/addi/ph2\%20database.html} [Access date: 21-Jun-2022]} is a database acquired at the Dermatology Service of Hospital Pedro Hispano, Portugal. This dataset contains 200 images of melanocytic lesions, including 160 in the Nev class and 40 in the Mel class, with a $768 \times 560$ pixel resolution. The \textbf{IAD}\footnote{\url{https://espace.library.uq.edu.au/view/UQ:229410} [Access date: 21-Jun-2022]} dataset has $700 \times 447$ RGB pixels of images but does not have segmentation masks. However, the authors in \citep{celebi2010robust} provide ground truth masks to continue the research study. The \textbf{Dermaquest} dataset has 137 images for the binary SLS. The SLC is a binary classification task with 76 samples in the Mel class and 61 in the Nev class. The \textbf{IRMA} dataset is unlisted but available under special request to the authors and was created by the Department of Medical Informatics, RWTH Aachen University. It comprises 747 dermoscopic images with a resolution of $512 \times 512$ pixels, of which 187 images are in the Mel class, and 560 are in the Nev class.

An overview of these datasets with their sample distributions for SLS and SLC tasks in Table~\ref{tab:data_distribution} reveals that the SLC class ranges from binary to eight classes, with most binary tasks. Per-class sample figures indicate a vast disparity, especially for ISIC-19 and ISIC-20. Such an uneven class distribution causes a supervised SLC system to favor the overrepresented class if this is not considered during training \citep{hasan2021dermo}. It is very challenging for automated SLS and SLC systems to be generic while training with fewer samples. \MKH{Section~\ref{Data_scarcity_and_imbalance_problem} explores different ways in the past literature, ranging from 2011 to 2022, to mitigate these challenges. The following section~\ref{Datasets_utilization_and_trends} looks at the frequency of dataset usage in SLA literature from 2011 to 2022.}

\subsection{Datasets' Utilization and Trends}
\label{Datasets_utilization_and_trends}
\MKH{Table~\ref{tab:dataset_wise_papers_seg} illustrates the usage of different datasets in 356 SLS articles.} Some articles are assigned to multiple columns of datasets because the authors used multiple datasets in their papers. It can be observed that researchers employed publicly available IAD and other private datasets (in the last column) to validate image analysis (and/or computer vision)-based SLS methods between 2011 and 2015, when automated DL models were still not popular for the SLS task. Recalling Fig.~\ref{fig:year_wise_paper}, the publication numbers after 2016 have significantly increased; one of the underlying reasons is the release of ISIC datasets, as reflected in Table~\ref{tab:dataset_wise_papers_seg} (most of the articles are in the first four columns after 2016). \MKH{A similar pattern of the usage frequency of various SLC datasets in the 238 SLC articles is noticed in Table~\ref{tab:dataset_wise_papers_class}.}


{\tiny
\begin{longtable}[c]{l|p{1.6cm}|p{2cm}|p{2cm}|p{2.2cm}|p{1.6cm}|p{1.6cm}|p{1.4cm}}
\caption{\MKH{The number of SLS articles each year on various publicly available datasets in Table~\ref{tab:data_distribution}.}}
\label{tab:dataset_wise_papers_seg}\\ \Xhline{1pt}
\textbf{Year} &
  \textbf{ISIC-16} &
  \textbf{ISIC-17} &
  \textbf{ISIC-18} &
  \textbf{PH2} &
  \textbf{IAD} &
  \textbf{Dermaquest} &
  \textbf{Others} \\ \Xhline{1pt}
\endfirsthead
\multicolumn{8}{c}%
{{\bfseries Table \thetable\ Continued from previous page}} \\ \Xhline{1pt}
\textbf{Year} &
  \textbf{ISIC-16} &
  \textbf{ISIC-17} &
  \textbf{ISIC-18} &
  \textbf{PH2} &
  \textbf{IAD} &
  \textbf{Dermaquest} &
  \textbf{Others} \\ \Xhline{1pt}
\endhead


2022 & \citep{sun2022acfnet, gu2022net, feng2022slt, le2022antialiasing, feng2022bla, singh2022empirical, song2022res, shamsolmoali2022salient, khouloud2022w, khan2022skin, dong2022tc, hafhouf2022improved, dai2022ms, csahin2022robust, kazaj2022u, kaur2022skin, kaur2022automatic, cao2022icl, wu2022fat, wang2022cascaded, malik2022hybrid, malik2022novel}
& \citep{tran2022fully, thapar2022novel, sun2022acfnet, liu2022ncrnet, li2022mhau, gu2022net, feng2022slt, alahmadi2022semi, alahmadi2022multiscale, al2022weakly, le2022antialiasing, singh2022empirical, song2022res, stofa2022u, aghdam2022attention, fan2022egfnet, shu4141765active, shamsolmoali2022salient, ruan2022malunet, ren2022serial, khouloud2022w, khan2022skin, khan2022ensemble, dong2022tc, barin2022automatic, hafhouf2022improved, dai2022ms, csahin2022robust, ramadan2022dgcu, ramadan2022cu, kaur2022skin, kaur2022automatic, chen2022skin, zhang2022dynamic, zhang2022dense, wu2022fat, pennisi2022skin, jiang2022seacu, hu2022net, basak2022mfsnet, barzegar2022skin, wang2022skin, wang2022cascaded, malik2022hybrid, malik2022novel, liu2022skin}
& \citep{thapar2022novel, lin2022quality, li2022mhau, gulzar2022skin, feng2022slt, alahmadi2022semi, alahmadi2022multiscale, kosgiker2022significant, feng2022bla, singh2022empirical, aghdam2022attention, ahmed2022new, akyel2022linknet, shamsolmoali2022salient, ruan2022malunet, khouloud2022w, khan2022skin, dong2022tc, alhudhaif2022novel, anand2022fusion, barin2022automatic, han2022hwa, dong2022learning, dai2022ms, zuo2022efficient, zhou2022superpixel, rehman2022machine, ramadan2022dgcu, ramadan2022cu, kaur2022skin, kaur2022automatic, cao2022icl, zhang2022dynamic, wu2022fat, nour2022skin, joseph2022preprocessing, jiang2022seacu, hu2022net, basak2022mfsnet, wang2022cascaded, malik2022hybrid, malik2022novel, liu2022skin}
& \citep{tran2022fully, thapar2022novel, gu2022net, alahmadi2022semi, alahmadi2022multiscale, al2022weakly,le2022antialiasing, singh2022empirical, song2022res, aghdam2022attention, akyel2022linknet, shamsolmoali2022salient, khouloud2022w, khan2022ensemble, hafhouf2022improved, dong2022learning, dai2022ms, zhou2022superpixel, rehman2022machine, ramadan2022dgcu, ramadan2022cu, kaur2022skin, chen2022skin, cao2022icl, zhang2022dense, wu2022fat, joseph2022preprocessing, hu2022net, basak2022mfsnet, wang2022cross, wang2022cascaded, malik2022hybrid, malik2022novel}
&
& 
& \citep{li2022mhau, lee2022progressive, zhao2022self, mehmood2022k, bhakta2022tsalli, wang2022cross} \\ \hline

2021 & \citep{khan2021skin, wang2021knowledge, phan2021skin, jiang2021approximated, chauhan2021multi, wang2021boundary, tong2021ascu, dong2021fac, tang2021afln}
& \citep{saini2021b, kosgiker2021segcaps, khan2021skin, abhishek2021matthews, tao2021attention, adegun2021probabilistic, ding2021efficient, kaur2021deep, chowdary2021automated, xiao2021prior, jin2021cascade, jignesh2021automated, chen2021mt, wibowo2021lightweight, wang2021knowledge, wang2021focal, phan2021skin, hussain2021recu, chauhan2021multi, bagheri2021skin_2, bagheri2021skin, tong2021ascu, mirikharaji2021d, liu2021skin, garg2021skin, dong2021fac, le2021modified, tang2021afln, araujo2021convolutional, gangwar2021study}
& \citep{yacin2021deep, saini2021b, ali2021skin, khan2021skin, chowdary2021automated, xie2021semi, reddy2021handling, qamar2021dense, jin2021cascade, jignesh2021automated, jiang2021residual, yang2021deep, wibowo2021lightweight, mu2021channel, jiang2021approximated, hussain2021recu, wang2021boundary, arora2021automated, dong2021fac, li2021digital, tang2021afln, araujo2021convolutional, arora2021skin}
& \citep{yacin2021deep, singh2021slicaco, saini2021b, krishna2021mlrnet, kosgiker2021segcaps, khan2021skin, abhishek2021matthews, tao2021attention, adegun2021probabilistic, ding2021efficient, kaur2021deep, dayananda2021skin, das2021skin, xiao2021prior, reddy2021handling, jiang2021residual, yang2021deep, chen2021mt, wibowo2021lightweight, jiang2021approximated, imtiaz2021efficient, bagheri2021skin_2, wang2021boundary, tong2021ascu, mirikharaji2021d, garg2021skin, gajera2021improving}
&
& \citep{bagheri2021skin_2, mirikharaji2021d}
& \citep{peter2021internet, filali2021graph}  \\ \hline

2020           & \citep{osadebey2020evaluation, santos2020skin, huang2020skin, khan2020frequency, sanjar2020improved, wang2020cascaded, hafhouf2020modified, salih2020skin, nathan2020lesion, ramella2020automatic, tang2020imscgnet, jayapriya2020hybrid, hawas2020oce, ribeiro2020less, al2020automatic, pour2020transform, xie2020skin, hasan2021dermoexpert, hasan2021dermo}
& \citep{rout2020transition, deng2020weakly, qin2020asymmetric, anjum2020deep, santos2020skin, nampalle2020efficient, wang2020cascaded, nathan2020lesion, abhishek2020illumination, thanh2020adaptive, wei2020attentive, tang2020imscgnet, jayapriya2020hybrid, ribeiro2020less, jiang2020skin, qiu2020inferring, pour2020transform, kaymak2020skin, wu2020automated, hasan2020dsnet, azad2020attention, zafar2020skin, ozturk2020skin, xie2020skin, xie2020mutual, mahbod2020effects, shan2020automatic, lei2020skin, hasan2021dermoexpert, hasan2021dermo}
& \citep{anjum2020deep, setiawan2020image, kamalakannan2020self, wang2020cascaded, wang2020donet, nathan2020lesion, saha2020leveraging, zhu2020asnet, tang2020imscgnet, zhang2020kappa, salih2020skin, ribeiro2020less, wu2020automated, azad2020attention, lei2020skin, hasan2021dermoexpert}
& \citep{osadebey2020evaluation, deng2020weakly, iranpoor2020skin, hajabdollahi2020simplification, nampalle2020efficient, khan2020frequency, wang2020cascaded, wang2020donet, salih2020skin, justin2020skin, pereira2020dermoscopic, nathan2020lesion, abhishek2020illumination, sayed2020novel, tang2020imscgnet, rizzi2020skin, al2020automatic, jiang2020skin, qiu2020inferring, hasan2020dsnet, azad2020attention, zafar2020skin, ozturk2020skin, xie2020skin, xie2020mutual, shan2020automatic}
& \citep{pereira2020dermoscopic}
& \citep{bansal2020improved, parida2020transition, hajabdollahi2020simplification, pillay2020macroscopic}
& \citep{sivaraj2020detecting, ganesan2020hsl, liu2020multi, low2020automating}

\\ \hline

2019           
& \citep{agilandeeswari2019skin, javed2019intelligent, khan2019construction, tang2019multi, singh2019fca, wang2019bi, huang2019skin, seeja2019deep, khan2019skin, zhang2019automatic, baghersalimi2019dermonet, tang2019efficient}
& \citep{thanh2019adaptive, hasan2019comparative, saini2019detector, rawas2019hcet, alfaro2019brief, al2019deep, soudani2019image, javed2019intelligent, goyal2019skin, wang2019automated, ma2019light, liu2019enhanced, de2019skin, song2019dense, jiang2019decision, adegun2019deep, sarker2021slsnet, bi2019improving, ninh2019skin, ribeiro2019handling, singh2019fca, wang2019bi, liu2019skin, thanh2019skin, tu2019dense, khan2019skin, zhang2019automatic, unver2019skin, bisla2019skin, baghersalimi2019dermonet, tschandl2019domain, wei2019attention, tang2019efficient}
& \citep{shan2019improving, saini2019detector, salih2019skin, cui2019ensemble, wang2019automated, shahin2019deep, ali2019supervised, wu2019skin, sarker2021slsnet, ali2019skin, lameski2019skin, ribeiro2019handling, singh2019fca, hasan2019skin, khan2019skin, bisla2019skin, canalini2019skin}
& \citep{xie2020mutual, aljanabi2019various, ooi2019interactive, bingol2019entropy, brahmbhatt2019skin, saini2019detector, abdullah2019deep, rawas2019hcet, salih2019skin, javed2019intelligent, khan2019construction, yang2019sampling, liu2019enhanced, tu2019dense, unver2019skin, bisla2019skin, baghersalimi2019dermonet, goyal2019skin, tang2019efficient}
& \citep{filali2019improved} 
& \citep{filali2019multi}  
& \citep{sengupta2019segmentation, bhakta2019tsalli, dash2019pslsnet} 
\\\hline

2018           
& \citep{riaz2018active, burdick2018rethinking, he2018dense, kolekar2018skin, olugbara2018segmentation, akram2018skin, khan2018implementation, li2018dense, aljanabi2018skin, li2018skin} 
& \citep{vesal2018multi, li2018semi, vesal2018skinnet, aljanabi2018skin, mirikharaji2018deep, venkatesh2018deep, yuan2017automatic, jaisakthi2018automated, he2018dense, zeng2018multi, li2018skin, li2018dense, chen2018multi, nguyen2018isic, ammar2018learning, youssef2018deep, khan2018implementation, ross2018effects, patino2018automatic, navarro2018accurate} 
& \citep{jiang2018skin, guth2018skin, xu2018automatic, wang2018skin, bi2018improving, qian2018detection, bissoto2018deep} 
& \citep{riaz2018active, ammar2018learning, olugbara2018segmentation, youssef2018deep, louhichi2018skin, khan2018implementation, hu2018skin, vesal2018multi, nasir2018improved, patino2018automatic, jaisakthi2018automated, aljanabi2018skin, akram2018skin}  
&   
& \citep{chakkaravarthy2018automatic, luo2018fast} 
& \citep{ahmed2018segmentation, salih2018skin} \\\hline

2017           & \citep{lynn2017segmentation, pour2017automated, bozorgtabar2017investigating, he2017skin, yuan2017automatic, bozorgtabar2017skin, bi2017semi}
& \citep{ramachandram2017skin, he2017skin, alvarez2017k, qi2017global, jaisakthi2017automatic, yuan2017automatic, mishra2017deep, lin2017skin} 
&    
& \citep{pardo2017automated, yuan2017automatic} 
&  
& \citep{nasr2017dense, martinez2017pigmented, gupta2017adaptive, agarwal2017automated} 
& \citep{george2017automatic} \\\hline

2016           
& \citep{bozorgtabar2016sparse, majtner2016improving, hassan2016skin, kasmi2016biologically}   
&  
& 
&  
&   \citep{khalid2016segmentation, joseph2016skin, bozorgtabar2016sparse, bi2016automated, pennisi2016skin}  
& \citep{bi2016automated, sagar2016color, jafari2016skin}
&   \citep{ortega2016statistical, azehoun2016novel, azmi2016abcd}  \\ \hline

2015           &  &  &      & \citep{torkashvand2015automatic} &    &  &   \citep{rashid2015novel, pereira2015adaptive, trabelsi2015skin, yasmin2015improved, khattak2015maximum} \\ \hline

2014           &  &  &  &     &  &  &     \citep{abbas2014combined, al2014automatic, lezoray2014graph, ch2014two, jyothilakshmi2014detection, masood2014integrating} \\ \hline

2013           &  &  &      &  & \citep{amelio2013skin} &   &      \citep{nisar2013color, wu2013automatic, abbas2013improved} \\ \hline

2012           &  &  &      &  & \citep{khakabi2012multi, madooei2012automated}  &      & \citep{ivanovici2012color, he2012automatic} \\ \hline

2011           &  &  &      &  & \citep{schaefer2011colour, zhou2011gradient} &      & \citep{wong2011automatic, li2011estimating} \\ \Xhline{1pt}

\textbf{Total} & \textbf{83 (15.4\,\%)}  & \textbf{166 (30.7\,\%)} & \textbf{103 (19.0\,\%)}   &  \textbf{121 (22.4\,\%)}   & \textbf{12 (2.2\,\%)}   & \textbf{16 (3.0\,\%)} & \textbf{39 (7.3\,\%)} \\ \Xhline{1pt}
\end{longtable}}

{\tiny
\begin{longtable}[c]{l|p{1.4cm}|p{1.7cm}|p{1.8cm}|p{1.45cm}|p{1.5cm}|p{1.3cm}|p{1.4cm}|p{1.44cm}}
\caption{\MKH{The number of SLC articles each year on various publicly available datasets in Table~\ref{tab:data_distribution}.}}
\label{tab:dataset_wise_papers_class}\\ \Xhline{1pt}
\textbf{Year} &
  \textbf{ISIC-16} &
  \textbf{ISIC-17} &
  \textbf{ISIC-18} &
  \textbf{ISIC-19} &
  \textbf{PH2} &
  \textbf{IAD} &
  \textbf{Dermaquest} &
  \textbf{Others} \\ \Xhline{1pt}
\endfirsthead
\multicolumn{8}{c}%
{{\bfseries Table \thetable\ Continued from previous page}} \\ \Xhline{1pt}
\textbf{Year} &
  \textbf{ISIC-16} &
  \textbf{ISIC-17} &
  \textbf{ISIC-18} &
  \textbf{ISIC-19} &
  \textbf{PH2} &
  \textbf{IAD} &
  \textbf{Dermaquest} &
  \textbf{Others} \\ \Xhline{1pt}
\endhead


2022 & \citep{khan2022skin, hosny2022refined, khouloud2022w, afza2022hierarchical, he2022deep}
& \citep{khan2022skin, ngo2022skin, shen2022low, thapar2022novel, batista2022classification, hosny2022refined, khouloud2022w, nakai2022dpe, wei2022dual, wan2022mslanet, deng2022efficient, nakai2022enhanced, he2022deep, khan2022ensemble, shan2022automatic}
& \citep{yue2022towards, afza2022multiclass, hoang2022multiclass, khan2022skin, popescu2022skin, thapar2022novel, aldhyani2022multi, batista2022classification, foahom2022end, hosny2022refined, khouloud2022w, nakai2022dpe, nguyen2022skin, samsudin2022skin, shetty2022skin, wei2022dual, afza2022hierarchical, alptekin2022analysis, anand2023fusion, bozkurt2022skin, deng2022efficient, nakai2022enhanced, nancy2022impact, mohanty2022integrated, sarker2022transslc, shan2022automatic, somfai2022handling, qian2022skin}
& \citep{yue2022towards, ayas2022multiclass, chabi2022towards, hoang2022multiclass, ngo2022skin, wang2022ssd, zhuang2022cs, wei2022dual, nigar2022deep, dong2022learning, hsu2022hierarchy, somfai2022handling}
& \citep{thapar2022novel, hosny2022refined, khouloud2022w, rasel2022convolutional, afza2022hierarchical, he2022deep, khan2022ensemble, shan2022automatic, somfai2022handling}
&
& \citep{hosny2022refined}
& \citep{serrano2022clinically, sahin2022human, hosny2022refined, pundhir2022towards, wang2022adversarial, tang2022fusionm4net, yilmaz2022mobileskin, camacho2022multi, piatek2022analysis, lihacova2022multi, somfai2022handling} \\ \hline

2021 & \citep{ding2021deep, nersisson2021dermoscopic, khan2021skin, balabantaray2021melanoma, khan2021pixels, mahbod2021investigating}
& \citep{zanddizari2021new, ding2021deep, bian2021skin, wang2021multi, khan2021skin, carvalho2021multimodal, khan2021pixels, mahbod2021investigating}
& \citep{wang2021unlabeled, redha2021skin, liu2021multiscale, bayasi2021culprit, yao2021single, shetty2022skin, reddy2021handling, melbin2021integration, jain2021deep, bansal2021skin, yacin2021deep, samanta2021skin, rahman2021approach, khan2021skin, reimers2021conditional, thurnhofer2021skin, hu2021toporesnet, calderon2021bilsk, arshad2021computer, sevli2021deep}
& \citep{zanddizari2021new, sun2021skin, santos2021transfer, yao2021single, cullell2021convolutional, bdair2021semi, tian2021mixed, shahabi2021performance, rahman2021approach, bdair2021fedperl, xiao2021boosting}
& \citep{sun2021skin, moldovanu2021skin, bayasi2021culprit, samia2021skin, reddy2021handling, melbin2021integration, khan2021skin, imtiaz2021efficient, khan2021pixels}
& 
& \citep{bayasi2021culprit, bansal2021skin}
& \citep{pereira2021skin_2, bayasi2021culprit, melbin2021integration, krohling2021smartphone, bdair2021semi, bansal2021skin, mukherjee2021transfer, peter2021internet} \\ \hline

2020          
& \citep{wu2020skin, wei2020automatic, yilmaz2020benign, rodrigues2020new, kwasigroch2020neural, kamalakannan2020self}
& \citep{dhivyaa2020skin, mahbod2020effects, wu2020skin, mahbod2020effects, yilmaz2020benign, kwasigroch2020self, afza2020skin, akram2020multilevel, xie2020mutual, kamalakannan2020self, anjum2020deep, yildirim2020pre}
& \citep{dhivyaa2020skin, rieger2020interpretations, liu2020automatic, wu2020multi, mporas2020color, sun2020comparative, salian2020skin, chaturvedi2020skin, miglani2020skin, mahbod2020transfer, yilmaz2020benign, rahman2020transfer, qin2020gan, guha2020performance, harangi2020assisted, hassan2020skin, kamalakannan2020self, jibhakate2020skin, anjum2020deep, almaraz2020melanoma}
& \citep{zhuang2020cs, molina2020classification, ahmed2020skin, muckatira2020properties, nunnari2020study, bagchi2020learning, anjum2020deep}
& \citep{gessert2020skin, salian2020skin, rodrigues2020new, afza2020skin, akram2020multilevel, xie2020mutual}
& \citep{ghalejoogh2020hierarchical, bi2020multi}
& \citep{pereira2020skin}
& \citep{damian2020feature, thomsen2020deep, goceri2020comparative, yan2020scalable}
\\ \hline

2019           
& \citep{filali2019texture, wang2019mutual, khan2019skin, khan2019multi, khan2019construction, chatterjee2019integration}
& \citep{veltmeijer2019integrating, tschandl2019diagnostic, xue2019robust, zheng2019relation, wang2019automated, mahbod2019skin, mahbod2019fusing, kulhalli2019hierarchical, khan2019skin, khan2019multi, chatterjee2019integration, bisla2019skin, tschandl2019domain, serte2019wavelet, serte2019gabor, al2019deep, albahar2019skin, sadeghi2020using}
& \citep{gessert2019skin, milton2019automated, rashid2019skin, yoon2019generalizable, wang2019automated, kulhalli2019hierarchical, khan2019skin, khan2019multi, chung2019toporesnet, chatterjee2019integration, bisla2019skin, van2019quantifying, kassani2019depthwise, ali2019skin, aldwgeri2019ensemble}
& \citep{eddine2019skin, guissous2019skin}
& \citep{yoon2019generalizable, khan2019construction, chatterjee2019integration, aljanabi2019various}
& \citep{filali2019improved, chatterjee2019integration} 
& \citep{yoon2019generalizable}  
& \citep{fisher2019classification, monisha2019classification, yoon2019generalizable}\\\hline

2018           
& \citep{zhang2018skin, ur2018classification, wahba2018novel, khan2018implementation} 
& \citep{pham2018deep, thandiackal2018structure, khan2018implementation, harangi2018skin, chen2018multi, filali2018study, sadeghi2018users} 
& \citep{thandiackal2018structure, xie2018multi, li2018skinlesion, lee2018wonderm, kitada2018skin, namozov2018adaptive, bissoto2018deep, pan2018residual} 
&  
& \citep{pham2018deep, nasir2018improved, dos2018robust}  
&  
& \citep{liao2018deep} 
& \citep{songpan2018improved, yap2018multimodal, van2018visualizing, navarro2018webly} \\\hline

2017           
& \citep{wahba2017combined, satheesha2017melanoma, lynn2017segmentation, lopez2017skin}
& \citep{murphree2017transfer, mirunalini2017deep, jia2017skin, devries2017skin, danpakdee2017classification} 
& 
&  
& \citep{satheesha2017melanoma} 
&  \citep{satheesha2017melanoma} 
& \citep{mahdiraji2017skin, lopez2017skin, haofu2017deep} 
& \citep{filali2017multiscale, al2017classification} \\\hline

2016           
& \citep{majtner2016combining}   
& 
& 
& 
& \citep{chakravorty2016dermatologist} 
& \citep{liao2016skin}  
& \citep{liao2016skin, pomponiu2016deepmole}
& \citep{farooq2016automatic, liao2016skin}  \\ \hline

2015       
& 

&

& 

&  

& 

&  

& 
& \citep{di2015hierarchical, mahmoud2014hybrid, sirakov2015skin} \\ \hline

2014      
& 

&

& 

& 

&

& \citep{celebi2014automated} 
& 
& \citep{jamil2014comparative, surowka2014optimal} \\ \hline

2013    
& 

& 

&   

& 

& 

& 

&  

& \citep{she2013lesion, rosado2013prototype, cavalcanti2013macroscopic} \\ \hline

2012        
& 

&  

&  

& 

& 

& \citep{mete2012skin}
& \citep{amelard2012extracting}    
& \citep{ballerini2012non} \\ \hline

2011    
& 

&  

& 

& 

& 

& 

&  

& \citep{she2011skin, ramlakhan2011mobile} \\ \Xhline{1pt}

\textbf{Total} & \textbf{31 (9.7\,\%)} & \textbf{62 (19.5\,\%)} & \textbf{90 (28.3\,\%)} & \textbf{32 (10.0\,\%)} & \textbf{33 (10.4\,\%)}  & \textbf{13 (4.1\,\%)}   & \textbf{12 (3.8\,\%)} &  \textbf{45 (14.2\,\%)} \\ \Xhline{1pt}
\end{longtable}}

\begin{figure}[!ht]
  \centering
\subfloat[\MKH{Per year SLS articles using a specific dataset}]{\includegraphics[width=15cm, height= 2.6cm]{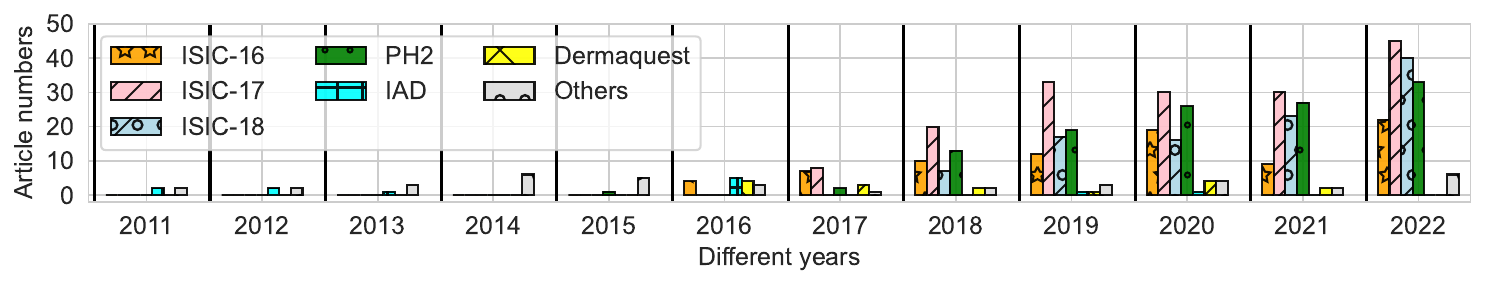}} \\ 
\subfloat[\MKH{Per year SLC articles using a specific dataset}]{\includegraphics[width=15cm, height= 2.6cm]{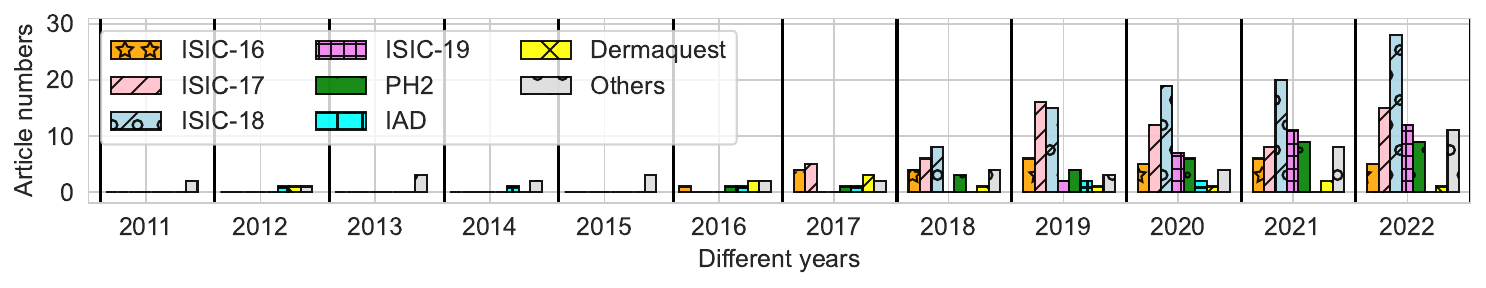}}
\caption{\MKH{The summary of the dataset's utilization from 2011 to 2022 for the SLS (top) and SLC (bottom) on different publicly available datasets, demonstrating the most valuable datasets in the past years.}}
\label{fig:data_used_per_year}
\end{figure}

In Fig.~\ref{fig:data_used_per_year}, a summary of various dataset utilization in the specified 594 articles is presented, showing the frequencies of dataset applications. As illustrated in both images, the earlier articles applied relatively small and non-public datasets to their SLA system, limiting the research community's ability to reproduce their findings.
As obtaining annotated medical images is extremely expensive \citep{harangi2018skin, hasan2021dermoexpert}, few researchers were interested in working on the SLA until 2015. However, this obstacle was overcome to some extent after introducing the ISIC datasets in 2016, as reflected in Fig.~\ref{fig:data_used_per_year} (a) and (b). 
Despite being introduced after the PH2, IAD, Dermaquest, and other earlier private datasets, the ISICs datasets were extensively used in numerous articles, as seen in Fig.~\ref{fig:data_used_per_year} and the last rows of Table~\ref{tab:dataset_wise_papers_seg} and Table~\ref{tab:dataset_wise_papers_class}. 

\MKH{Although the ISIC-18 has more samples than other SLS datasets, it was not used like as ISIC-17 dataset. This trend has been noticed in the last six years (2017-2022) (see Fig.~\ref{fig:data_used_per_year} (a)). Since ISIC-18 lacks independent test and validation sets, it necessitates time-consuming cross-validation to describe the results. A robustly trained SLS model may only sometimes be produced by random validation and test set selection. On the other hand, the ISIC-16 for SLS also has fewer training examples and no validation images. Significantly few articles \citep{camacho2022multi, krishna2021mlrnet, das2021skin, xiao2021prior} used the ISIC-19 SLS dataset in the past. Therefore, the choice of ISIC-17 to build a robust SLS model is preferred by researchers, as this trend is reflected in the last twelve years' scenario.} Again, Table~\ref{tab:dataset_wise_papers_class} and Fig.~\ref{fig:data_used_per_year} (b) demonstrate that the authors have favored datasets with more class numbers for constructing an SLC model, which led to an increased usage of the ISIC-18 dataset. The SLC models are challenged by the increased class number and unbalanced class sample distribution, and researchers have attempted to create trustworthy SLC models that are free from specific class bias \citep{hasan2021dermo, hasan2021dermoexpert}.

It is also noticed from Fig.~\ref{fig:data_used_per_year} that the PH2, Dermaquest, IAD, and other private datasets were also used in some articles even after 2016, although they have fewer sample images. The discernible reason is that some authors trained and evaluated their models on ISIC \citep{hasan2020dsnet, al2018skin}, where they further evaluated their SLS and SLC models on those datasets to uncover guaranteed robustness, as they were neither utilized during the training nor validation phases. \MKH{However, among PH2, Dermaquest, IAD, and other private datasets, PH2 was most commonly applied in the past literature. Thus, our review confirms that PH2 has been commonly employed as an external dataset in the last twelve years to validate the results obtained from the ISIC training.}

\section{Preprocessing and Augmentations}
\label{Preprocessing_Augmentations}
Preprocessing seeks to improve image data quality, suppress distortions, or enhance additional processing features, while augmentation creates new training examples from the existing data distribution. Identifying those optimal procedures requires a basic understanding of the problem, data collection, and system environment. This is a complex task because procedures work satisfactorily in some circumstances but not in others. Therefore, this article summarizes the applied preprocessing (in section~\ref{Preprocessing}) and augmentation (in section~\ref{Data_scarcity_and_imbalance_problem}) techniques. The approaches to mitigating the class bias problem (addressed in section~\ref{Datasets_explain}) are also surveyed in section~\ref{Data_scarcity_and_imbalance_problem}.

\subsection{Preprocessing}
\label{Preprocessing}
Image preprocessing eliminates noise and enhances the quality of the original image by removing uncorrelated information and surplus background portions for further processing. An appropriate selection of preprocessing approaches can considerably increase the accuracy of the intended system \citep{tushar2019brain, hasan2021dermo, hasan2021dermoexpert, joseph2022preprocessing}. Examining selected articles reveals that skin lesion images contain many types of noise and artifacts: markers, body hairs \& veins, body fibers, air bubbles, reflections, non-uniform lighting, rolling lines, shadows, non-uniform vignetting, artificial landmarks, and patient-specific effects like lesion textures, colors, diverse shape \& size of lesion area \citep{hasan2020dsnet}. \MKH{Table~\ref{tab:Preprocessing} highlights all the preprocessing methods employed for SLA in the past twelve years with their short descriptions and corresponding articles.}

{\scriptsize
\begin{longtable}[c]{p{2cm}|p{8.6cm}p{4.6cm}}
\caption{\MKH{Commonly employed preprocessing in the last twelve years, scrutinizing a total of 594 articles: 356 for SLS and 238 for SLC. }}
\label{tab:Preprocessing}\\
\Xhline{1pt}

\textbf{Methods} & \textbf{Remarks} & \textbf{Employed articles} \\ \Xhline{1pt}
\endfirsthead
\multicolumn{3}{c}%
{{\bfseries Table \thetable\ Continued from previous page}} \\
\Xhline{1pt}
\textbf{Methods} & \textbf{Remarks} & \textbf{Employed articles} \\ \Xhline{1pt}
\endhead
Hair removal      &    The lesion's boundary and texture information are often occluded due to the presence of hair, leading to over-segmentation and weak pattern analysis \citep{abbas2011hair}. Therefore, an automatic hair removal method \citep{abbas2011hair} is necessitated, preserving all the lesion features.       &  \citep{wahba2017combined, lynn2017segmentation, jia2017skin, al2017classification, farooq2016automatic, mahmoud2014hybrid, cavalcanti2013macroscopic, wahba2018novel, lee2018wonderm, filali2018study, chatterjee2019integration, ghalejoogh2020hierarchical, kamalakannan2020self, almaraz2020melanoma, mporas2020color, madooei2012automated, al2014automatic, lezoray2014graph, sagar2016color, pennisi2016skin, majtner2016improving, khalid2016segmentation, jafari2016skin, hassan2016skin, jaisakthi2017automatic, alvarez2017k, riaz2018active, ross2018effects, olugbara2018segmentation, nasir2018improved, jaisakthi2018automated, akram2018skin, ammar2018learning, khan2018implementation, sengupta2019segmentation, yang2019sampling, hasan2019skin, unver2019skin, hasan2019comparative, bingol2019entropy, sivaraj2020detecting, justin2020skin, santos2020skin, zafar2020skin, khan2020frequency, moldovanu2021skin, rahman2021approach, yacin2021deep, al2022weakly, basak2022mfsnet, csahin2022robust, fawzy2022high, kosgiker2022significant, singh2022empirical, thapar2022novel, garg2021skin, imtiaz2021efficient, kosgiker2021segcaps, li2021digital, singh2021slicaco}         \\ \hline

Normalization    &  Subtraction of mean RGB values computed over each image or whole training dataset to exclude poor contrast issues, which also deals with the various lighting conditions in the skin images \citep{mahbod2020transfer, mahbod2019fusing}.      &     \citep{lopez2017skin, devries2017skin, pomponiu2016deepmole, van2018visualizing, fisher2019classification, mahbod2019fusing, chatterjee2019integration, al2019deep, guissous2019skin, kassani2019depthwise, eddine2019skin, mahbod2019skin, ghalejoogh2020hierarchical, mahbod2020transfer, jibhakate2020skin, bagchi2020learning, yildirim2020pre, schaefer2011colour, madooei2012automated, pereira2015adaptive, sagar2016color, alvarez2017k, yuan2017automatic, jaisakthi2018automated, ammar2018learning, liu2019skin, goyal2019skin, zafar2020skin}     \\ \hline
 
Standardization      &  Appliances of the mean and standard deviation of RGB values to scale all images to the same range to decrease biasing from different sources \citep{jibhakate2020skin}.      &     \citep{mahbod2019fusing, chatterjee2019integration, kassani2019depthwise, jibhakate2020skin, lameski2019skin, zafar2020skin, low2020automating}     \\ \hline

Median filter        &  Filtering an image by placing the median value in the input window at the center of that window to lessen impulsive, salt-and-pepper, or sudden random noise.      &    \citep{al2017classification, mahmoud2014hybrid, ghalejoogh2020hierarchical, damian2020feature, jibhakate2020skin, abbas2013improved, abbas2014combined, al2014automatic, rashid2015novel, aljanabi2018skin, santos2020skin, khan2020frequency, jia2017skin, ch2014two}      \\ \hline
        
Remove light reflections        &   Devices' light reflections are eliminated by applying morphological closing and erosion. A non-linear median filter is also helpful for removing light reflection and other tiny dots in the background outside the lesion area \citep{al2014automatic}.       &     \citep{al2017classification, abbas2013improved, al2014automatic, sagar2016color, pennisi2016skin, jafari2016skin, alvarez2017k, hasan2019skin}     \\ \hline
        
Sharpening filter       &    The sharpening spatial filter removes blurring, improving the definition of fine detail and sharpening edges that are not clearly defined in the original given image.     &      \citep{farooq2016automatic}    \\ \hline
        
Wiener filter      &   It is a low pass linear filter, usually applied in the frequency domain, for images degraded by additive noise, blurring, and constant power additive noise.      &    \citep{mahmoud2014hybrid}      \\ \hline
        
Gabor filter       &  A Gabor filter is a bandpass filter and can be defined as a sinusoidal plane of particular frequency and orientation, modulated by a Gaussian envelope.        &     \citep{mahmoud2014hybrid}      \\ \hline
        
Histogram equalization      &   It improves the contrast of an image by utilizing its histogram, spreading out the most frequent pixel intensity values, or stretching out the image's intensity range.      &   \citep{mahmoud2014hybrid, pennisi2016skin}        \\ \hline
        
Elimination of shading      &   It is induced by imaging non-flat skin surface and light-intensity falloff towards the edges of the skin image, causing color degradation and poor segmentation results \citep{madooei2012automated}.      &   \citep{cavalcanti2013macroscopic, bagchi2020learning, goyal2019skin, madooei2012automated}        \\ \hline
        
Mean filter      &   It is a smoothing method to overcome the noise effect by reducing the intensity variation between neighboring pixels, a circular or square neighbor.       &   \citep{madooei2012automated, torkashvand2015automatic, martinez2017pigmented, agarwal2017automated, sengupta2019segmentation, rosado2013prototype}    \\ \hline
        
Automatic color equalization        & This method enhances both color information and contrast by applying two main stages, including chromatic or spatial adjustment and dynamic tone reproduction scaling \citep{schaefer2011colour}.         &   \citep{rosado2013prototype, guissous2019skin, eddine2019skin, mahbod2020effects, damian2020feature, yildirim2020pre, schaefer2011colour, madooei2012automated, alvarez2017k}        \\ \hline

Contrast enhancement       &  It adjusts the relative brightness and darkness of lesions to improve their visibility. The contrast or tone of the skin image can be modified by mapping the gray levels in the image to new values through a gray-level transform.     &   \citep{nasir2018improved, khan2018implementation, filali2018study, khan2019construction, khan2019multi, almaraz2020melanoma, afza2020skin, guha2020performance, schaefer2011colour, abbas2014combined, torkashvand2015automatic, yasmin2015improved, khalid2016segmentation, hassan2016skin, agarwal2017automated, ross2018effects, olugbara2018segmentation, nasir2018improved, jaisakthi2018automated, hu2018skin, ammar2018learning, khan2018implementation, khan2019skin, sengupta2019segmentation, khan2019construction, hasan2019skin, sivaraj2020detecting, pereira2020dermoscopic, mahbod2020effects, al2022weakly, kosgiker2022significant, csahin2022robust, malik2022hybrid}        \\ \hline
        
Dark region removal      &  The black corners having nearly the same lesion's intensity due to a round circular lens can be excluded by applying binary masks of the dark corners obtained from the OTSU's thresholding \citep{khalid2016segmentation}.        &   \citep{gessert2019skin, yang2019sampling, abbas2013improved, khalid2016segmentation, khan2018implementation, hasan2019comparative, bingol2019entropy}    \\ \hline

Gamma correction      &    It controls the overall brightness that is not adequately corrected, seeming either bleached out or too dark.       &   \citep{albahar2019skin, sagar2016color}    \\ \hline

Color space transformation      &  It is the translation of the representation of a color from one basis to another. In general, CIE L*a*b*, CIE L*u*v*, YCrCb (Y color component has most of the image details), and HSV are remarkably practiced in literature \citep{hassan2016skin}.      &   \citep{khan2019multi, almaraz2020melanoma, yildirim2020pre, jyothilakshmi2014detection, torkashvand2015automatic, pereira2015adaptive, sagar2016color, hassan2016skin, pardo2017automated, lin2017skin, agarwal2017automated, qian2018detection, yuan2017automatic, olugbara2018segmentation, ahmed2018segmentation, liu2019skin, jiang2020skin, pereira2020dermoscopic, ozturk2020skin, khan2020frequency, gangwar2021study, kaur2022skin, mehmood2022k, ramya2021segmentation, ramadan2022cu, ramadan2022dgcu}    \\ \hline

Bias field correction     &  It adjusts the bias field signal before the next processing, reducing intensity heterogeneity.        &   \citep{torkashvand2015automatic}    \\ \hline      

Gaussian filter      &  A Gaussian filter blurs an image using a Gaussian function to decrease noise and detail, similar to a mean filter.        &   \citep{almaraz2020melanoma, guha2020performance, torkashvand2015automatic, hassan2016skin, chakkaravarthy2018automatic, pereira2020dermoscopic, peter2021internet, khouloud2022w}    \\ \hline
     
CLAHE$^\dag$       &    CLAHE is a variant of adaptive histogram equalization and applied to enhance foggy skin images' perceptibility level.      &   \citep{jaisakthi2017automatic}    \\ \Xhline{1pt}
\multicolumn{3}{l}{$^\dag$CLAHE: Contrast Limited Adaptive Histogram Equalization}

\end{longtable}}

\begin{figure}[!ht]
  \centering
\includegraphics[width=10.6cm, height= 2.75cm]{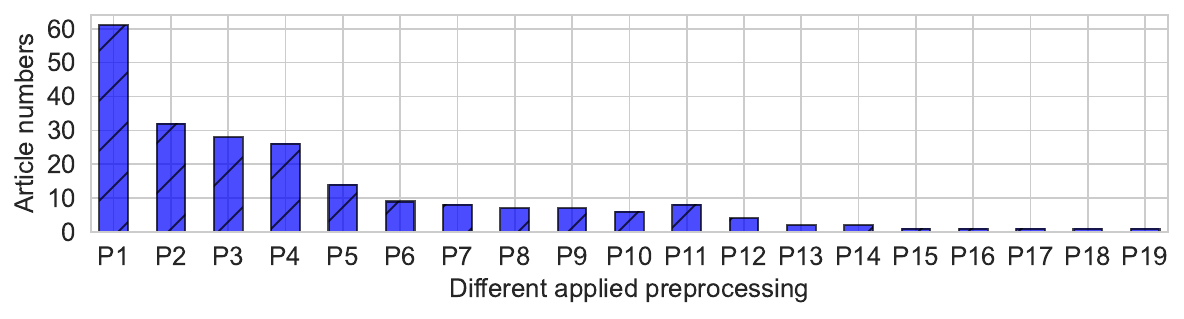}
\caption{\MKH{The number of articles employing different preprocessing, where the preprocessing P1 to P19, respectively, indicate Hair removal, Contrast enhancement, Normalization, Color space transformation, Median filter, Automatic color equalization, Remove light reflections, Standardization, Dark region removal, Mean filter, Gaussian filter, Elimination of shading, Histogram equalization, Gamma correction, Sharpening filter, Wiener filter, Gabor filter, Bias field correction, Contrast Limited Adaptive Histogram Equalization.}}
\label{fig:preprocessing}
\end{figure}

Describing the detailed theories of all preprocessing techniques is not the objective of this paper; instead, we focus on their efficacy in the SLA domain. However, citations for such theories are given in Table~\ref{tab:Preprocessing}. 
\MKH{Fig.~\ref{fig:preprocessing} displays the utilization frequencies of various preprocessing techniques used in 594 SLA articles from 2011 to 2022, revealing that hair removal, contrast enhancement, normalization, color space transformation, and median filtering are massively acknowledged top-5 preprocessing methods used throughout $27.9\,\%$, $14.6\,\%$, $12.8\,\%$, $11.9\,\%$, and $6.4\,\%$ SLA articles, respectively.} 
\MKH{Due to their high usage, these five preprocessing techniques are likely to be highly effective for SLA and can be grouped as \textbf{\textit{High-frequency}} techniques. They can thus be the benchmark for SLA preprocessing for future research. 
Other preprocessing methods like automatic color equalization, removal of light reflections, standardization, dark region removal, mean filter, and Gaussian filtering are employed in the same number of SLA articles, approximately $2.7\,\% \sim 4.1\,\%$ of articles. They could be categorized as \textbf{\textit{Medium-frequency}} preprocessing strategies for SLA. The remaining elimination of shading, histogram equalization, gamma correction, sharpening filter, wiener filter, Gabor filter, bias field correction, and contrast limited adaptive histogram equalization could be organized as \textbf{\textit{Low-frequency}} preprocessing schemes due to their less common usage in the last twelve years.
With the help of Fig.~\ref{fig:preprocessing}, this finding would assist researchers in choosing a preprocessing strategy for lesion analysis. In addition, numerous authors utilized grayscale images instead of the provided RGB images. Such a conversion can be obtained in many ways, for instance, by averaging RGB channels and taking a single channel from different color spaces. Eq.~\ref{eq:color_conversion} demonstrates the most frequent grayscale conversion technique.
\begin{equation}
\label{eq:color_conversion}
    Y = 0.299 \times R + 0.587 \times G + 0.114 \times B,
\end{equation}
where R, G, and B are the weighted summations of red, green, and blue pixels, respectively, and Y is the grayscale luminance value.}

\subsection{Augmentation and Imbalance Problem}
\label{Data_scarcity_and_imbalance_problem}
Data augmentation raises training data samples to minimize model overfitting. Affine transformations and color alteration are the basic augmentation techniques, while texture transformation preserves contours, shading, lines, strokes, and areas of information \citep{mikolajczyk2018data}. Recently, Generative Adversarial Network (GAN) \citep{mikolajczyk2018data} is a new technology for unsupervised image creation utilizing the min-max strategy and is beneficial for text-to-image synthesis, superresolution, image-to-image translation, blending, and inpainting. \MKH{However, the survey of the selected 594 articles returns the following augmentations in the past from 2011 to 2022 for the SLA task:} 
Rotation \citep{wang2020donet, qiu2020inferring, tang2020imscgnet, shan2020automatic, sanjar2020improved, nathan2020lesion, mahbod2020effects, hasan2020dsnet, zhang2020kappa, xie2020skin, low2020automating, lei2020skin, xie2020mutual, wu2020automated, anjum2020deep, al2020automatic, abhishek2020illumination, tang2019multi, tang2019efficient, soudani2019image, liu2019skin, lameski2019skin, brahmbhatt2019skin, wei2019attention, salih2019skin, saini2019detector, ali2019supervised, ali2019skin, al2019deep, tschandl2019domain, de2019skin, zeng2018multi, xu2018automatic, he2018dense, guth2018skin, vesal2018skinnet, venkatesh2018deep, chen2018multi, li2018semi, ross2018effects, qian2018detection, yuan2017automatic, nasr2017dense, mishra2017deep, bozorgtabar2017investigating, murphree2017transfer, lopez2017skin, jia2017skin, devries2017skin, zhang2018skin, li2018skin, lee2018wonderm, kitada2018skin, harangi2018skin, mahbod2019skin, wang2019mutual, mahbod2019fusing, chatterjee2019integration, serte2019wavelet, serte2019gabor, kassani2019depthwise, gessert2019skin, fisher2019classification, zhuang2020cs, yildirim2020pre, nunnari2020study, harangi2020assisted, thomsen2020deep, liu2020automatic, gessert2020skin, dhivyaa2020skin, rodrigues2020new, rahman2020transfer, kwasigroch2020neural, hassan2020skin, ahmed2020skin, anjum2020deep, arshad2021computer, bansal2021skin, bian2021skin, calderon2021bilsk, cullell2021convolutional, liu2021multiscale, mahbod2021investigating, rahman2021approach, redha2021skin, samanta2021skin, samia2021skin, santos2021transfer, shahabi2021performance, sun2021skin, thurnhofer2021skin, wang2021multi, zanddizari2021new, aldhyani2022multi, batista2022classification, bozkurt2022skin, deng2022efficient, serrano2022clinically, foahom2022end, he2022deep, lihacova2022multi, nakai2022dpe, nakai2022enhanced, nguyen2022skin, nigar2022deep, popescu2022skin, qian2022skin, rasel2022convolutional, sahin2022human, shan2022automatic, shetty2022skin, wang2022adversarial, wei2022dual, yilmaz2022mobileskin, zhuang2022cs, abhishek2021matthews, adegun2021probabilistic, ali2021skin, bagheri2021skin, bagheri2021skin_2, chowdary2021automated, das2021skin, ding2021efficient, hussain2021recu, jin2021cascade, liu2021skin, mu2021channel, phan2021skin, qamar2021dense, redha2021skin, saini2021b, singh2021slicaco, tang2021afln, wang2021focal, wibowo2021lightweight, xie2021semi, yang2021deep, ahmed2022new, alhudhaif2022novel, cao2022icl, chen2022skin, csahin2022robust, dai2022ms, dong2022learning, fan2022egfnet, feng2022bla, gu2022net, hafhouf2022improved, han2022hwa, hu2022net, kaur2022automatic, kaur2022skin, kazaj2022u, le2022antialiasing, lin2022quality, liu2022ncrnet, mehmood2022k, nour2022skin, ruan2022malunet, song2022res, tran2022fully, wang2022skin, wu2022fat, zhou2022superpixel, zuo2022efficient}, 
Horizontal flipping \citep{wang2020donet, qiu2020inferring, pour2020transform, jiang2020skin, tang2020imscgnet, shan2020automatic, zhu2020asnet, sanjar2020improved, nathan2020lesion, mahbod2020effects, hasan2020dsnet, zhang2020kappa, xie2020skin, low2020automating, lei2020skin, xie2020mutual, wu2020automated, abhishek2020illumination, tang2019multi, tang2019efficient, soudani2019image, liu2019skin, lameski2019skin, brahmbhatt2019skin, sarker2021slsnet, jiang2019decision, bi2019improving, wei2019attention, wang2019bi, salih2019skin, saini2019detector, baghersalimi2019dermonet, ali2019skin, al2019deep, tschandl2019domain, de2019skin, zeng2018multi, youssef2018deep, guth2018skin, vesal2018skinnet, venkatesh2018deep, li2018semi, ross2018effects, qian2018detection, yuan2017automatic, ramachandram2017skin, pour2017automated, nasr2017dense, mishra2017deep, bi2017semi, murphree2017transfer, lopez2017skin, jia2017skin, zhang2018skin, li2018skin, lee2018wonderm, kitada2018skin, harangi2018skin, chen2018multi, mahbod2019skin, wang2019mutual, mahbod2019fusing, kassani2019depthwise, gessert2019skin, fisher2019classification, zhuang2020cs, muckatira2020properties, harangi2020assisted, thomsen2020deep, liu2020automatic, gessert2020skin, kwasigroch2020self, rodrigues2020new, rahman2020transfer, kwasigroch2020neural, hassan2020skin, ahmed2020skin, arshad2021computer, balabantaray2021melanoma, calderon2021bilsk, ding2021deep, khan2021skin, mahbod2021investigating, shetty2022skin, zanddizari2021new, le2022antialiasing, pennisi2022skin}, 
Vertical flipping \citep{qiu2020inferring, pour2020transform, jiang2020skin, tang2020imscgnet, shan2020automatic, zhu2020asnet, nathan2020lesion, mahbod2020effects, hasan2020dsnet, zhang2020kappa, xie2020skin, low2020automating, lei2020skin, xie2020mutual, wu2020automated, abhishek2020illumination, tang2019multi, tang2019efficient, liu2019skin, lameski2019skin, sarker2021slsnet, jiang2019decision, bi2019improving, wei2019attention, wang2019bi, salih2019skin, saini2019detector, baghersalimi2019dermonet, ali2019skin, al2019deep, de2019skin, yuan2017automatic, youssef2018deep, guth2018skin, vesal2018skinnet, venkatesh2018deep, ross2018effects, qian2018detection, ramachandram2017skin, pour2017automated, nasr2017dense, mishra2017deep, bi2017semi, murphree2017transfer, lopez2017skin, jia2017skin, zhang2018skin, li2018skin, lee2018wonderm, kitada2018skin, harangi2018skin, chen2018multi, wang2019mutual, tschandl2019domain, kassani2019depthwise, gessert2019skin, fisher2019classification, muckatira2020properties, harangi2020assisted, thomsen2020deep, liu2020automatic, gessert2020skin, kwasigroch2020self, rodrigues2020new, rahman2020transfer, kwasigroch2020neural, hassan2020skin, ahmed2020skin, arshad2021computer, calderon2021bilsk, le2022antialiasing, li2022mhau, pennisi2022skin}, 
Scaling \citep{qiu2020inferring, shan2020automatic, mahbod2020effects, wu2020automated, wang2019bi, ali2019skin, xu2018automatic, vesal2018skinnet, venkatesh2018deep, li2018semi, yuan2017automatic, bozorgtabar2017investigating, lopez2017skin, mahbod2019fusing, gessert2019skin, gessert2020skin, dhivyaa2020skin, bi2020multi, nathan2020lesion, zhang2020kappa, low2020automating, al2020automatic, xie2020mutual, ali2019supervised, guth2018skin, chen2018multi, li2018skin, thomsen2020deep, salian2020skin, rahman2020transfer, hassan2020skin, cullell2021convolutional, sahin2022human, somfai2022handling, wang2022ssd, bansal2021skin, calderon2021bilsk, rahman2021approach, samia2021skin, aldhyani2022multi, alptekin2022analysis, batista2022classification, bozkurt2022skin, nguyen2022skin, popescu2022skin, yilmaz2022mobileskin, nakai2022dpe, nakai2022enhanced, bagheri2021skin, bagheri2021skin_2, das2021skin, jin2021cascade, mu2021channel, phan2021skin, qamar2021dense, wang2021boundary, wang2021focal, wang2021knowledge, yang2021deep, akyel2022linknet, alhudhaif2022novel, gu2022net, han2022hwa, kaur2022automatic, kaur2022skin, lin2022quality, mehmood2022k, nour2022skin, zhang2022dynamic, adegun2021probabilistic, cao2022icl, feng2022bla, hafhouf2022improved, hu2022net, wang2022skin, csahin2022robust, le2022antialiasing}, 
Region cropping \citep{wang2020donet, lei2020skin, xie2020mutual, wu2020automated, bi2019improving, saini2019detector, baghersalimi2019dermonet, qian2018detection, pour2017automated, mishra2017deep, bi2017semi, jia2017skin, li2018skin, kitada2018skin, pan2018residual, chen2018multi, rashid2019skin, fisher2019classification, harangi2020assisted, mahbod2020transfer, liu2020automatic, bi2020multi, salian2020skin, bian2021skin, ding2021deep, santos2021transfer, sun2021skin, deng2022efficient, nakai2022dpe, nakai2022enhanced, nigar2022deep, sarker2022transslc, wang2022adversarial, wei2022dual, ali2021skin, chen2021mt, liu2021skin, redha2021skin, saini2021b, tang2021afln, wang2021knowledge, yang2021deep, alhudhaif2022novel, dai2022ms, mehmood2022k}, 
Shifting \citep{hasan2020dsnet, zhang2020kappa, xie2020mutual, tang2019efficient, lameski2019skin, salih2019skin, ali2019supervised, ali2019skin, de2019skin, yuan2017automatic, xu2018automatic, vesal2018skinnet, venkatesh2018deep, lopez2017skin, devries2017skin, chatterjee2019integration, fisher2019classification, thomsen2020deep, dhivyaa2020skin, kwasigroch2020neural, hassan2020skin, ahmed2020skin, cullell2021convolutional, bansal2021skin, calderon2021bilsk, aldhyani2022multi, nguyen2022skin, popescu2022skin, somfai2022handling, tang2022fusionm4net, yilmaz2022mobileskin, le2021modified, wang2021focal, akyel2022linknet, cao2022icl, wang2022skin},
Contrast adjustment \citep{mahbod2020effects, low2020automating, wu2020automated, ribeiro2020less, al2020automatic, jiang2019decision, wei2019attention, saini2019detector, de2019skin, guth2018skin, luo2018fast, ramachandram2017skin, gessert2019skin, gessert2020skin, rodrigues2020new, somfai2022handling, saini2021b, yang2021deep, dong2022learning, feng2022bla, wu2022fat},
Shearing \citep{nathan2020lesion, zhang2020kappa, xie2020mutual, saini2019detector, ali2019supervised, qian2018detection, murphree2017transfer, wang2019mutual, mahbod2019fusing, gessert2019skin, rahman2020transfer, rahman2021approach, redha2021skin, samanta2021skin, samia2021skin, sahin2022human, shetty2022skin, adegun2021probabilistic, ali2021skin, das2021skin, saini2021b, singh2021slicaco, cao2022icl},
Horizontal and vertical or both flipping \citep{jiang2020skin, hasan2020dsnet, zhang2020kappa, al2020automatic, chen2018multi, li2018skin, zhuang2020cs, nunnari2020study, bi2020multi, bansal2021skin, bian2021skin, rahman2021approach, samanta2021skin, samia2021skin, santos2021transfer, shahabi2021performance, sun2021skin, thurnhofer2021skin, wang2021multi, aldhyani2022multi, ayas2022multiclass, bozkurt2022skin, deng2022efficient, dong2022learning, serrano2022clinically, foahom2022end, he2022deep, lihacova2022multi, nakai2022dpe, nakai2022enhanced, nguyen2022skin, nigar2022deep, qian2022skin, sarker2022transslc, shetty2022skin, somfai2022handling, tang2022fusionm4net, wang2022adversarial, wang2022ssd, wei2022dual, yilmaz2022mobileskin, zhuang2022cs, abhishek2021matthews, ali2021skin, arora2021automated, bagheri2021skin, bagheri2021skin_2, chen2021mt, chowdary2021automated, das2021skin, ding2021efficient, hussain2021recu, jin2021cascade, khan2021skin, le2021modified, liu2021skin, mu2021channel, phan2021skin, qamar2021dense, redha2021skin, sarker2021slsnet, singh2021slicaco, tang2021afln, tong2021ascu, wang2021boundary, wang2021focal, wang2021knowledge, yang2021deep, akyel2022linknet, ahmed2022new, alhudhaif2022novel, cao2022icl, chen2022skin, dai2022ms, dong2022learning, feng2022bla, gu2022net, hafhouf2022improved, hu2022net, kazaj2022u, le2022antialiasing, lin2022quality, liu2022ncrnet, mehmood2022k, nour2022skin, pennisi2022skin, ruan2022malunet, song2022res, tran2022fully, wu2022fat, zhang2022dynamic, zhou2022superpixel, zuo2022efficient},
Adaptive histogram equalization \citep{mahbod2020effects, wu2020automated, al2020automatic, sarker2021slsnet, wu2019skin, saini2019detector, pomponiu2016deepmole, dhivyaa2020skin, salian2020skin, bansal2021skin, liu2021multiscale, samia2021skin, sun2021skin, alptekin2022analysis, ayas2022multiclass, bozkurt2022skin, dong2022learning, sahin2022human, wang2022ssd, wei2022dual, sarker2022transslc, somfai2022handling, saini2021b, yang2021deep},
Gaussian noises \citep{ribeiro2020less, al2020automatic, youssef2018deep, chen2018multi, yuan2017automatic, pomponiu2016deepmole, rashid2019skin, sun2021skin, dong2022learning, somfai2022handling, adegun2021probabilistic, arora2021automated, le2022antialiasing},
RGB to HSV transformation \citep{jiang2020skin, wei2019attention, de2019skin, ramachandram2017skin, ayas2022multiclass, sahin2022human, wei2022dual, somfai2022handling, bansal2021skin, sun2021skin, somfai2022handling, wang2022ssd, wei2022dual, yang2021deep, song2022res}, 
Elastic distortion \citep{tang2019efficient, pomponiu2016deepmole, al2020automatic, bissoto2018deep, xie2021semi, alhudhaif2022novel, nour2022skin},
Adding noises (salt or pepper noises) \citep{wu2020automated, pomponiu2016deepmole, rashid2019skin},
Color jittering \citep{bi2019improving, wei2019attention, kwasigroch2020self}, 
Gamma correction \citep{sarker2021slsnet, rodrigues2020new}, 
GAN \citep{qin2020gan, rashid2019skin},
Whitening \citep{xie2020mutual}, 
and Dihedral transformation \citep{guth2018skin}. 
These typically applied augmentations are briefly discussed and explained in Table~\ref{tab:Augmentations}. 

{\scriptsize
\begin{longtable}[c]{p{2.4cm}|p{13.2cm}}
\caption{\MKH{Typically engaged augmentations in the past, surveying a total of 594 articles: 356 for SLS and 238 for SLC.}}
\label{tab:Augmentations}\\
\Xhline{1pt}
\textbf{Method} & \textbf{Details descriptions} \\ \Xhline{1pt}
\endfirsthead
\multicolumn{2}{c}%
{{\bfseries Table \thetable\ Continued from previous page}} \\
\Xhline{1pt}
\textbf{Method} & \textbf{Details descriptions}  \\ \Xhline{1pt}
\endhead
Rotation (A1)     &  Rotate the image coordinates of ($x_1, y_1$) by an angle of $\theta$ around ($x_0, y_0$), resulting the coordinates of ($x_2, y_2$) where $x_2 = cos(\theta)\times (x_1-x_0)+ sin(\theta)\times (y_1-y_0)$ and $y_2 = -sin(\theta)\times (x_1-x_0)+ cos(\theta)\times (y_1-y_0)$.      \\ \hline

Horizontal flipping (A2) & Change the image by a mirror-reversal of an original across a vertical axis, where the left side switches to the right side and vice versa.  \\\hline

Vertical flipping (A3) & Modify an image with a mirror-reversal of the original across a horizontal axis, where the top side switches to the bottom side and vice versa. \\\hline

Scaling (A4) & Enlarge or reduces the image's physical size by changing the number of pixels it contains, changing the size of the contents of the image and resizing the canvas accordingly. \\\hline

Region cropping (A5) & A data augmentation technique that picks a random subset of the original image containing more salient information about the region of interest.  \\\hline

Shifting (A6) &  Translation of an image in up, down, left, or right, along with any combination of the above direction, where every point of the object must be moved in the same direction and for the same distance. \\\hline

Contrast adjustment (A7) & Remap image intensity values to the full display range of the data type, sharpening differences between black and white. It can either make an image more vivid or mute the tones for a more subdued feel. \\\hline

Shearing (A8) & Shift one part of an image, a layer, a selection, or a path to a direction and the other part to the opposite direction; for example, a horizontal shearing will shift the upper part to the right and the lower part to the left, resulting in a diamond from a given rectangle.  \\\hline

Both flipping (A9) & Modify an image with a mirror-reversal of the original across both the vertical and horizontal axes, where the top side switches to the bottom side and then the left side switches to the right side and vice versa.  \\\hline

Histogram Equalization (A10) & Adjust the contrast of an image by using its histogram, spreading out the most frequent pixel intensity values, or stretching out the image's intensity range. It allows the image's areas with lower contrast to gain a higher contrast.  \\\hline

Gaussian noises (A11) & A type of mean spatial filtering that produces a new image by altering the structural details of an input image. \\\hline

HSV (A12) & Provide a numerical readout of the image corresponding to the color names contained therein, abstracting the color (hue) by separating it from saturation and pseudo-illumination. \\\hline

Elastic distortion (A13) & Generate a coarse displacement grid with a random displacement for each grid point that is then interpolated to compute a displacement for each pixel. Finally, the input image is then deformed using displacement vectors and spline interpolation. \\\hline

Adding noises (A14) & An impulse noise caused by sharp and sudden disturbances in the image signal that presents as sparsely occurring white and black pixels.  \\\hline

Color jittering (A15) & A type of image data augmentation that randomly changes the brightness, contrast, and saturation of the image. It also adds random noise to the image.  \\\hline

Gamma correction (A16) & A nonlinear operation that encodes and decodes luminance or tristimulus values in video or still image as $V_{in} = A\times V_{out}^\gamma$, where $\gamma$ and $A$ are the raised power and multiplied factors, respectively. \\\hline

Whitening (A17) & A whitening transformation converts a vector of random variables with a known covariance matrix into a set of new variables whose covariance is the identity matrix, which has widely been adopted to remove redundancy by making adjacent pixels less correlated.  \\\hline

Dihedral transformation (A18) & A linear transformation that includes rotations and reflections of the images in the eight possible directions or angles of a dihedron. \\\hline

GAN (A19) & Produce new data samples from a given random noise from a latent space and deliver unique images that mimic the feature distribution of the original dataset. \\\Xhline{1pt}


\end{longtable}}

\begin{figure}[!ht]
  \centering
\includegraphics[width=15cm, height=4.5cm]{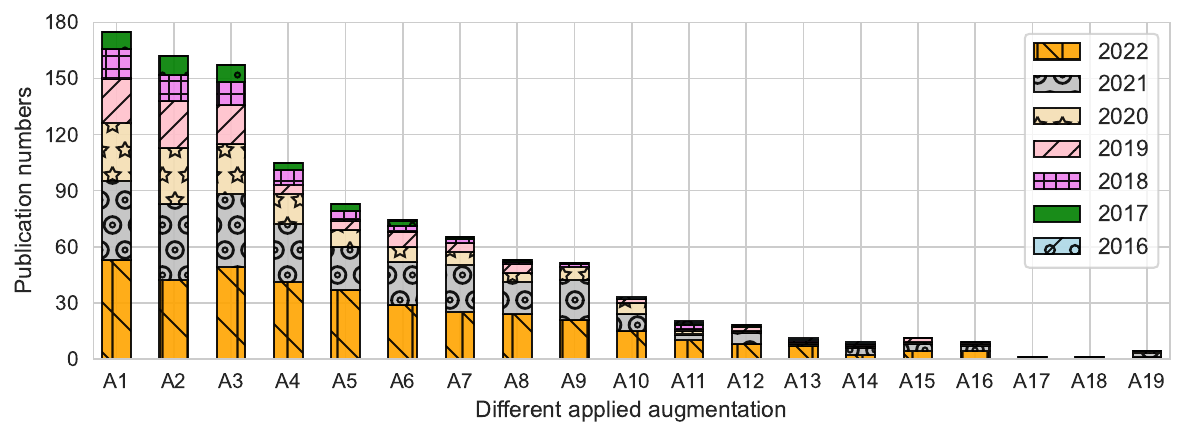}
\caption{\MKH{The number of SLS and SLC articles using augmentations A1 to A19 (defined in Table~\ref{tab:Augmentations}) in the past literature.}}
\label{fig:Augmentaion}
\end{figure}

\MKH{Fig.~\ref{fig:Augmentaion} indicates the frequency of these augmentations in the past twelve years' articles. It illustrates that rotation, horizontal flipping, vertical flipping, scaling, region of interest cropping, and shifting are the top-6 SLA augmentations, used throughout respectively in $29.5\,\%$, $27.3\,\%$, $26.4\,\%$, $17.7\,\%$, $14.0\,\%$, and $12.5\,\%$ of the total of 594 articles.}
\MKH{Notably, these six most common augmentations (\textbf{\textit{High-frequency}}) were more frequently operated in 2022, and their deployments have dropped from 2022 to 2016.} Again, contrast adjustment, shearing, both axis flipping, histogram equalization, and adding Gaussian noise can be classified as the \textbf{\textit{Medium-frequency}} augmentation methods due to their medium utilization frequency. The remaining augmentation techniques were less commonly applied in the past twelve years and could be termed as \textbf{\textit{Low-frequency}}. 

\MKH{Furthermore, a close inspection of Fig.~\ref{fig:Augmentaion} demonstrates that none of the augmentations were applied before 2016. It is also remarkable that the number of articles with augmentation applications has significantly increased trends from 2016 to 2022.} Recalling Fig.~\ref{fig:year_wise_paper} shows that the number of SLA articles published after 2016 has enormously increased after DL's engagement, especially using CNN methods. Such an advancement in CNN models for SLA tasks promotes the need for diverse augmentations, as CNNs rely on a large amount of data to be robust and effective. Besides, Fig.~\ref{fig:Augmentaion} in conjunction with Table~\ref{tab:Augmentations} and Fig.~\ref{fig:year_wise_paper} demonstrates that geometric augmentations, i.e., spatial information transformation, are more common in CNN-based learning systems. In contrast, color or texture augmentations are more common in manual feature-based ML and computer vision algorithms. Although the usage of GAN \citep{zunair2020melanoma} is a recent trend in other domains, it has yet to be widely employed in SLA in the last twelve years (see Fig.~\ref{fig:Augmentaion}). The generator and discriminator need to be in sync with each other, and the learning process of GANs may miss underlying anatomical structures and attributes. It could be mitigated by incorporating anatomical constraints for making realistic images.

The big picture of the class sample distributions of all datasets in Table~\ref{tab:data_distribution} reveals that class sample images are imbalanced, which could bias the classifier towards the class with more samples \citep{hasan2021dermo}. This is especially true in skin lesion datasets with few manually annotated training examples. However, SLA researchers attempted numerous methods to reduce class bias. The authors in \citep{nunnari2020study, yan2020scalable, mahbod2020transfer, mahbod2020effects, gessert2020skin, bagchi2020learning, ahmed2020skin, zhang2018skin, mahbod2019fusing, kassani2019depthwise, gessert2019skin, albahar2019skin, ayas2022multiclass, krohling2021smartphone, yao2021single, zhuang2022cs, carvalho2021multimodal, yue2022towards, qian2022skin, yue2022towards, qian2022skin} rewarded more extra consideration to the class with minority samples, estimating the class weight using a portion of $W_n=N_n/N$, where $W_n$, $N$, and $N_n$ separately indicate the $n^{th}$-class wight, the total number of samples, and the sample in the $n^{th}$-class. Some authors acquired more photographs and integrated them with the minority class to balance it with the majority class \citep{devries2017skin, bisla2019skin, aldwgeri2019ensemble}. The augmentations in Table~\ref{tab:Augmentations} are sometimes applied to the minority class to increase its representative sample, as in \citep{yildirim2020pre, imtiaz2021efficient, qin2020gan, thomsen2020deep, salian2020skin, anjum2020deep, hassan2020skin, li2018skin, kulhalli2019hierarchical, eddine2019skin, chatterjee2019integration, bisla2019skin, serte2019wavelet, kassani2019depthwise}. A SMOTE oversampling technique was applied in \citep{almaraz2020melanoma}, demonstrating an improvement in the differentiation of melanoma and benign lesion images. To handle imbalanced classes, an ensemble method using three classifiers with a linear plurality voting was employed in \citep{molina2020classification}, where the other two models can overcome the bias from any candidate model. Lastly, the authors in \citep{zhuang2020cs, muckatira2020properties, ghalejoogh2020hierarchical, rahman2020transfer, chakravorty2016dermatologist, lopez2017skin, lee2018wonderm, navarro2018webly, yoon2019generalizable, murphree2017transfer} employed random oversampling and undersampling. The first technique randomly adds minority class samples to the training dataset. In contrast, the latter undersampling chooses examples from the majority class and deletes them from the training dataset. To sum up, class weighting, minority class augmentations, oversampling, and undersampling are the most frequent strategies for overcoming class imbalance in SLA tasks. These strategies can help researchers construct a generic SLA framework.

\section{Segmentation Techniques}
\label{Segmentation_Techniques}
In recent years, segmentation algorithms for SLA have been improved \citep{kassem2021machine} using different approaches. These approaches are categorized into four groups to contrast them: \textbf{edge-based SLS}, \textbf{region-based SLS}, \textbf{threshold-based SLS},\textbf{ AI-based SLS}, and \textbf{other SLS}. The following five paragraphs briefly introduce those clusters of SLS techniques. Then, we provide our insightful discussions regarding them from the 356 SLS papers in the rest of the paragraphs.

\paragraph{\textbf{Edge-based SLS}} 
Edge-based SLS methods look for edge pixels and link them to produce image contours, which can be manual or automatic. The manual application utilizes the trackpad to delineate lesion boundaries. \MKH{In contrast, the automatic method employs edge detection algorithms like the watershed algorithm \citep{chakkaravarthy2018automatic}, active contours \citep{gangwar2021study, marosan2021automated, nasir2018improved, riaz2018active, jyothilakshmi2014detection, ivanovici2012color, cavalcanti2013macroscopic}, canny edge detector \citep{yasmin2015improved, rashid2015novel}, and multi-direction gradient vector flow snake model \citep{cavalcanti2013macroscopic}.} In this segmentation, an edge filter is applied to the image, pixels are classified as edge or non-edge based on the filter output, and pixels not separated by an edge are assigned to the same class. \MKH{Active contours are the most common edge-based SLS method out of the 356 papers written in the last twelve years.}

\paragraph{\textbf{Region-based SLS}}
In these systems, images are divided into regions or groups of comparable pixels based on their attributes, assuming neighboring pixels should have the same value. Each pixel in a region is compared to its neighbors and clustered based on specific conditions. \MKH{This form of SLS covers iterative region-based \citep{louhichi2018skin, schaefer2011colour, singh2021slicaco}, iterative stochastic region-merging \citep{wong2011automatic}, mean shift-based gradient vector flow \citep{zhou2011gradient}, K-means \& fuzzy C-means clustering \citep{he2012automatic, khakabi2012multi, navarro2018accurate, jaisakthi2018automated, jaisakthi2017automatic, lynn2017segmentation, alvarez2017k, agarwal2017automated, gupta2017adaptive, george2017automatic, nisar2013color, ch2014two, lin2017skin, azehoun2016novel, trabelsi2015skin, masood2014integrating, veeramuthu2021comparative, das2021skin, reddy2021handling, gangwar2021study, garg2021skin, joseph2022preprocessing}, and Eikonal-based region growing clustering \citep{lezoray2014graph}.} Lastly, this article suggests that K-means \& fuzzy C-means clustering are the most common region-based SLS methods, which would be viewed as a representation of the region-based SLS approach.

\paragraph{\textbf{Threshold-based SLS}}
Threshold-based SLS can be classified as point-based or pixel-based segmentation, depending on the threshold estimation approaches, and commonly suffers from difficulty in estimating effective thresholds due to dermoscopic artifacts \citep{hasan2020dsnet}. \MKH{OTSU \citep{sivaraj2020detecting, parida2020transition, rout2020transition, bansal2020improved, thanh2020adaptive, abhishek2020illumination, sengupta2019segmentation, yang2019sampling, rawas2019hcet, ahmed2018segmentation, pardo2017automated, sagar2016color, joseph2016skin, hassan2016skin, madooei2012automated, wu2013automatic, cavalcanti2013macroscopic, ramya2021segmentation, gangwar2021study, joseph2022preprocessing}, histogram estimation \citep{pereira2015adaptive, jaisakthi2017automatic, gupta2017adaptive}, morphological operations \citep{akram2018skin, ahmed2018segmentation, abbas2013improved}, optimal color channel-based empirical threshold estimation \citep{abbas2014combined}, and mean pixel intensity level-based threshold estimation \citep{al2014automatic} are examples of this cluster of techniques.} Notably, the OTSU thresholding technique is the most common threshold-based SLS strategy in the selected 356 SLS papers, setting it to stand apart from other techniques. This technique's automatic and faster threshold estimate could be the reason for its popularity.

\paragraph{\textbf{AI-based SLS}} 
AI models, especially CNNs, make it possible to build an end-to-end supervised model without having to manually extract features \citep{lecun2015deep, krizhevsky2012imagenet}, and they have been very successful in many areas of medical imaging: arrhythmia detection \citep{yildirim2018arrhythmia,hannun2019cardiologist,acharya2017deep}, skin lesion segmentation and classification \citep{hasan2020dsnet, esteva2017dermatologist, codella2017deep, hasan2021dermoexpert, hasan2021dermo}, breast cancer detection \citep{celik2020automated, cruz2014automatic, hasan2020automatic}, brain disease classification \citep{talo2019convolutional}, pneumonia detection from CXR images \citep{rajpurkar2017chexnet}, fundus image segmentation \citep{tan2017automated, hasan2021drnet}, minimally invasive surgery \citep{hasan2021detection}, lung segmentation \citep{gaal2020attention}, \textit{etc}. \MKH{This category of methods applies computational intelligence techniques to the segmentation process, including genetic algorithms \citep{sayed2020novel, amelio2013skin}, fully convolutional neural networks \citep{xie2021semi, zhang2022dynamic, barzegar2022skin, alahmadi2022semi, bagheri2021skin, dayananda2021skin, kaur2021deep, redha2021skin, krishna2021mlrnet, chowdary2021automated, adegun2021probabilistic, dong2021fac, le2021modified, yacin2021deep, lee2022progressive, kazaj2022u, barin2022automatic, song2022res, gangwar2021study, marosan2021automated, wibowo2021lightweight, gajera2021improving, phan2021skin, qamar2021dense, li2022mhau, alahmadi2022multiscale, liu2022ncrnet, liu2022skin, sun2022acfnet, stofa2022u, tran2022fully, chen2022skin, wang2022skin, ren2022serial, wu2022fat, gulzar2022skin, feng2022slt, dong2022tc, csahin2022robust, ramadan2022dgcu, zhang2022dense, saini2021b, kosgiker2021segcaps, khan2021skin, chen2021nl, chen2021mt, yang2021deep, jin2021cascade, jiang2021residual, chauhan2021multi, xiao2021prior, jiang2021approximated, imtiaz2021efficient, hussain2021recu, arora2021skin, arora2021automated, wang2021knowledge, wang2021focal, mu2021channel, mirikharaji2021d, ramadan2022cu, kaur2022skin, wang2022cross, kaur2022automatic, jiang2022seacu, camacho2022multi, malik2022novel, han2022hwa, basak2022mfsnet, lin2022quality, hafhouf2022improved, gu2022net, anand2022fusion, feng2022bla, dong2022learning, ahmed2022new, peter2021internet, tang2021afln, liu2021skin, dai2022ms, wang2021boundary, tong2021ascu, tao2021attention, jignesh2021automated, hu2022net, lameski2019skin, soudani2019image, singh2019fca,liu2019skin,liu2019enhanced, dash2019pslsnet, qian2018detection, li2018dense, kolekar2018skin, jiang2018skin, xu2018automatic, yuan2017automatic, youssef2018deep, mirikharaji2018deep, venkatesh2018deep, ur2018classification, guth2018skin, nguyen2018isic, he2018dense, wang2018skin, vesal2018multi, luo2018fast, chen2018multi, ammar2018learning, burdick2018rethinking, bissoto2018deep, bi2018improving, zeng2018multi, ross2018effects, qi2017global, lin2017skin, bozorgtabar2017skin, bozorgtabar2017investigating, mishra2017deep, nasr2017dense, pour2017automated, he2017skin, jafari2016skin, ramachandram2017skin, li2018semi, tang2019multi, shahin2019deep, tang2019efficient}, expectation-maximization \citep{majtner2016improving}, rule-based algorithms \citep{bozorgtabar2016sparse}, Co-operative neural network \citep{schaefer2011colour}, and artificial bee colony \citep{aljanabi2018skin}.} Individual scrutiny of the specified 356 SLS reveals that fully convolutional neural networks are the most frequently employed in the SLS challenge, establishing them as the characteristic AI-based SLS methods.

\paragraph{\textbf{Other SLS methods}}
\label{others_SLS}
Other approaches include probabilistic maximum-a-posteriori \citep{khan2018implementation, li2011estimating}, Markov random field \citep{khattak2015maximum, salih2018skin, torkashvand2015automatic}, Delaunay Triangulation \citep{pennisi2016skin}, Grap-cut \citep{jaisakthi2018automated}, Cellular Automata \citep{bi2016automated}, Wavelet transform \citep{hu2018skin, khalid2016segmentation}, and optimized color feature \citep{khan2019skin, olugbara2018segmentation}. They are grouped as other SLS task approaches from 2011 to 2022, like four other segmentation groups.

\begin{figure}[!ht]
\centering
\subfloat[]{\includegraphics[width=15cm, height=3cm]{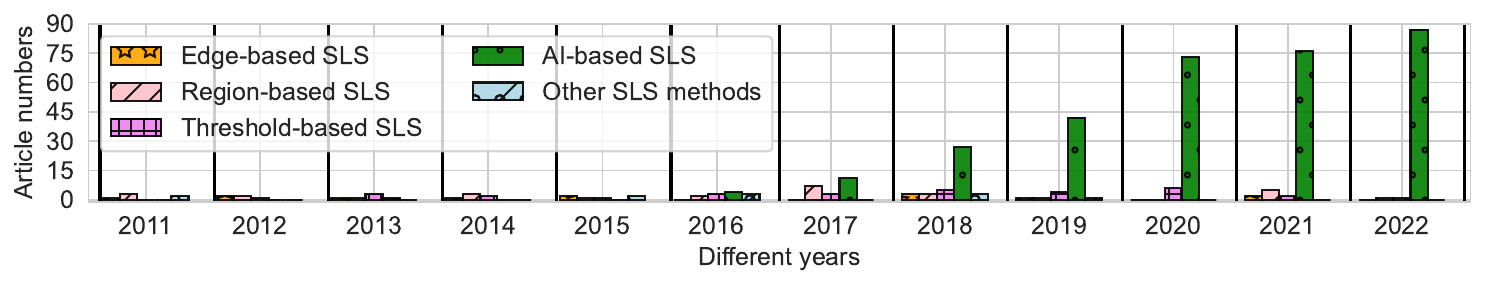}} \\
\subfloat[]{\includegraphics[width=6cm, height=5.5cm]{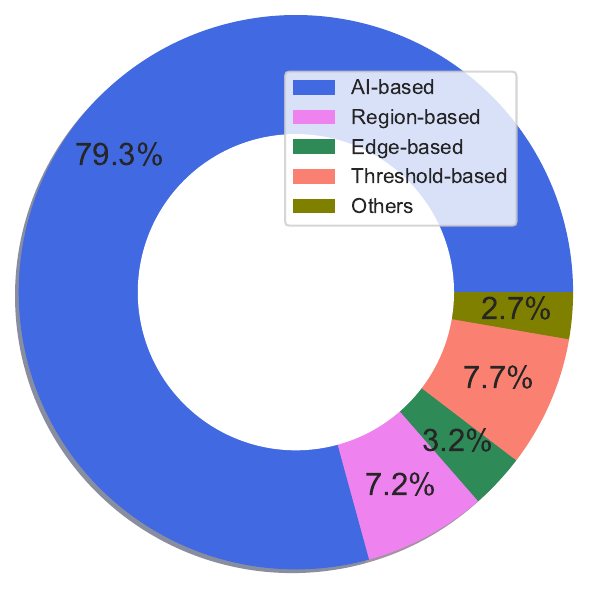}}
\caption{\MKH{(a) The number of articles employed different SLS methods, consulting the commonly applied techniques and (b) the percentage donut chart of the SLS articles for the five SLS categories.}}
\label{fig:sls_yearwise_method}
\end{figure}

Fig.~\ref{fig:sls_yearwise_method} (a) demonstrates year-wise SLS article numbers in the past, bestowing the frequency of various SLS techniques covered in the previous five paragraphs.
AI-based SLS methods have been most extensively employed over the last twelve years, notably after 2016. \MKH{Again, the application frequencies of the five SLS categories are shown in Fig.~\ref{fig:sls_yearwise_method} (b), highlighting that $79.3\,\%$ of publications used AI-based SLS.} Fig.~\ref{fig:year_wise_paper} and Fig.~\ref{fig:sls_yearwise_method} (a \& b) collectively demonstrate that the employment of AI-based SLS models has become crucial to the SLS task as the number of AI-based SLS articles has been expanding tremendously since 2016. Even in the last few years, the application of remaining SLS methods, except AI-based methods, has significantly reduced. This AI-based SLS category would be considered a \textbf{\textit{High-frequency}} SLS method. Thus, this SLS approach will likely continue its dominance in the future. Threshold-and region-based SLS can be designated \textbf{\textit{Medium-frequency}} SLS techniques, while the remaining SLS methods can be considered as \textbf{\textit{Low-frequency}} SLS approaches (see Fig.~\ref{fig:sls_yearwise_method} (b)). Discussions on the details of SLS techniques are not the focus of this article. For such discussions, the reader is referred to a previous review article of \citet{kassem2021machine}. \MKH{However, our analysis reveals that $87.2\,\%$ of AI-based SLS articles employed DL methods, in particular, deep CNNs.} We thus review the details of this particular CNN technique in the following paragraph.

In order to build an end-to-end CNN-based SLS system, a model consists of two essential components: the encoder and the decoder \citep{hasan2020dsnet}. The former component, the encoder, comprises convolutional and subsampling layers and is responsible for automatic feature extraction. The convolutional layers are applied to construct the feature maps, whereas the subsampling layers are employed to achieve spatial invariance by decreasing the maps' resolution. This reduction in resolution leads to an extension of the field of view of the feature map, which in turn makes extracting more salient features easier and minimizes the computational cost \citep{hasan2021detection}. \MKH{This is notably observed from the chosen 356 SLS article that many authors have employed different variants of pre-trained (on ImageNet, PASCAL-VOC, MS-COCO, etc) CNN models in the encoders like AlexNet \citep{barin2022automatic, pour2017automated, khan2019skin}, Xception \citep{tang2019efficient, azad2020attention}, VGG \citep{kaur2021deep, redha2021skin, ramachandram2017skin, qi2017global, zhang2019automatic, chen2018multi, burdick2018rethinking, soudani2019image, low2020automating, shahin2019deep, lameski2019skin, khan2019skin, huang2019skin, jayapriya2020hybrid}, ResNet \citep{kazaj2022u,basak2022mfsnet, song2022res,barin2022automatic, lee2022progressive, zafar2020skin, qamar2021dense, luo2018fast, tschandl2019domain, al2019deep, jiang2019decision, iranpoor2020skin, he2018dense, guth2018skin, soudani2019image, bisla2019skin, zhu2020asnet, huang2020skin, low2020automating, wei2020attentive, anjum2020deep, kamalakannan2020self}, LeNet \citep{kamalakannan2020self}, Inception \citep{wu2019skin, goyal2019skin, nampalle2020efficient, yacin2021deep, lee2022progressive}, InceptionResNet \citep{jayapriya2020hybrid}, MobileNet \citep{wibowo2021lightweight, gajera2021improving}, and DenseNet \citep{li2018semi, qamar2021dense, phan2021skin, shan2020automatic, hasan2020dsnet, zeng2018multi, wei2019attention, lei2020skin, nampalle2020efficient}. Remarkably, it is revealed that $37.3\,\%$, $23.7\,\%$, and $15.3\,\%$ of the articles used ResNet, VGG, and DenseNet, respectively, for transfer learning (backbone in the encoder).} Again, the latter component in the CNN-based SLS system, the decoder, semantically projects the distinctive lower-resolution features learned by the encoder onto the pixel space of higher resolution to achieve a dense pixel-wise classification. The semantic segmentation networks have similar encoder designs, but they vary mainly in their decoder mechanisms concerning how the discriminating features are projected onto the pixel space. However, the significantly decreased feature maps due to subsampling often undergo spatial resolution loss, which introduces coarseness, less edge information, checkerboard artifacts, and over-segmentation in semantically segmented masks \citep{hasan2020dsnet}. To resolve these problems, skip connections in a U-Net were proposed in \citep{ronneberger2015u}, which allowed the decoder to recover the relevant features learned at each stage of the dropped encoder due to pooling. Similarly, the features at various coarseness levels of the encoder in a Fully Convolutional Network (FCN) were fused in \citep{long2015fully} to refine the segmentation. However, when the deconvolution kernel size is not divisible by the up-scaling factor, a deconvolution overlap occurs as the number of low-resolution features that contribute to a single high-resolution feature is not consistent across the high-resolution feature map \citep{odena2016deconvolution}. Due to this deconvolution overlap, checkerboard artifacts may appear in the segmented mask. Therefore, a full-resolution convolution network was designed in \citep{al2018skin}, excluding subsampling layers in the encoder to conserve the spatial information of the feature maps. However, the subsampling of feature maps is exceptionally desirable to be employed on CNN due to the several positive perspectives, as previously mentioned. A complete survey and review on those decoder mechanisms and skip (shortcut) connection can be found in \citep{mirikharaji2022survey}.

\MKH{Many authors in \citep{saini2021b, kosgiker2021segcaps, khan2021skin, chen2021nl, chen2021mt, yang2021deep, jin2021cascade, jignesh2021automated, jiang2021residual, chauhan2021multi, xiao2021prior, jiang2021approximated, imtiaz2021efficient, hussain2021recu, arora2021skin, arora2021automated, wibowo2021lightweight, wang2021knowledge, wang2021focal, mu2021channel, mirikharaji2021d, gajera2021improving, marosan2021automated, ramadan2022dgcu, ramadan2022cu, kaur2022skin, wang2022cross, kaur2022automatic, jiang2022seacu, camacho2022multi, malik2022novel, han2022hwa, basak2022mfsnet, lin2022quality, hafhouf2022improved, gu2022net,barin2022automatic, anand2022fusion, feng2022bla, dong2022learning, ahmed2022new, krishna2021mlrnet, adegun2021probabilistic, chowdary2021automated} have customized CNN networks by exploiting different segmentations architectures like U-Net, FCN, and SegNet \citep{csahin2022robust}. It is notable that many recent works applied different attention mechanisms (channel, spatial, and/or atrous attention) in the CNN-based SLS methods \citep{tong2021ascu, chowdhury2021exploring, barata2021explainable, pennisi2022skin, nour2022skin, tao2021attention, jignesh2021automated, arora2021automated, wibowo2021lightweight, mu2021channel, mirikharaji2021d, hu2022net, le2021modified, dong2021fac, zhang2022dense, wang2022skin, ren2022serial, chen2022skin, tran2022fully, liu2022skin, sun2022acfnet, stofa2022u, liu2022ncrnet, li2022mhau, alahmadi2022multiscale}. Some authors also proposed semi-supervised CNN-based SLS techniques \citep{xie2021semi, zhang2022dynamic, barzegar2022skin, alahmadi2022semi}. Recently, transformer models were also applied for automated lesion segmentation \citep{wang2021boundary, wu2022fat, gulzar2022skin, feng2022slt, dong2022tc}. Often, the authors ensembled multiple SLS models or outputs to produce a refined lesion segmentation \citep{nampalle2020efficient, anjum2020deep, lei2020skin, qiu2020inferring, saha2020leveraging, canalini2019skin, goyal2019skin, liu2019skin, bi2018improving, bissoto2018deep, chen2018multi, guth2018skin, qian2018detection, alvarez2017k, yuan2017automatic, liu2021skin, dai2022ms, song2022res, redha2021skin, qamar2021dense, peter2021internet, tang2021afln}. For example, a bagging-type ensemble approach was executed in \citep{yuan2017automatic} to combine the outputs of various FCNs while improving the image segmentation performance on the testing images, multi-scale ensembles \citep{liu2021skin, dai2022ms, song2022res}, and adaptive ensembles \citep{tang2021afln}. Again, the authors in \citep{qian2018detection, chen2018multi, liu2019skin} ensembled different outputs at the post-processing stage to increase the segmentation results. Firstly, they segmented an input image by augmenting it to generate various outputs. Then reverse operations were performed on the results that were finally averaged to get a lesion mask. DeeplabV3 \citep{redha2021skin, bagheri2021skin} and Mask R-CNN \citep{bagheri2021skin, ahmed2022new} were employed in \citep{goyal2019skin} to propose three ensemble model variants. Ensemble-ADD combines the results from both models, Ensemble-Comparison-Large picks the larger segmented area by comparing the number of pixels in the output of both methods, and Ensemble-Comparison-Small picks the smaller area from the output. Lastly, a pixel-wise majority voting ensemble for lesion segmentation was proposed and implemented in \citep{qiu2020inferring}.} Also, the authors in many articles have incorporated different post-processing methods to refine the segmented lesion masks, as explained in Table~\ref{tab:Postprocessing}.

{\scriptsize
\begin{longtable}[c]{p{13cm}|p{2.6cm}}
\caption{\MKH{Commonly employed post-processing for SLS from 2011 to 2022, scrutinizing a total of 356 for lesion segmentation.}}
\label{tab:Postprocessing}\\
\Xhline{1pt}
\textbf{Details post-processing method} & \textbf{Articles} \\ \Xhline{1pt}
\endfirsthead
\multicolumn{2}{c}%
{{\bfseries Table \thetable\ Continued from previous page}} \\
\Xhline{1pt}
\textbf{Details post-processing method} & \textbf{Articles}  \\ \Xhline{1pt}
\endhead

Conditional random fields, as a statistical modeling method for structured prediction, considering neighboring pixels' information in the new pixel prediction. & \citep{qiu2020inferring, wu2019skin, baghersalimi2019dermonet, tschandl2019domain, de2019skin, luo2018fast}  \\\hline

The binary lesion masks are followed by morphological dilation operations. The region closer to the image center is picked, followed by unwanted components like corner effect removal and filling of the small holes.  & \citep{pour2020transform}  \\\hline

The best cluster from all the predicted clusters is determined by calculating the mean value for each cluster and selecting the maximum mean value. Then morphology operations, for example, opening, closing, and filling holes are applied to enhance the segmented lesion masks. 	 & \citep{sayed2020novel} \\\hline
 
Subtraction of smaller or misclassified areas and sharp fragments from the mask's borders. 	& \citep{santos2020skin, ammar2018learning} \\\hline
  
Iterative self-organizing data analysis technique was employed to threshold the output probability map from the sigmoid activator of their DSNet to retrieve the lesion mask. &  \citep{hasan2020dsnet} \\\hline

Simply morphological smoothing and extracting the largest connected component from the predicted binary lesion mask.	&  \citep{saha2020leveraging, huang2019skin, schaefer2011colour, jafari2016skin, mishra2017deep, gupta2017adaptive} \\\hline
 
Watershed-based postprocessing feeds high-confidence pixel classifications as seeds into the watershed algorithm. & \citep{low2020automating} \\\hline
  
Several consecutive morphological filtering, such as dilation, erosion, closing, and opening, for enhancing the predicted lesion boundary. & \citep{rizzi2020skin, yang2019sampling, unver2019skin, lameski2019skin, javed2019intelligent, hasan2019comparative, goyal2019skin, nguyen2018isic, yuan2017automatic, abbas2014combined, ch2014two, alvarez2017k, martinez2017pigmented, ramya2021segmentation, joseph2022preprocessing} \\\hline

A convex hull operation follows the morphological opening process of the predicted binary mask for lesion shape approximation.  & \citep{ribeiro2020less} \\\hline
 
Region merging and morphological operations are applied to obtain the final lesion mask. 	  & \citep{ramella2020automatic} \\\hline
  
Fill in the small holes in the predicted binary lesion mask.  & \citep{gangwar2021study, tang2019efficient, bissoto2018deep, lin2017skin, jafari2016skin, alvarez2017k, mishra2017deep, gupta2017adaptive} \\\hline

The output probability map is subjected to three consecutive post-processing actions constructing the final mask: thresholding the probability map at 0.5, choosing the largest connected component, and finally, filling the small holes. & \citep{shahin2019deep, riaz2018active} \\\hline
 
k-means clustering and flood-fill techniques are combined to get the final lesion mask. Then, holes are sealed using a hole-filling algorithm. & \citep{jaisakthi2018automated} \\\hline
  
The central lesion location is identified, then hole filling and morphological closing are applied to fill tiny holes and islands. For the final segmented mask, a Gaussian mask looks at how closely connected regions are to the image's center.  	&  \citep{hu2018skin} \\\hline
 
A dual-threshold method to generate a binary mask, where a relatively high threshold (= 0.8) is devoted to determining the lesion center, and a lower threshold (= 0.5) is applied to the output map. After filling small holes with morphological dilation, the final lesion mask is decided. & \citep{yuan2017automatic, xu2018automatic} \\\hline
  
The largest connected binary objects are kept and joined. Then, morphological operations are performed to patch gaps and remove excess skin. Finally, lesion masks are smoothed using a convolution filter. & \citep{khalid2016segmentation, csahin2022robust} \\\hline

A graph-cut algorithm is applied to the output score maps to fine-tune the predicted masks. & \citep{bozorgtabar2017investigating} \\\Xhline{1pt}
\end{longtable}}

The above table reveals that morphological operations, such as dilation, erosion, closing, and opening, have been the workhorse and are massively employed for post-processing segmented lesion masks \citep{ramya2021segmentation, pour2020transform, sayed2020novel,rizzi2020skin, yang2019sampling, unver2019skin, lameski2019skin, javed2019intelligent, joseph2022preprocessing, hasan2019comparative, goyal2019skin, nguyen2018isic, yuan2017automatic, abbas2014combined, ch2014two, alvarez2017k, martinez2017pigmented, hu2018skin}. Small holes and disconnected island regions in the segmented lesion masks were resolved in various ways in the 356 SLS articles. For example, hole-filling algorithms can be used to fill tiny holes \citep{gangwar2021study, tang2019efficient, bissoto2018deep, lin2017skin, jafari2016skin, alvarez2017k, mishra2017deep, gupta2017adaptive}, and island regions can be removed by preserving the target region(s) using the region property techniques \citep{saha2020leveraging, huang2019skin, schaefer2011colour, jafari2016skin, mishra2017deep, gupta2017adaptive}. In some articles \citep{qiu2020inferring, wu2019skin, baghersalimi2019dermonet, tschandl2019domain, de2019skin, luo2018fast}, the authors have employed Conditional random field techniques to refine the segmented masks from the coarse output masks.

\section{Classification Techniques}
\label{Classification_Techniques}
Computational SLC methods related to the lesion features are founded on the ABCD(E) rule \citep{melbin2021integration, chowdhury2021exploring, imtiaz2021efficient, reimers2021conditional}, pattern analysis, the seven-point checklist, and Menzies' method. Those methods are illustrations of clinical techniques used for the prognosis and diagnosis of image-based skin cancer \citep{oliveira2018computational}, color, diameter (or differential structures in the case of dermoscopic images), and evolution (or elevation) features, according to the standards delivered in \citep{oliveira2018computational}. The other traditional methods of the SLC have been reviewed in detail in \citep{oliveira2018computational}. However, those SLC methods require a dermatologist for the naked-eye assessment, which may incur subjectivity associated with human evaluations and human errors. Dermatologists' manual examination is usually monotonous, time-consuming, and subjective. The accuracy of manual assessment can also depend on the reviewer's experience and workload. Hence, CAD systems have been developed to avoid the above-mentioned limitations. This article focuses on reviewing automated CAD systems. There has been a significant advancement in developing automated lesion classification algorithms and approaches for SLA in the recent past \citep{kassem2021machine}. Some methods were based on manual feature engineering with ML models, but DL-based automatic feature learning schemes are becoming more popular \citep{kassem2021machine}. These two main approaches are surveyed and reviewed below.

\subsection{ML-based SLC}
\label{ML_SLC}
As noted previously, ML-based SLC strategies depend on successfully engineering the manual lesion features that are first described and reviewed in section~\ref{Lesionfeatures}. Then, the employed classifiers are inspected in section~\ref{Lesionclassifiers}.

\subsubsection{Lesion Features}
\label{Lesionfeatures}
Lesion attributes (or features) can be extracted either globally or locally to acquire category information. Most works explore the global features of the lesion, for instance, extracting features from all segmented regions \citep{chakravorty2016dermatologist, wahba2017combined, filali2017multiscale, nasir2018improved, khan2018implementation, seeja2019deep, mahbod2020effects}. Some examinations have employed local characteristics, allowing the characterization of a diverse lesion region. Lesion features can generally be organized into different categories: shape, color variation, texture analysis, and other miscellaneous. These lesion features are summarized in Table~\ref{tab:SLC_features}.

{\scriptsize
\begin{longtable}{p{13.1cm}p{2.5cm}}
\caption{\MKH{Different pigmented skin lesion features from macroscopic and dermoscopic images in the past from 2011 to 2022.}}
\label{tab:SLC_features}\\
\Xhline{1pt}
\textbf{Different lesion attributes} & \textbf{Articles} \\ \Xhline{1pt}
\endfirsthead
\multicolumn{2}{c}%
{{\bfseries Table \thetable\ Continued from previous page}} \\
\Xhline{1pt}
\textbf{Different lesion attributes} & \textbf{Articles} \\ \Xhline{1pt}
\endhead
\multicolumn{2}{c}{\textbf{Shape-based lesion attributes}} \\ \Xhline{1pt}
Convex hull to estimate notched and ragged edges &   \citep{ramlakhan2011mobile} \\ \hline
                                         
Circularity Index ($(4\pi A)/P^2$) for border's irregularity, where A and P are the lesion contour's area and perimeter, respectively. & \citep{ramlakhan2011mobile, cavalcanti2013macroscopic, almaraz2020melanoma} \\\hline   

Hull/Contour Ratio to measure the raggedness or spikiness of the lesion border & \citep{ramlakhan2011mobile, cavalcanti2013macroscopic}  \\\hline 

The lesion boundary comes from a snake-based edge detection
technique, segmenting the image into skin area $A_s$ and lesion area $A_l$. The average skin pattern isotropies in the skin ($m_s$) and lesion ($m_l$) areas are calculated as $m_s=\frac{1}{N_s}\sum_{(i,j)\in A_s}I(i,j)$ and $m_l=\frac{1}{N_l}\sum_{(i,j)\in A_l}I(i,j)$, where $N_s$ and $N_l$ are the number of sub-images in the skin and lesion areas. & \citep{she2011skin, she2013lesion} \\\hline

Clinical border irregularity features, delineating a skin lesion into eight segments. & \citep{amelard2012extracting}  \\\hline
                                         
Morphologically fine irregularities feature from the image I, as $f^B=\frac{I_{c}-I_l}{I_l}+\frac{I_{l}-I_o}{I_l}$, where $I_l$, $I_c$, and $I_o$ respectively denote the original lesion area, ROI's closing, and ROI's opening.     & \citep{amelard2012extracting}  \\\hline

Coarse irregularities from the perimeters of the low-frequency border and the original border as $f^B = \frac{|P_{lesion}-P_{low}|}{P_{lesion}}$, where $P_{lesion}$ and $P_{low}$ are the lengths of the perimeter of the original and low-frequency border (details in \citep{amelard2012extracting}).  & \citep{amelard2012extracting} \\\hline

Structural irregularities from the Fourier descriptors as $f^B=\sum_{u=0}^{N-1}\big(|C(u)|-|\bar{C(u)}| \big)^2$, where $C$ is the Fourier coefficient (details in \citep{amelard2012extracting}).  & \citep{melbin2021integration, amelard2012extracting}  \\\hline

Asymmetry features were extracted by a shrinking active contour model to find major and minor axes, vertical and horizontal dash lines, of the lesion boundary.  & \citep{mete2012skin, imtiaz2021efficient, piatek2022analysis}   \\\hline

The asymmetry feature by computing the principal and secondary axes of inertia, where the axes of the image were aligned with the axes of inertia, allowing a better assessment of the lesion symmetry in terms of geometry and internal structures.      & \citep{rosado2013prototype}   \\\hline

The border feature for finding the abrupt ending of pigment pattern in two peripheral regions: inside and outside the lesion, using the Euclidian Distance Transform.  & \citep{rosado2013prototype}   \\\hline

The asymmetry characterization features such as the ratio between the lesion area and its bounding box area, equivalent diameter ($4A/(L_1\pi$), the ratio between the principal axes ($L_2/L_1$), the ratio between sides of the lesion bounding box, the ratio between the lesion perimeter ($p$) and its area ($A$), $(B_1 -B_2)/A$ [$B_1$ and $B_2$ are the areas in each side of axis $L_1$ or $L_2$], and $B_1/B_2$ ratios concerning the axis $L_1$ or $L_2$.   & \citep{cavalcanti2013macroscopic, farooq2016automatic, lynn2017segmentation, wahba2018novel, fisher2019classification, almaraz2020melanoma}  \\\hline

Boundary Irregularity description features like average and variance gradient magnitudes of the pixels in the extended lesion rim in each of the three channels ($i=1,2,3$); average and variance of the $R\,(=1,2,...,8)$ $\mu_{R,i}$ values in each of the three channels. $\mu_{R,i}$ is the mean of 8 different symmetric regions, obtained by rotating orthogonal axes by 45 degrees.   & \citep{cavalcanti2013macroscopic, lynn2017segmentation, melbin2021integration}  \\\hline

Ulnar variance measures the relative length of articular surfaces of some particular radius and image asymmetry.  & \citep{farooq2016automatic}   \\\hline

Solidity, asymmetry index, extent, diameter, circularity, eccentricity, aspect ratio, structural similarity, and the ratio of the major to the minor axis as structural features.  & \citep{chakravorty2016dermatologist, chatterjee2019integration, damian2020feature, dhivyaa2020skin}  \\\hline
 
2D and 3D shape features (details in \citep{satheesha2017melanoma}).  & \citep{satheesha2017melanoma}           \\ \Xhline{1pt}

\multicolumn{2}{c}{\textbf{Color-based lesion attributes}} \\ \Xhline{1pt}

R, G, B colors means ($\mu=(\mu_R, \mu_G, \mu_B)$) and their covariance matrices ($\Sigma$) (details in \citep{ballerini2012non}). &           \citep{ramlakhan2011mobile, ballerini2012non, melbin2021integration, peter2021internet}        \\ \hline

Six color features from two-stage detection algorithms: color clustering using a mean-shift algorithm and color supervised classification based on a dataset of reference RGB codes. & \citep{rosado2013prototype}  \\\hline

Maximum, minimum, mean, and variance of the intensities of the pixels inside the lesion segment in the color variation channel and each of the three original channels; ratios between the mean values of the original three channels, for example, $\mu_R/\mu_G$, assuming only pixels inside the
lesion segment.    & \citep{cavalcanti2013macroscopic, satheesha2017melanoma, lynn2017segmentation, wahba2018novel, nasir2018improved, khan2018implementation, seeja2019deep, yildirim2020pre, ghalejoogh2020hierarchical, almaraz2020melanoma}  \\ \hline

The variance, skewness, and entropy as color-related features.    & \citep{farooq2016automatic, khan2018implementation, seeja2019deep, yildirim2020pre, almaraz2020melanoma, dhivyaa2020skin} \\ \hline

Boundary color value differences for each channel as $f=\frac{1}{N}\sum_{i=1}^N(V_i-V_m)^2$, where $V_i$, $V_m$, and $N$ are the value of $i^{th}$ boundary pixel, the mean value of the boundary pixel, and the number of boundary pixels.   & \citep{mahdiraji2017skin} \\ \hline

Boundary color clustering features, clustering a lesion with different k groupings (details in \citep{mahdiraji2017skin}).   & \citep{mahdiraji2017skin} \\ \hline

Color features of six different colors like white, red, light brown, dark brown, blue-gray, and black (estimation details in \citep{filali2018study}).   & \citep{filali2018study, filali2019improved} \\ \hline

Histogram-based features (color distribution)   & \citep{barata2015melanoma, rastgoo2015automatic, schaefer2014ensemble, shahabi2021performance, moldovanu2021skin,shetty2022skin, peter2021internet, nersisson2021dermoscopic} \\ \hline

Color asymmetry & \citep{schaefer2014ensemble} \\\hline

Kullback-Leibler divergence of the color distribution (each channel separately) from the two halves along each axis.                    &   \citep{chakravorty2016dermatologist}               \\ \Xhline{1pt}

\multicolumn{2}{c}{\textbf{Texture-based lesion attributes}} \\ \Xhline{1pt}

Twelve generalized cooccurrence matrices features like energy, contrast, correlation, entropy, homogeneity, inverse difference moment, cluster shade, cluster prominence, max probability, autocorrelation, dissimilarity and variance.     & \citep{ballerini2012non, nersisson2021dermoscopic, shahabi2021performance, shetty2022skin}  \\ \hline

Five differential structures based on texture relevant for the detection of melanoma: homogeneous areas, streaks, dots, globules, and pigment
network.    & \citep{rosado2013prototype} \\ \hline

Textural variation in the channel as maximum, minimum, mean, and variance of the intensities of the pixels inside the lesion segment.    & \citep{cavalcanti2013macroscopic, lynn2017segmentation}  \\ \hline

Texture-base Gray Level Co-occurrence Matrix (GLCM) features that include mean, correlation, homogeneity, contrast, energy, dissimilarity, kurtosis, variance, entropy, maximum probability, inverse difference, angular second moment, and standard deviation. & \citep{farooq2016automatic, filali2017multiscale, danpakdee2017classification, songpan2018improved, filali2018study, filali2019improved, chatterjee2019integration, monisha2019classification, abbas2019efficient, ghalejoogh2020hierarchical, almaraz2020melanoma, nersisson2021dermoscopic,serrano2022clinically, peter2021internet, samsudin2022skin}     \\ \hline

Haralick texture features using gray-tone spatial-dependence matrices (details in \citep{satheesha2017melanoma}).  & \citep{satheesha2017melanoma, shetty2022skin, shahabi2021performance} \\\hline

SFTA features (details in \citep{nasir2018improved}).  & \citep{nasir2018improved} \\\hline

SURF, SIFT, and ORB features & \citep{yildirim2020pre} \\\hline

Texture-based RSurf features & \citep{majtner2016combining} \\ \hline

Bi-dimensional Empirical Mode Decomposition (BEMD), BEMD-Riesz, Gray-level Difference Method (GLDM), and combined BEMD-Riesz with GLDM.    & \citep{wahba2017combined, melbin2021integration}  \\ \hline

Fractal dimensions and GLDM features (details in \citep{wahba2018novel}). & \citep{wahba2018novel, chatterjee2019integration, melbin2021integration, moldovanu2021skin} \\ \hline

Histogram of oriented gradients features (details in \citep{nasir2018improved}).  & \citep{nasir2018improved, khan2018implementation, filali2019improved, seeja2019deep} \\\hline

Gabor wavelets features (details in \citep{filali2019improved}).  & \citep{filali2019improved, seeja2019deep, nersisson2021dermoscopic, deng2022efficient} \\\hline

Fractal-based regional texture analysis-based texture features (details in \citep{chatterjee2019integration}).  & \citep{chatterjee2019integration} \\\hline

Edge and local edge Histogram features (details in \citep{seeja2019deep}).  & \citep{seeja2019deep} \\\hline

Texture feature-based on fractional Poisson to estimate the structure of regions in an image (details in \citep{al2017classification}). & \citep{al2017classification})              \\ \hline

 Coarseness features that measure of different angle of texture representation.                     &  \citep{farooq2016automatic}                 \\ \Xhline{1pt}

\multicolumn{2}{c}{\textbf{Other (miscellaneous) lesion attributes}} \\ \Xhline{1pt}

Local Binary Patterns (LBP) features    &  \citep{majtner2016combining, filali2019improved, seeja2019deep, monisha2019classification, yildirim2020pre, pereira2020skin, shahabi2021performance, samsudin2022skin, nersisson2021dermoscopic} \\ \hline

Manual information & \citep{giotis2015med} \\\hline

Model-based features. & \citep{garnavi2012computer, oliveira2018computational})              \\ \Xhline{1pt}

\end{longtable}
}

Shape features consider the lesion's asymmetry or border's irregularity, dividing the lesion region into two sub-regions by an axis of symmetry to analyze the area's similarity by overlapping the two sub-regions of the lesion along the axis. Then, the asymmetry index is estimated by the difference between the two sub-regions of the lesion, for example, with the XOR operation between them \citep{oliveira2018computational}. Sometimes, geometrical measurements from the segmented lesion area are computed to assess the lesion's asymmetry and border irregularity, which include the area of the lesion, aspect ratio, compactness, perimeter, greatest diameter, shortest diameter, equivalent, convex hull, eccentricity, solidity, rectangularity, entropy measures, circularity index, and irregularity index (see details in Table~\ref{tab:SLC_features}) \citep{chakravorty2016dermatologist, chatterjee2019integration, damian2020feature, dhivyaa2020skin}. Out of the 238 SLC articles, 33 articles ($13.9\,\%$) employed lesion shape features in the last twelve years for the SLC task.

The RGB color space is commonly employed to represent the colors of skin lesions. Other color spaces have also been applied in order to acquire more specific information about a lesion's colors, such as normalized RGB, HSV, HVC, CMY, YUV, I1/2/3, Opp, IiN, JCh, L*C*H, CIEXYZ, CIELAB, and CIELUV (see details in \citep{oliveira2018computational}). Several statistical measures such as minimum, maximum, average, standard deviation, skewness, and variance are widely applied to feature extraction from skin lesion images, computing each color channel of the lesion region using one or several color models \citep{maglogiannis2015enhancing, ramlakhan2011mobile, ballerini2012non, cavalcanti2013macroscopic, satheesha2017melanoma, lynn2017segmentation, wahba2018novel, nasir2018improved, ghalejoogh2020hierarchical, almaraz2020melanoma, farooq2016automatic, khan2018implementation, seeja2019deep, yildirim2020pre, dhivyaa2020skin}. Furthermore, these measures may also be applied to other regions associated with the lesion's border to identify a sharp transition, indicating malignancy. Skin lesion features based on relative colors have been proposed to assess color features in different regions associated with the lesion. The relative color consists of comparing each pixel value of the lesion to the average color value of the surrounding skin. Likewise, the use of this feature may present advantages such as compensating for the variability in the image's color caused by uneven illumination and varying equalization in skin color across individuals. The occurrence of possible primary colors present in skin lesions has also been analyzed through the quantification of the number or percentage of pixels within the segmented area for each of the primary colors \citep{abbas2013pattern}. Out of the 238 SLC articles, 34 SLC articles ($14.3\,\%$) employed lesion color features in the past SLC task.

Texture analysis is generally used to discriminate between benign and malignant lesions by estimating their structure's roughness, encompassing descriptors like statistical-, model-, and filter-based methods \citep{oliveira2018computational}. Among the various statistical-based texture descriptors applied in the literature, the Gray Level Co-occurrence Matrix (GLCM) has been one of the most commonly utilized \citep{farooq2016automatic, filali2017multiscale, danpakdee2017classification, songpan2018improved, filali2018study, filali2019improved, chatterjee2019integration, monisha2019classification, abbas2019efficient, ghalejoogh2020hierarchical, almaraz2020melanoma}. GLCM is a statistical measure that computes the joint probability of occurrence of gray levels considering two pixels spatially separated by a fixed vector. Several measures may be computed based on the GLCM, such as mean, correlation, homogeneity, contrast, energy, dissimilarity, kurtosis, variance, entropy, maximum probability, inverse difference, angular second moment, and standard deviation \citep{farooq2016automatic, filali2017multiscale, danpakdee2017classification, songpan2018improved, filali2018study, filali2019improved, chatterjee2019integration, monisha2019classification, abbas2019efficient, ghalejoogh2020hierarchical, almaraz2020melanoma}. Model-based texture descriptors have also been proposed to assess the skin lesion's texture, such as fractal dimensional \citep{wahba2018novel, chatterjee2019integration}, auto-regression, and Markov random fields. Among these, fractal dimension has been applied with the box-counting method, one of the most commonly used methods since it is simple and effective \citep{wahba2018novel, chatterjee2019integration}. Wavelet, Fourier, and Gabor transform \citep{filali2019improved, seeja2019deep}, and the Scale-invariant Feature Transform (SIFT) \citep{yildirim2020pre, nasir2018improved}, which are filter-based texture descriptors, have also been proposed for feature extraction of skin lesion images. Such descriptors allow the decomposition of the input image into parts in order to extract features from the structures of interest. Sobel, Hessian, Gaussian, and difference of Gaussian features have also been extracted based on the bank of Gaussian filters \citep{nasir2018improved, khan2018implementation, filali2019improved, seeja2019deep}. 46 articles ($19.3\,\%$) of 238 SLC articles utilized lesion texture characteristics in the past twelve years for the SLC task.

\begin{figure}[!ht]
\centering
\includegraphics[width=12cm, height=3.0cm]{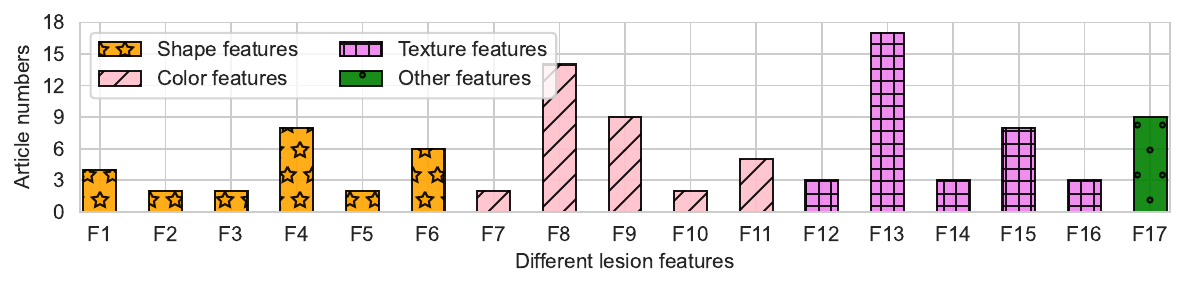}
\caption{\MKH{The frequency of utilization of various skin lesion attributes with their complementary number of employed articles for the SLC task. F1$\mapsto$ Circularity index, F2$\mapsto$ Contour ratio, F3$\mapsto$ Average skin pattern isotropies, F4$\mapsto$ Asymmetry characterization, F5$\mapsto$ Boundary irregularity description, F6$\mapsto$ Structural features, F7$\mapsto$ RGB colors means and covariances, F8$\mapsto$ Maximum, minimum, mean, and variance of the intensities, F9$\mapsto$ Variance, skewness, and entropy as color-related features, F10$\mapsto$ White, red, light brown, dark brown, blue-gray, and black color features, F11$\mapsto$ Color histogram-based features, F12$\mapsto$ Textural variation in the channels, F13$\mapsto$ Gray level co-occurrence matrix, F14$\mapsto$ Fractal dimension, F15$\mapsto$ Histogram of oriented gradients, F16$\mapsto$ Gabor wavelets features, and F17$\mapsto$ Local binary patterns.}}
\label{fig:Feature_Used}
\end{figure}

Some authors combined the color and texture features in some articles \citep{ballerini2012non, giotis2015med, sadeghi2012global, monisha2019classification} to construct a distance measure between a test image and a database image, using color covariance-based features and the Bhattacharyya distance metric \citep{ballerini2012non}. Although texture-based lesion features are slightly more commonly exploited, most articles apply all three types of features to represent the feature vector with all categories of lesion characteristics, as various feature types expose various properties of the skin lesion. Fig.~\ref{fig:Feature_Used} displays the most generally applied (at least in two articles) lesion characteristics from Table~\ref{tab:SLC_features} in the past from 2011 to 2022 for the SLC task.  
A close observation in Fig.~\ref{fig:Feature_Used} indicates a wide range of attributes from each type of feature category, revealing that Asymmetry characterization (F4), Maximum, minimum, mean, and variance of the intensities (F8), Gray level co-occurrence matrix (F13), and Local binary patterns (F17) are the most common shape, color, texture, and other types of lesion attributes, respectively. These lesion features can be considered to be the representative lesion characteristics of those attribute categories (\textbf{\textit{High-frequency}}) used in SLC, and their popularity likely indicates more substantial effectiveness. Again, structural features, skewness, and entropy as color-related features and histograms of oriented gradients are the next most commonly used (\textbf{\textit{Medium-frequency}}) lesion characteristics of shape-, color-, and texture-based lesion attributes, whereas the other remaining attributes in Fig.~\ref{fig:Feature_Used} are applied in a few articles (\textbf{\textit{Low-frequency}}) in the past.

\subsubsection{Lesion Classifiers}
\label{Lesionclassifiers}
After feature extraction, a feature selection step is crucial \citep{hasan2021associating}, and has been employed for the SLC task to determine the most relevant features and reduce the dimensionality of the feature space \citep{melbin2021integration, iyatomi2010classification, ghalejoogh2020hierarchical, yildirim2020pre, pereira2020skin, wahba2018novel, chatterjee2019integration, mohanty2022integrated, batista2022classification}. Moreover, such features may influence the performance of the classification process, i.e., render it slower. Several benefits are associated with the application of feature selection schemes \citep{oliveira2018computational} such as reducing the feature extraction time, decreasing the classification complexity, improving the classification accuracy rate, lowering training-testing time, simplifying the understanding, and visualizing the data. The feature selection process has the following steps: feature subset selection, feature subset evaluation, stopping criterion, and validation procedure \citep{oliveira2018computational}, which is then heeded by lesion classification step(s). 

The classification phase consists of recognizing and interpreting the information about the skin lesions based on extracted and selected lesion features. The classification process is generally accomplished by randomly dividing (or K-folding) the available image samples into training and testing sets. The training step consists of developing a classification model to be employed by one or more classifiers based on the samples of the training set. Each sample comprises features extracted from a provided image and its corresponding class value, which are applied as input data to the classifier for the learning process. The testing step measures the model's accuracy learned by the training step over the test set. From 2011 to 2022, the commonly used SLC models were Support Vector Machine (SVM) \citep{melbin2021integration, shetty2022skin, samia2021skin, nancy2022impact, mohanty2022integrated, batista2022classification, thapar2022novel, mete2012skin, amelard2012extracting, patil2014detection, sirakov2015skin, chakravorty2016dermatologist, farooq2016automatic, al2017classification, danpakdee2017classification, filali2017multiscale, lynn2017segmentation, ozkan2017skin, satheesha2017melanoma, wahba2017combined, hardie2018skin, khan2018implementation, nasir2018improved, songpan2018improved, wahba2018novel, abbas2019efficient, chatterjee2019integration, filali2019improved, javed2019intelligent, khan2019construction, ghalejoogh2020hierarchical, guha2020performance}, K-nearest Neighbors (KNN) \citep{moldovanu2021skin, shetty2022skin, batista2022classification, afza2022multiclass, ramlakhan2011mobile, ballerini2012non, cavalcanti2013macroscopic, patil2014detection, di2015hierarchical, chakravorty2016dermatologist, al2017classification, lynn2017segmentation, ozkan2017skin, fisher2019classification, afza2020skin, ghalejoogh2020hierarchical}, AdaBoost (AdB) \citep{shahabi2021performance, imtiaz2021efficient, chakravorty2016dermatologist, al2017classification, satheesha2017melanoma, mporas2020color}, Decision Tree (DT) \citep{shetty2022skin,nancy2022impact, mohanty2022integrated, chakravorty2016dermatologist, ozkan2017skin, dhivyaa2020skin}, Random Forest (RF) \citep{shetty2022skin, imtiaz2021efficient, shahabi2021performance, batista2022classification, dhivyaa2020skin, mporas2020color, rosado2013prototype}, Artificial Neural Network (ANN) \citep{farooq2016automatic, ozkan2017skin, nancy2022impact, samsudin2022skin, moldovanu2021skin}, Multilayer Perceptron Neural Network (MPNN) \citep{ghalejoogh2020hierarchical, serrano2022clinically, mohanty2022integrated,batista2022classification, ayas2022multiclass, dong2022learning}, Linear Discriminant Analysis (LDA) \citep{khan2019construction, shetty2022skin}, Quadratic Discriminant Analysis (QDA) \citep{khan2019construction}, Naive Bayes (NB) \citep{chakravorty2016dermatologist, shetty2022skin, imtiaz2021efficient,mohanty2022integrated, afza2022multiclass, afza2022hierarchical}, K-means Clustering (KMC) \citep{celebi2014automated}, Probabilistic Neural Network (PNN) \citep{jayapal2014skin}, Feed Forward Back Propagation Neural Network (FFBPNN) \citep{patil2014detection}, Ensemble Binary Classifiers (EBC) \citep{surowka2014optimal}, and Elman Neural Network (ENN) \citep{ghalejoogh2020hierarchical}.

\begin{figure}[!ht]
\centering
\includegraphics[width=6cm, height=5cm]{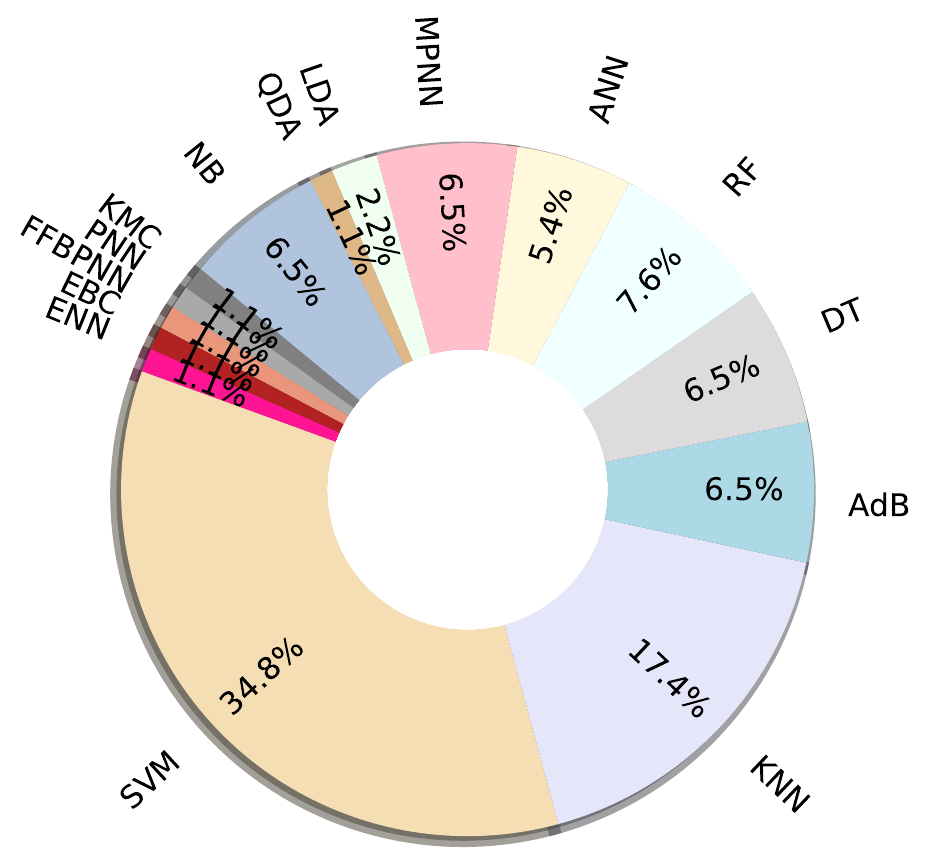}
\caption{\MKH{The pie chart of the percentage of SLC articles employing various ML models on manual lesion features in the past from 2011 to 2022.}}
\label{fig:Pie_slc_ML_methods}
\end{figure}

\MKH{Fig.~\ref{fig:Pie_slc_ML_methods} demonstrates the percentage of publications from 2011 to 2022 that use ML models, revealing that the top ML models are SVM and KNN, which account for $34.8\,\%$ and $17.4\,\%$ of publications, respectively.} 
\MKH{These two ML models were applied in $52.2\,\%$ of articles; therefore, they can be regarded as the representative ML classification models (\textbf{\textit{High-frequency}}) for SLC. This is followed by AdB, DT, RF, ANN, MPNN, and NB classifiers, which were the next most popular models (\textbf{\textit{Medium-frequency}}) and accounted for $39.0\,\%$ of SLC articles. The other ML models, such as LDA, QDA, KMC, PNN, FFBPNN, EBC, and ENN, were less significant (\textbf{\textit{Low-frequency}}), accounting for $8.8\,\%$ of SLC articles.} Quick calculation time, a simple algorithm to interpret, versatile usefulness for regression and classification, and evolution with new data are potential reasons for the high frequency of usage for SVM and KNN models.

\subsection{DL-based SLC}
\label{DL_SLC}
As discussed in the earlier section, ML-based SLC requires extensive feature engineering with comprehensive parameter tuning to achieve better and more robust performance, which is often challenging due to the appearance of various intrinsic and extrinsic noises in dermoscopic images (see details in \citep{hasan2020dsnet, hasan2021dermo}). On the other hand, the CNN-based DL-SLC approaches provide excellent SLC results and boost diagnostic procedure rates while being end-to-end systems. Nowadays, CNN-based segmentation networks have been widely applied to medical images, outperforming traditional image processing methods relying on manual features \citep{hasan2021drnet}. \MKH{However, our review of the 238 SLC articles demonstrates that most authors employed pre-trained (on ImageNet, PASCAL-VOC, MS-COCO, etc.) CNN models and fine-tuned them with the skin lesion datasets. The most commonly employed pre-trained CNN models found in the SLC literature from 2011 to 2022 are: MobileNet \citep{yilmaz2022mobileskin, jain2021deep, bansal2021skin,batista2022classification, pundhir2022towards, wang2022ssd, sahin2022human, popescu2022skin, yilmaz2022mobileskin, miglani2020skin, thurnhofer2021skin, salian2020skin, rodrigues2020new, almaraz2020melanoma, dos2018robust, guissous2019skin}, variants of EfficientNet \citep{sarker2022transslc, zhuang2022cs, shen2022low, serrano2022clinically, mohanty2022integrated, sahin2022human, somfai2022handling,hsu2022hierarchy, pundhir2022towards, santos2021transfer, zanddizari2021new, sun2021skin, alptekin2022analysis, miglani2020skin, mahbod2020transfer, mahbod2020effects, gessert2020skin, zhuang2020cs, gessert2019skin}, DenseNet \citep{rahman2021approach, nguyen2022skin, zanddizari2021new, nakai2022dpe, shan2022automatic, bansal2021skin, samia2021skin, khan2021skin, popescu2022skin, yue2022towards, liu2021multiscale, santos2021transfer, wang2021multi, mahbod2021investigating, yue2022towards, hsu2022hierarchy,batista2022classification,pundhir2022towards, rodrigues2020new, rahman2020transfer, liu2020automatic, jibhakate2020skin, hassan2020skin, almaraz2020melanoma, wu2020multi, akram2020multilevel, molina2020classification, li2018skin, lee2018wonderm, gessert2019skin, guissous2019skin, rashid2019skin}, SeReNeXt-50 \citep{yue2022towards, he2022deep, nakai2022enhanced, nakai2022dpe, mahbod2020transfer}, varients of VGGNet \citep{pundhir2022towards, nakai2022dpe} like VGG16 \citep{nunnari2021overlap, barata2021improving, santos2021transfer,bansal2021skin, calderon2021bilsk, samanta2021skin, serrano2022clinically,rasel2022convolutional, mohanty2022integrated, khan2022ensemble, batista2022classification, goceri2020comparative, jia2017skin, guissous2019skin, salian2020skin, rodrigues2020new, kwasigroch2020neural, nunnari2020study, kamalakannan2020self, jibhakate2020skin, bagchi2020learning, almaraz2020melanoma, thomsen2020deep, guha2020performance, harangi2018skin, chen2018multi, bissoto2018deep, khan2019skin, lopez2017skin, mahbod2019skin, mahbod2019fusing} and VGG19 \citep{zanddizari2021new, santos2021transfer, jain2021deep, balabantaray2021melanoma, samanta2021skin, rasel2022convolutional, batista2022classification, goceri2020comparative, kwasigroch2020neural, dos2018robust}, LeNet-5 \citep{kamalakannan2020self, namozov2018adaptive, albahar2019skin}, GoogleNet \citep{thurnhofer2021skin, rasel2022convolutional, sahin2022human, popescu2022skin, goceri2020comparative, yilmaz2020benign, bagchi2020learning, harangi2020assisted, harangi2018skin}, PNASNet-5-Large \citep{zhuang2020cs, milton2019automated}, Squeeze-and-Excitation Networks (SENet) \citep{gessert2020skin, kitada2018skin, zhuang2020cs, milton2019automated, rasel2022convolutional}, AlexNet \citep{rasel2022convolutional, khan2022ensemble, popescu2022skin, yilmaz2020benign, bagchi2020learning, harangi2018skin, chen2018multi, khan2019skin, majtner2016combining, liao2016skin, mahbod2019skin, mahbod2019fusing}, Xception \citep{zanddizari2021new, jain2021deep, bansal2021skin, calderon2021bilsk, rahman2021approach, yilmaz2022mobileskin,sarker2022transslc, batista2022classification, zhuang2022cs, popescu2022skin, zhuang2020cs, rodrigues2020new, rahman2020transfer, almaraz2020melanoma, ahmed2020skin}, different versions of Inception model like Inception-V2 \citep{zanddizari2021new, santos2021transfer,jain2021deep, bozkurt2022skin, sarker2022transslc, nguyen2022skin, zhuang2022cs, popescu2022skin}, Inception-V3 \citep{zanddizari2021new, batista2022classification,zhuang2022cs, jain2021deep,bansal2021skin, serrano2022clinically, sarker2022transslc, goceri2020comparative, rodrigues2020new, zhuang2020cs, almaraz2020melanoma, harangi2020assisted, gessert2019skin, guissous2019skin, kulhalli2019hierarchical, mirunalini2017deep, devries2017skin}, and Inception-V4 \citep{zhuang2022cs, zhuang2020cs, pham2018deep, milton2019automated}, variants of ResNet like ResNet18 \citep{mahbod2021investigating, yue2022towards, rasel2022convolutional, nigar2022deep, sahin2022human}, ResNet101 \citep{liu2021multiscale,wang2021multi, khan2021skin, arshad2021computer, rasel2022convolutional, sarker2022transslc, popescu2022skin, goceri2020comparative, zhuang2020cs, rodrigues2020new, rahman2020transfer, liu2020automatic, yan2020scalable, jibhakate2020skin, anjum2020deep, chung2019toporesnet, khan2019multi, xue2019robust, mahbod2019fusing}, ResNet152 \citep{zanddizari2021new, santos2021transfer, deng2022efficient, zhuang2022cs, santos2021transfer, deng2022efficient, zhuang2022cs, zhuang2020cs}, ResNet50 \citep{nguyen2022skin, yang2018classification, liu2021multiscale, krohling2021smartphone, jain2021deep,cullell2021convolutional, bansal2021skin, arshad2021computer, calderon2021bilsk, mahbod2021investigating, rasel2022convolutional, mohanty2022integrated, he2022deep, qian2022skin, nakai2022enhanced, hsu2022hierarchy, batista2022classification, pundhir2022towards, nakai2022dpe, wang2022ssd, popescu2022skin, afza2022hierarchical, yilmaz2020benign, qin2020gan, nunnari2020study, kamalakannan2020self, jibhakate2020skin, bagchi2020learning, almaraz2020melanoma, muckatira2020properties, yap2018multimodal, zhang2018skin, liao2018deep, li2018skin, thandiackal2018structure, pan2018residual, harangi2018skin, dos2018robust, chen2018multi, yoon2019generalizable, fisher2019classification, al2019deep, gessert2019skin, rashid2019skin, serte2019wavelet, bisla2019skin, khan2019multi, haofu2017deep, mahbod2019skin, tschandl2019domain, mahbod2019fusing}, ResNeXt with webly supervised learning \citep{gessert2020skin, navarro2018webly, gessert2019skin, rahman2021approach}, and SE-ResneXt101 \citep{rahman2021approach, hsu2022hierarchy, zhuang2020cs, gessert2019skin}, NASNet \citep{yilmaz2022mobileskin, sahin2022human, zanddizari2021new, afza2022multiclass, zhuang2022cs, zhuang2020cs, rodrigues2020new, ahmed2020skin}, and a mixture of Inception and ResNet (InceptionResNet-V2 \citep{zanddizari2021new, santos2021transfer,jain2021deep, bozkurt2022skin, sarker2022transslc, nguyen2022skin, zhuang2022cs, popescu2022skin, zhuang2020cs, rodrigues2020new, ahmed2020skin, akram2020multilevel, guha2020performance, milton2019automated}).}

\MKH{Moreover, some authors in \citep{rasel2022convolutional, aldhyani2022multi} experimented on various CNN models by changing different activation functions. In the past, different attention-based modules were also integrated with CNN models for the SLC task \citep{nakai2022enhanced, wei2022dual, wang2022adversarial, ding2021deep, sarker2022transslc}. \citet{bdair2021semi} presented federated learning for SLC to train decentralized models in a privacy-preserved fashion relying on labeled data on the client side. The patient-specific metadata also has been incorporated with CNN models for the SLC task \citep{sun2021skin, pundhir2022towards, li2020fusing}. Furthermore, a few authors have proposed new networks, a modification of the current models, or hybrid methods dedicated to the SLC task, for example, in \citep{xiao2021boosting,sevli2021deep, wang2021unlabeled, reddy2021handling, hu2021toporesnet, khan2021pixels, mukherjee2021transfer, yilmaz2022mobileskin, qian2022skin, aldhyani2022multi, hosny2022refined, tang2022fusionm4net, anand2022fusion, camacho2022multi, hasan2021dermoexpert, hasan2021dermo, wu2020skin, bi2020multi, mahbod2020transfer, sun2020comparative, bian2021skin, bayasi2021culprit, peter2021internet, zhang2019attention}. Some of them are briefly illustrated in the next paragraph.} 

An end-to-end hybrid-CNN classifier with a two-level ensembling was presented and designed in \citep{hasan2021dermoexpert}. A channel-wise concatenated 2D feature map was proposed to enhance the first-level ensembling’s depth information. In contrast, the authors also proposed aggregating the various outputs from different fully connected layers in the second-level ensemble, learning more discriminating features with limited training samples. The authors in \citep{wu2020skin} proposed a densely connected convolutional network with an Attention and Residual learning method for the SLC task, where each ARDT block consists of dense blocks, transition blocks, and attention and residual modules. Compared to a residual network with the same number of convolutional layers, the size of the parameters of the densely connected network proposed was reduced by half. The improved densely connected network added an attention mechanism and residual learning after each dense and transition block without additional parameters. The authors of \citep{bi2020multi} presented a hyper-connected CNN (HcCNN) to classify skin lesions, having an additional hyper-branch that hierarchically integrates intermediary image features, unlike existing multi-modality CNNs. The hyper-branch enabled the network to learn more complex combinations between the images at early and late stages. They also coupled the HcCNN with a multi-scale attention block to prioritize semantically meaningful and subtle regions in the two modalities across various image scales. A CNN-based framework for simultaneous detection and recognition of skin lesions was proposed in \citep{hasan2021dermo}, named Dermo-DOCTOR, consisting of two encoders. The feature maps from the two encoders were fused channel-wise, called the fused feature map. This fused feature map was utilized for decoding in the detection sub-network, concatenating each stage of two encoders’ outputs with corresponding decoder layers to retrieve the lost spatial information due to pooling in the encoders. For the recognition sub-network, the outputs of three fully connected layers, utilizing feature maps of two encoders and fused feature maps, were aggregated to obtain a final SLC class. The authors in \citep{mahbod2020transfer} developed a three-level fusion scheme named multi-scale multi-CNN (MSM-CNN). They trained CNN models with cropped images at a fixed size at level one. At level two, they also fused the results from the individual networks trained on the six different image sizes (i.e., $224\times 224$, $240 \times 240$, $260 \times 260$, $300 \times 300$, $380 \times 380$, and $450 \times 450$). At the third and final fusion level, the authors fused the predicted probability vectors of the various architectures to yield the final classification result. The final MSM-CNN classification was thus derived from 90 ($5 \times 6 \times 3$) sub-models. Lastly, the authors in \citep{sun2020comparative} constructed a CNN-based SLC model that is loosely based on a baseline vanilla CNN network. The basic structure was two convolutional layers followed by a max-pooling layer containing fewer filters to reduce the training time. Additionally, their network included more dropout layers and fewer fully connected neurons to combat the overfitting present when replicating the original network.

\MKH{This review reveals that some authors, such as in \citep{shetty2022skin, serrano2022clinically, mohanty2022integrated, redha2021skin, rodrigues2020new, liu2020automatic, kwasigroch2020self, yildirim2020pre, nunnari2020study, almaraz2020melanoma, guha2020performance, molina2020classification, ur2018classification, li2018skin, ali2019skin, seeja2019deep, chung2019toporesnet, khan2019multi, khan2019skin, kulhalli2019hierarchical, majtner2016combining, samia2021skin, nersisson2021dermoscopic, khan2021skin, nancy2022impact, batista2022classification, thapar2022novel, sahin2022human, afza2022hierarchical, dong2022learning}, combined the CNN's automated lesion features and manual features (described in Table~\ref{tab:SLC_features}) for the SLC task to enhance the classification results in the past. They either extracted CNN attributes and merged them with manual features and then categorized them using typical ML model(s) (as mentioned in~\ref{Lesionclassifiers}) or manually extracted lesion features and stacked them in a particular CNN layer and then classified them using a Fully-connected layer. Furthermore, single CNN models are often indirectly limited when trained with highly variable and distinctive image datasets with limited samples. The authors of many SLC articles \citep{rahman2020transfer, mahbod2020transfer, mahbod2020effects, bagchi2020learning, ahmed2020skin, harangi2020assisted, harangi2018skin, milton2019automated, mahbod2019skin, mahbod2019fusing, thurnhofer2021skin, santos2021transfer, redha2021skin, bansal2021skin, mahbod2021investigating,rahman2021approach, serrano2022clinically, khan2022ensemble, hsu2022hierarchy, nguyen2022skin, popescu2022skin, hoang2022multiclass} mitigated this difficulty by applying the ensemble techniques. There are many analyses on the design of these ensemble techniques; for instance, whether the individual ensemble's candidate model should be trained first and then aggregation should be performed, such as in \citep{gessert2020skin, goyal2019skin, ha2020identifying, harangi2018skin, lee2018wonderm, mahbod2020transfer, pacheco2019skin, shahin2018deep}? Despite their approaches' satisfactory results, such an ensemble is tedious and time-consuming, requiring excessive time and resources for training and testing, as each model is independently trained and tested. Therefore, an end-to-end ensemble strategy that mitigated those weaknesses in \citep{gessert2020skin, goyal2019skin, ha2020identifying, harangi2018skin, lee2018wonderm, mahbod2020transfer, pacheco2019skin, shahin2018deep, liu2021multiscale} without compromising the state-of-the-art SLC results as demonstrated in \citep{hasan2021dermoexpert}.}

\begin{figure}[!ht]
\centering
\includegraphics[width=12cm, height=3cm]{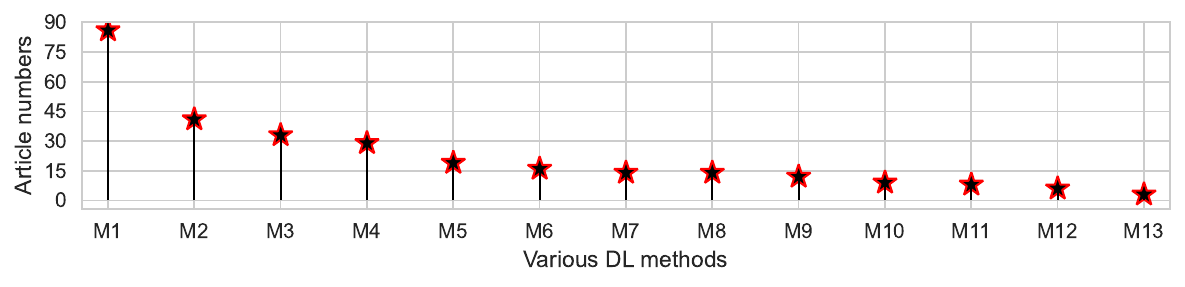}
\caption{\MKH{A discrete usage frequency plot of various DL models for SLC, where M1 to M13 denotes ResNet, VGG, DenseNet, InceptionNet, EfficientNet, MobileNet, InceptionResNet, Xception, AlexNet, GoogleNet, NASNet, SENet, and LeNet-5, respectively.}}
\label{fig:Pie_slc_DL_methods}
\end{figure}

The frequency of usage of various DL models in the last twelve years from the specified 238 SLC articles is shown in Fig.~\ref{fig:Pie_slc_DL_methods}. 
\MKH{It can be observed in Fig.~\ref{fig:Pie_slc_DL_methods} that ResNet, VGG, DenseNet, InceptionNet, and EfficientNet are the top-5 most SLC methods in those 238 SLC articles, which account for $36.1\,\%$, $17.2\,\%$, $13.9\,\%$, $12.2\,\%$, and $8.0\,\%$ of the total SLC article, respectively. These techniques are thus the most significant CNN-based SLC approaches (\textbf{\textit{High-frequency}}) in the past 238 SLC articles. The other architectures are employed in $1.3\,\% \sim 6.7\,\%$ of the SLC articles, where the MobileNet, InceptionResNet, Xception, AlexNet, and GoogleNet are separately used in $3.8\,\% \sim 6.7\,\%$ articles and would be thought of as the \textbf{\textit{Medium-frequency}} CNN models for SLC. Other CNN models in Fig.~\ref{fig:Pie_slc_DL_methods} are assumed to be \textbf{\textit{Low-frequency}} CNN-based SLC models in the past.} 

\begin{figure}[!ht]
\centering
\includegraphics[width=12cm, height=3.5cm]{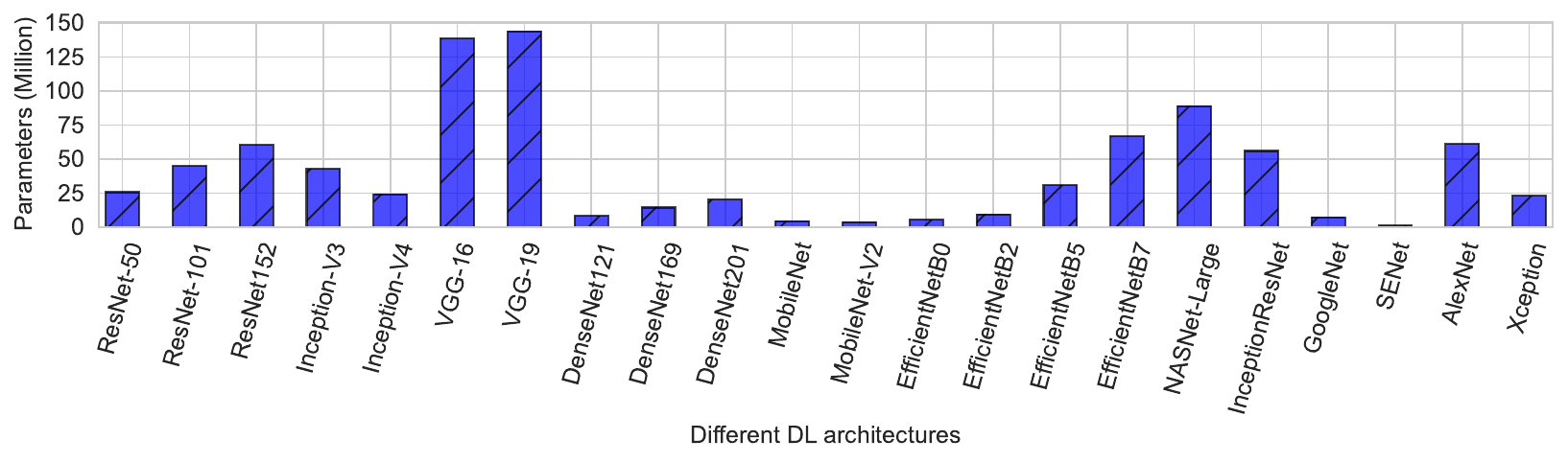}
\caption{The bar plot of diverse CNN architectures with their complementary number of learning parameters employed for the SLC task in the past from 2011 to 2022.}
\label{fig:Model_weights}
\end{figure}

\MKH{Further analysis reveals that out of the $36.1\,\%$ using ResNet models, the ResNet-50 and ResNet-101 are applied to $21.4\,\%$ and $8.0\,\%$ of SLC articles, respectively. Out of the $17.2\,\%$ using VGG architectures, the VGG-16 and VGG-19 are applied in the $12.2\,\%$ and $4.2\,\%$ articles. Again, the Inception-V3 is more commonly employed than the Inception-V4 in $7.1\,\%$ (out of $12.2\,\%$) SLC articles.} Similar patterns are also visible in all other networks; that is, the smaller network of each type of architecture is most commonly employed, for example, EfficientNetB0 than other EfficientNets, DenseNet121 than other DenseNets, and MobileNet-V2 than other MobileNets. This is summarized in Fig.~\ref{fig:Model_weights}.
This figure also illustrates different CNN architectures with their corresponding numbers of learning parameters. With larger parameter numbers, an increasing number of samples are needed in the training set, as DL requires large datasets to be effective. Both Fig.~\ref{fig:Pie_slc_DL_methods} and Fig.~\ref{fig:Model_weights} can demonstrate that smaller networks with fewer parameters have often been used for SLC in the last twelve years. Although this is likely a result of challenges in obtaining large sample sizes in medical imaging \citep{hasan2020dsnet, hasan2021dermo, hasan2021dermoexpert, harangi2018skin}, it is also likely preferred due to reduced computational costs. Further, several authors \citep{rahman2020transfer, mahbod2020transfer, mahbod2020effects, bagchi2020learning, ahmed2020skin, harangi2020assisted, harangi2018skin, milton2019automated, mahbod2019skin, mahbod2019fusing} could show that such smaller networks could be combined for ensemble design that performed very well even with fewer training samples.

\section{Training and Evaluation}
\label{Training_and_Evaluation}
This section surveys and explores the model training and evaluation protocols used for SLA tasks from 2011 to 2022. We investigate hyperparameter settings and training environments in sections~\ref{Hyperparameters_Settings} and~\ref{Training_Environments}, respectively, and analyze the evaluation strategies for SLA in section~\ref{Evaluation}. These issues are often explicitly addressed in these articles, which informed our review here.

\subsection{Hyperparameter Settings}
\label{Hyperparameters_Settings}
The identification of a good set of hyperparameters is essential for the robust performance of the CNN model \citep{hasan2021associating}. However, hyperparameters cannot be directly learned from regular training processes and must be tuned separately. Our review of the 594 SLA articles reveals several explicitly mentioned hyperparameters, which are summarized below. We also analyze the frequency of hyperparameter choices in past studies so as to inform future investigations.

\textbf{Batch size} is the number of images utilized to train a single forward and backward pass and is one of the essential hyperparameters. Larger or smaller batch size does not usually guarantee a better outcome, as there is a tradeoff between achieved results and computational resource availability \citep{kandel2020effect}. The authors in \citep{kandel2020effect} experimented on different batch sizes with fixed data and CNN architecture and revealed that it significantly impacted the network's performance. However, this review of 594 articles showed that the utilized batch sizes were 128 \citep{yao2021single, bozkurt2022skin}, 100 \citep{harangi2020assisted}, 96 \citep{van2018visualizing}, 64 \citep{aldhyani2022multi, unver2019skin, miglani2020skin, mirikharaji2021d}, 50 \citep{hosny2022refined}, 40 \citep{zhou2022superpixel}, 36 \citep{wei2020attentive}, 32 \citep{shetty2022skin, zhuang2022cs,shan2022automatic, nigar2022deep, cullell2021convolutional, rahman2021approach, calderon2021bilsk, das2021skin, gu2022net, osadebey2020evaluation, wang2021focal, xie2018multi, aldwgeri2019ensemble, yoon2019generalizable, kwasigroch2020self, zhuang2020cs}, 30 \citep{bayasi2021culprit}, 28 \citep{khan2021skin}, 26 \citep{burdick2017impact}, 24 \citep{liu2020automatic, wang2021boundary, aghdam2022attention}, 20 \citep{he2018dense, li2018dense, li2018skin, zhang2019automatic, haofu2017deep, liao2018deep, yan2020scalable}, 18 \citep{nampalle2020efficient}, 16 \citep{zanddizari2021new, ayas2022multiclass, bdair2021fedperl, thurnhofer2021skin, wang2021unlabeled, sun2021skin, rahman2021approach, balabantaray2021melanoma, hussain2021recu, hu2022net, jiang2022seacu, zhang2022dynamic, shamsolmoali2022salient, arora2021automated, tao2021attention, chowdary2021automated, xie2021semi, ammar2018learning, bissoto2018deep, chen2018multi, liu2019skin, shahin2019deep, anjum2020deep, hafhouf2020modified, saha2020leveraging, tang2020imscgnet, wu2020automated}, 15 \citep{wang2021multi}, 12 \citep{dong2021fac, dong2022tc, zhu2020asnet, kulhalli2019hierarchical}, 10 \citep{thurnhofer2021skin, sevli2021deep, calderon2021bilsk, li2018semi, wu2020skin, liu2022ncrnet}, 8 \citep{zanddizari2021new, popescu2022skin, ding2021deep, shahabi2021performance, khouloud2022w, dong2022learning, ren2022serial, ruan2022malunet, chen2021mt, zuo2022efficient, liu2021skin, tong2021ascu, tang2021afln, bagheri2021skin, ding2021efficient, phan2021skin, jiang2021residual, bagheri2021skin_2, saini2021b, qian2018detection, xu2018automatic, jiang2018skin, de2019skin, wu2019skin, xie2020skin, low2020automating, bagheri2020two, ozturk2020skin}, 6 \citep{qi2017global, feng2022bla}, 4 \citep{xiao2021boosting, hosny2022refined, nathan2020lesion, zhang2020kappa, xiao2021prior, al2022weakly, alahmadi2022semi}, 2 \citep{singh2019fca, mirikharaji2021d}, and 1 \citep{bi2019improving, huang2019skin, kaymak2020skin, dayananda2021skin}. 
\begin{figure}[!ht]
\centering
\includegraphics[width=15cm, height=3cm]{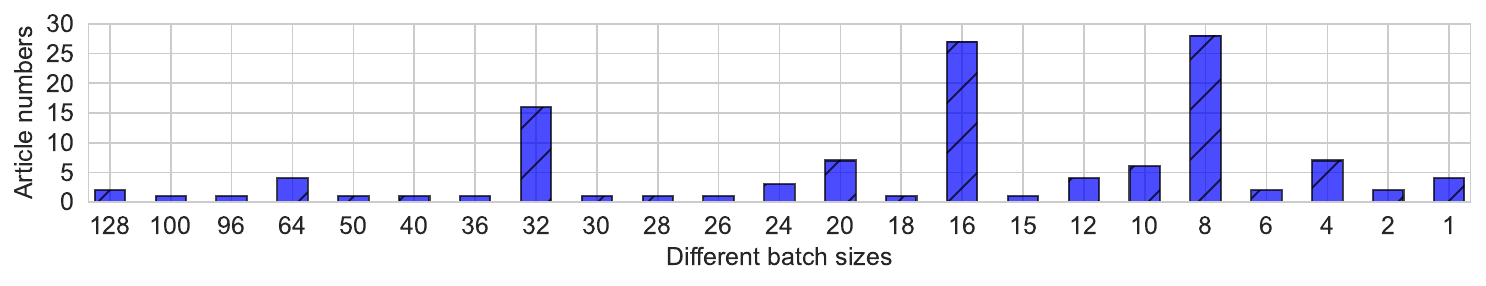}
\caption{\MKH{The number of articles that utilized various batch sizes in the past SLA articles.}}
\label{fig:BS_Comparison}
\end{figure}
\MKH{Fig.~\ref{fig:BS_Comparison} indicates that the most commonly employed ($78.7\,\%$) batch sizes lie between 8 and 32, where the topmost batch sizes of 8, 16, and 32 are applied to the $23.0\,\%$, $22.1\,\%$, and $13.1\,\%$ of the total articles, respectively. Those three topmost batch sizes follow the power of two ($2^n$) patterns, which is usually maintained for batch size selection. It is also observed that the larger batch sizes, like 64 and above, are used in merely $6.6\,\%$ of the articles, likely because they are associated with high computational resources. On the other hand, smaller batch sizes take a longer time to reach convergence during the training of the models. This likely explains why small batch sizes of 6 and below were applied to only $12.3\,\%$ of SLA articles.} 

\begin{figure}[!ht]
\centering
\includegraphics[width=12cm, height=3cm]{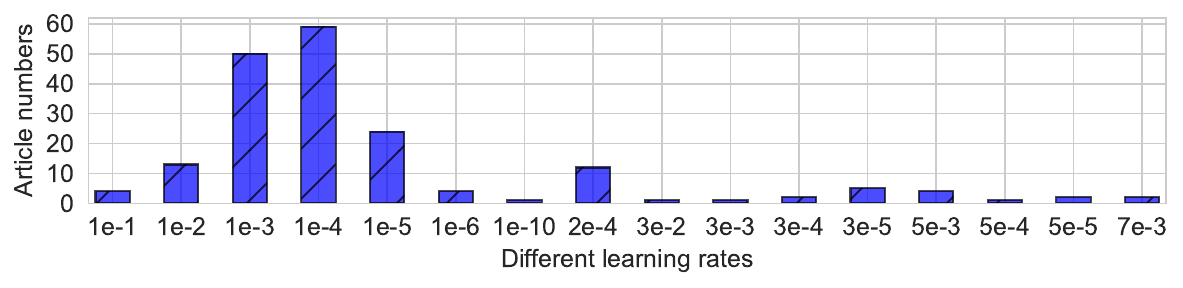}
\caption{\MKH{The number of articles that employed various LRs in the past 594-SLA papers.}}
\label{fig:LR_Comparison}
\end{figure}

\textbf{Learning Rate (LR)} ($\eta$) is a hyperparameter that controls how much the network's weights ($W$) are adjusted ($W_{new}=W_{old}-\eta \times \nabla$) concerning the loss gradient ($\nabla$). Too small of an LR functions a slowly converging training, while too large of an LR creates the training model diverge \citep{hasan2021dermoexpert}. Our review reveals that the employed LRs were 0.1 \citep {ali2019supervised, hasan2020dsnet, jiang2020skin, tao2021attention}, 0.01 \citep {mu2021channel, khan2021pixels, jiang2021residual, li2018semi, alfaro2019brief, iranpoor2020skin, wang2020donet, zhu2020asnet, danpakdee2017classification, devries2017skin, ozkan2017skin, namozov2018adaptive, ahmed2020skin}, 0.001 \citep {zanddizari2021new, aldhyani2022multi, nigar2022deep, wang2021unlabeled, liu2021multiscale, shetty2022skin, shahabi2021performance, shamsolmoali2022salient, han2022hwa, al2022weakly, jiang2022seacu, ruan2022malunet, zhou2022superpixel, bagheri2021skin, dayananda2021skin, das2021skin, wang2021boundary, bagheri2021skin_2, khan2021skin, qamar2021dense, qi2017global, ramachandram2017skin, luo2018fast, qian2018detection, venkatesh2018deep, vesal2018multi, zeng2018multi, brahmbhatt2019skin, anjum2020deep, bagheri2020two, liu2020multi, nampalle2020efficient, nathan2020lesion, osadebey2020evaluation, pour2020transform, xie2020skin, zafar2020skin, haofu2017deep, lee2018wonderm, liao2018deep, pan2018residual, songpan2018improved, xie2018multi, hassan2020skin, kwasigroch2020self, miglani2020skin, muckatira2020properties, qin2020gan, sun2020comparative, zhuang2020cs}, 0.0001 \citep {sun2021skin, shan2022automatic, zhuang2022cs, bozkurt2022skin, wan2022mslanet, ding2021deep, thurnhofer2021skin, sevli2021deep, wang2021knowledge, zhang2022dynamic, gu2022net, alahmadi2022semi, alahmadi2022multiscale, dong2022tc, liu2021skin, li2021digital, dong2021fac, xie2021semi, murphree2017transfer, feng2022bla, van2018visualizing, aldwgeri2019ensemble, eddine2019skin, kassani2019depthwise, kulhalli2019hierarchical, rashid2019skin, xue2019robust, yoon2019generalizable, milton2019automated, harangi2020assisted, liu2020automatic, mahbod2020transfer, wu2020skin, yan2020scalable, mishra2017deep, nasr2017dense, pour2017automated, jiang2018skin, ross2018effects, vesal2018skinnet, xu2018automatic, baghersalimi2019dermonet, bisla2019skin, canalini2019skin, de2019skin, liu2019enhanced, liu2019skin, tang2019multi, tschandl2019domain, wu2019skin, zhang2019automatic, deng2020weakly, hafhouf2020modified, kaymak2020skin, saha2020leveraging, tang2020imscgnet, wang2020cascaded, xie2020mutual, zhang2020kappa}, 0.00001 \citep {hosny2022refined, popescu2022skin, bayasi2021culprit, ayas2022multiclass, nguyen2022skin, shetty2022skin, rahman2021approach, ren2022serial, liu2022ncrnet, aghdam2022attention, dong2022learning, zuo2022efficient, chauhan2021multi, arora2021automated, ding2021efficient, wang2021focal, ammar2018learning, chen2018multi, he2018dense, li2018dense, li2018skin, saini2019detector, zhang2018skin, mirikharaji2021d}, 0.000001 \citep {chen2021mt, burdick2017impact, burdick2018rethinking, hu2022net}, 0.0000000001 \citep {bozorgtabar2017investigating}, 0.0002 \citep {tong2021ascu, wang2022skin, saini2021b, chen2021nl, lin2017skin, bi2019improving, jiang2019decision, singh2019fca, tu2019dense, wei2019attention, wei2020attentive, wu2020automated}, 0.03 \citep{kaur2021deep}, 0.003 \citep {ribeiro2020less}, 0.0003 \citep{hussain2021recu, jiang2021approximated}, 0.00003 \citep {somfai2022handling,tang2022fusionm4net, shan2020automatic, chowdary2021automated, phan2021skin}, 0.005 \citep {ozturk2020skin, kwasigroch2020neural, xiao2021prior, tang2021afln}, 0.0005 \citep {bi2020multi}, 0.00005 \citep {shahin2019deep, bdair2021fedperl}, and 0.007 \citep {cui2019ensemble, lei2020skin}. 
\MKH{Fig.~\ref{fig:LR_Comparison} displays that the LRs of scales between e-2 and e-5 were employed in $95.1\,\%$ of the total articles while the remaining $4.9\,\%$ articles applied other LRs of e-1, e-6, and e-10 scales. Specifically, the LRs of 0.001, 0.0001, and 0.00001 were the most popular in past SLA articles.}
This likely implied that most investigators found this range of LRs to be the most effective for developing the supervised SLA model. These results may imply that LRs that are too large or too small are not recommended.

\textbf{Loss function}, also known as the cost function or error function, is an objective function or criterion for minimizing during the supervised DL model's training. The scrutiny of the selected articles reveals that the most commonly employed loss function is cross-entropy for both the SLS and SLC tasks that are employed in \citep{chauhan2021multi, murphree2017transfer, xie2021semi, mirunalini2017deep, zhang2018skin, pham2018deep, dos2018robust, gessert2019skin, harangi2020assisted, xie2020mutual, gessert2020skin, bi2020multi, bagchi2020learning, kwasigroch2020self, kwasigroch2020neural, jibhakate2020skin, hassan2020skin, qin2020asymmetric, jiang2020skin, shan2020automatic, sanjar2020improved, zafar2020skin, xie2020skin, kamalakannan2020self, soudani2019image, liu2019skin, liu2019enhanced, canalini2019skin, bisla2019skin, zhang2019automatic, saini2019detector, tang2019efficient, guth2018skin, luo2018fast, chen2018multi, burdick2018rethinking, bi2018improving, ross2018effects, li2018semi, jiang2018skin, baghersalimi2019dermonet, yuan2017automatic, ramachandram2017skin, lin2017skin, li2018skin, qamar2021dense, bagheri2021skin, tong2021ascu, wang2021knowledge, chen2021mt, liu2021skin, zhang2022dynamic, dai2022ms, wang2021unlabeled, nigar2022deep, sun2021skin, ding2021deep, wang2021multi, shetty2022skin, cullell2021convolutional, nersisson2021dermoscopic, balabantaray2021melanoma, dong2022learning, tang2022fusionm4net, mukherjee2021transfer, he2022deep, foahom2022end, ayas2022multiclass,yue2022towards}. 
Other loss functions in the past were Softmax \citep{liao2016skin, wang2021interpretability, haofu2017deep, ali2019supervised, qi2017global, bozorgtabar2017investigating, tao2021attention}, Square error \citep{filali2017multiscale, shan2020automatic}, Logistic loss function \citep{liu2020automatic}, Tversky loss \citep{arora2021automated, tran2022fully, ramadan2022dgcu} and Focal loss \citep{yan2019melanoma, wang2021interpretability, wang2021focal, chowdary2021automated, bagheri2021skin_2, jiang2021approximated, arora2021automated, hussain2021recu, sun2021skin, yao2021single}. Although binary or categorical cross-entropy functions are widely employed as loss functions in SLS, they may lead to bias effects as the size of a lesion is drastically smaller than the size of the background \citep{hasan2021detection}. Therefore, some authors employed Dice Similarity Coefficient (DSC) (as in~Eq.~\ref{eq:dsc}) or Jaccard Index (JI) (as in~Eq.~\ref{eq:ji}), which are applied in \citep{li2021digital, saini2021b, thandiackal2018structure, wang2020donet, shan2020automatic, pereira2020dermoscopic, salih2020skin, al2020automatic, abhishek2020illumination, wu2019skin, wang2019bi, he2018dense, vesal2018skinnet, vesal2018multi, qian2018detection, kolekar2018skin, ahmed2018segmentation, wei2020attentive, tang2019multi, singh2019fca, shahin2019deep, dash2019pslsnet, sarker2021slsnet, tu2019dense, goyal2019skin, ninh2019skin, yuan2017automatic, nasr2017dense, dong2021fac, chowdary2021automated, wang2021boundary, phan2021skin, jiang2021approximated, ding2021efficient, bagheri2021skin, arora2021automated, hussain2021recu, adegun2021probabilistic, xiao2021prior, wibowo2021lightweight, dai2022ms, wang2022skin, jiang2022seacu, barin2022automatic}. 
\begin{equation} \label{eq:dsc}
L_{DSC}(y,\hat{y})= -1+\frac{2\times \displaystyle\sum_{i=1}^{N}y_i\times \hat{y_i}}{\displaystyle\sum_{i=1}^{N}y_i+\displaystyle\sum_{i=1}^{N}\hat{y_i}},
\end{equation}
\begin{equation} \label{eq:ji}
L_{JI}(y,\hat{y})= -1+\frac{\displaystyle\sum_{i=1}^{N}y_i\times \hat{y_i}}{\displaystyle\sum_{i=1}^{N}y_i+\displaystyle\sum_{i=1}^{N}\hat{y_i}-\sum_{i=1}^{N}y_i\times \hat{y_i}},
\end{equation}
\noindent where $y$, $\hat{y}$, and $N$ are the true label, predicted label, and the total number of pixels, respectively. In those two equations, the product of $y$ and $\hat{y}$ is the measure of similarity (intersection) between true and predicted masks. Again, some authors in \citep{ruan2022malunet, wu2022fat, lee2022progressive, al2022weakly, dong2022tc, zuo2022efficient, zhao2022self, bissoto2018deep, yang2021deep, tschandl2019domain, shan2020automatic, zhu2020asnet, nathan2020lesion, mahbod2020effects, hasan2020dsnet, hafhouf2020modified, saha2020leveraging, low2020automating, xie2020mutual, wu2020automated, ribeiro2020less, sarker2021slsnet, wei2019attention, de2019skin, venkatesh2018deep, chen2021mt, dong2022learning, ahmed2022new, zhang2022dense, wang2022skin, liu2022ncrnet} combined BCC and Eq.~\ref{eq:dsc} or Eq.~\ref{eq:ji} for further enhancing the lesion segmentation results, as mentioned in Eq.~\ref{eq:loss}.    
\begin{equation} \label{eq:loss}
L_{SLS}(y,\hat{y})= 1-\frac{\displaystyle\sum_{i=1}^{N}y_i\times \hat{y_i}}{\displaystyle\sum_{i=1}^{N}y_i+\displaystyle\sum_{i=1}^{N}\hat{y_i}-\sum_{i=1}^{N}y_i\times \hat{y_i}}-\frac{1}{N} \sum_{i=1}^{N}[y_i \log \hat{y_i} + (1-y_i)\log (1-\hat{y_i}) ],
\end{equation}
\noindent where $\log\hat{y_i}$ and $\log(1-\hat{y_i})$ are the measure of the log-likelihood of the pixel being lesion or not, respectively.
In summary, our review of loss function usage demonstrates that the BCC was widely applied for the SLC and SLS tasks, whereas Eq.~\ref{eq:dsc} or Eq.~\ref{eq:ji} or Eq.~\ref{eq:loss} were commonly applied for the SLS task in the past 594 articles.

\textbf{Optimizer} calculates the value of the model's parameters (weights), minimizing the error (cost function as described in the previous paragraph) and maximizing the predefined metric (see Table~\ref{tab:metrics_evaluation}) when mapping inputs to outputs. It widely affects the CNN models' accuracy and training speed \citep{hasan2021dermoexpert}. \MKH{In the past (2011-2022), the most commonly applied optimizers (with their corresponding articles) for the SLA models are Adam, also known as an adaptive optimizer, \citep {khouloud2022w, barin2022automatic, liu2022ncrnet, zhang2022dense, feng2022bla, al2022weakly, hu2022net, aghdam2022attention, dong2022tc, ren2022serial, ruan2022malunet, zhou2022superpixel, zuo2022efficient, chen2021mt, mu2021channel, hussain2021recu, chauhan2021multi, chauhan2021multi, xie2021semi, ding2021efficient, kaur2021deep, phan2021skin, chowdary2021automated, dong2021fac, li2021digital, wang2021focal, qamar2021dense, lin2017skin, nasr2017dense, ramachandram2017skin, yuan2017automatic, ammar2018learning, li2018dense, luo2018fast, qian2018detection, ross2018effects, venkatesh2018deep, vesal2018multi, vesal2018skinnet, xu2018automatic, baghersalimi2019dermonet, canalini2019skin, de2019skin, liu2019enhanced, liu2019skin, saini2019detector, shahin2019deep, singh2019fca, soudani2019image, tang2019multi, tu2019dense, wei2019attention, wu2019skin, hafhouf2020modified, iranpoor2020skin, lei2020skin, nathan2020lesion, pour2020transform, qin2020asymmetric, ribeiro2020less, shan2020automatic, tang2020imscgnet, wei2020attentive, wu2020automated, xie2020mutual, xie2020skin, zafar2020skin, zhang2020kappa, devries2017skin, bissoto2018deep, van2018visualizing, bisla2019skin, eddine2019skin, kassani2019depthwise, kulhalli2019hierarchical, mahbod2019fusing, milton2019automated, tschandl2019domain, yoon2019generalizable, gessert2020skin, jibhakate2020skin, liu2020automatic, mahbod2020transfer, muckatira2020properties, sun2020comparative, xie2020mutual, xiao2021prior, liu2021skin, shetty2022skin, bansal2021skin, tang2022fusionm4net, rahman2021approach, sevli2021deep, bdair2021fedperl, nguyen2022skin}, Stochastic Gradient Descent (SGD) \citep {sun2021skin, liu2021multiscale, hosny2022refined, ding2021deep, yao2021single, wang2021multi, nigar2022deep, zhuang2022cs, shamsolmoali2022salient, tang2021afln, dayananda2021skin, mirikharaji2021d, saini2021b, tao2021attention, bozorgtabar2017investigating, burdick2017impact, he2017skin, mishra2017deep, pour2017automated, qi2017global, yuan2017automatic, bissoto2018deep, burdick2018rethinking, chen2018multi, he2018dense, jiang2018skin, li2018semi, li2018skin, mirikharaji2018deep, zeng2018multi, brahmbhatt2019skin, cui2019ensemble, huang2019skin, jiang2019decision, ma2019light, soudani2019image, zhang2019automatic, jiang2020skin, nampalle2020efficient, osadebey2020evaluation, ozturk2020skin, pour2020transform, wang2020cascaded, zhu2020asnet, haofu2017deep, sousa2017araguaia, lee2018wonderm, liao2018deep, namozov2018adaptive, navarro2018webly, xie2018multi, mahbod2019fusing, xue2019robust, kwasigroch2020self, wu2020skin, zhuang2020cs}, Adadelta \citep {hasan2020dsnet}, RMSprop \citep{yao2021single, rahman2021approach, balabantaray2021melanoma}, and Nadam \citep {yao2021single, low2020automating}. Indeed, $96.6\,\%$ of the articles applied Adam ($60.5\,\%$) and SGD ($35.7\,\%$) optimizers. However, it has also appeared from the current review that the authors had fine-tuned optimizers' control parameters like the momentum in SGD, the decay rate in Adadelta, and the exponential decay rate for the first and second moments ($\beta_1$ and $\beta_2$) in Adam and Nadam. The variations of these parameters in the 594 SLA articles are summarized in Table~\ref{tab:control_params}.}
\begin{table}[!ht]
\caption{\MKH{The variations of different parameters in optimizer(s) like momentum, decay rates, $\beta_1$, and $\beta_2$ in the selected 594 articles.}}
\label{tab:control_params}
\scriptsize
\begin{tabular}{l|lp{12.2cm}}
\Xhline{1pt}
\textbf{Parameters}          & \textbf{Values} & \textbf{Corresponding articles} \\ \Xhline{1pt}
                    &       0.25           & \citep{danpakdee2017classification}                      \\ 
                    &   0.90               &  \citep{bozorgtabar2017investigating, burdick2017impact, he2017skin, pour2017automated, qi2017global, bissoto2018deep, burdick2018rethinking, he2018dense, li2018semi, li2018skin, zeng2018multi, cui2019ensemble, ma2019light, unver2019skin, zhang2019automatic, zhu2020asnet, wang2020cascaded, ozturk2020skin, nampalle2020efficient, lei2020skin, kaymak2020skin, jiang2020skin, haofu2017deep, murphree2017transfer, ozkan2017skin, lee2018wonderm, namozov2018adaptive, navarro2018webly, bisla2019skin, bi2020multi, kwasigroch2020self, wu2020skin, zhuang2020cs, tao2021attention, saini2021b, jiang2021residual, shamsolmoali2022salient, al2022weakly, alahmadi2022semi, nigar2022deep, hoang2022multiclass, hosny2022refined, zhuang2022cs, shan2022automatic}                      \\ 
                    &    0.95              &   \citep{huang2020skin}                     \\ 
\multirow{-5}{*}{Momentum} &   0.99               &   \citep{mirikharaji2018deep, huang2019skin, pour2020transform, mirikharaji2021d, dayananda2021skin}                     \\ \hline
                    &      0.001            &      \citep{he2018dense, unver2019skin, yoon2019generalizable}                  \\ 
                    &      0.0001            &  \citep{he2017skin, qi2017global, li2018skin, huang2020skin, ma2019light, zhu2020asnet, haofu2017deep, liao2018deep, xue2019robust, wu2020skin, aldwgeri2019ensemble, jiang2021approximated, wang2021unlabeled, shan2022automatic} 
                     \\ 
                    &      0.00001            &  \citep{huang2020skin, muckatira2020properties, dong2022learning, hosny2022refined}                     \\ 
                    &      0.000001            &   \citep{kaymak2020skin, chauhan2021multi}                     \\ 
                    &       0.00158           &   \citep{low2020automating}                     \\ 
                    &       0.5           &         \citep{wei2019attention}               \\ 
                    &        0.005          &     \citep{zeng2018multi, xie2020skin, al2022weakly, namozov2018adaptive, liu2021multiscale}                   \\ 
                    &         0.0005         &     \citep{huang2019skin, shamsolmoali2022salient, nampalle2020efficient, pour2020transform, pour2017automated, mirikharaji2018deep, jiang2021residual}                   \\                   
\multirow{-9}{*}{Decay rates} &    0.00005              &       \citep{bi2020multi, xie2021semi, tao2021attention}                 \\ \hline
                    &      0.50            &    \citep{singh2019fca, wang2022skin, wei2019attention, qin2020asymmetric, wei2020attentive, bisla2019skin}                    \\ 
                    &       0.60           &   \citep{kassani2019depthwise}                     \\  
\multirow{-3}{*}{$\beta_1$} &    0.90              &    \citep {rahman2021approach, sevli2021deep, bozkurt2022skin, li2018dense, bisla2019skin, pour2020transform, van2018visualizing, mahbod2019fusing, xie2021semi, jiang2021approximated, tong2021ascu}                    \\ \hline
                    &     0.990             &    \citep{bisla2019skin, van2018visualizing, mahbod2019fusing, xie2021semi}                    \\ 
                    &      0.995            &     \citep{kassani2019depthwise}                   \\ 
\multirow{-3}{*}{$\beta_2$} &   0.999               &   \citep{rahman2021approach, sevli2021deep, bozkurt2022skin, li2018dense, wang2022skin, singh2019fca, wei2019attention, pour2020transform, qin2020asymmetric, wei2020attentive, jiang2021approximated, tong2021ascu}                     \\ \Xhline{1pt}
\end{tabular}
\end{table}
Momentum increases the rate of convergence in gradient descent in the appropriate direction and dampens oscillations, while the decay rate continuously changes the LR values to construct an adaptive optimizer. On the other hand, $\beta_1$ and $\beta_2$ are the initial decay rates utilized when estimating the first and second moments of the gradient, which are multiplied by themselves (exponentially) at the end of each training step (batch). \MKH{However, it is noteworthy from the table that the most commonly applied momentum, decay rate, $\beta_1$, and $\beta_2$ in the past SLA articles were 0.90, 0.0001, 0.90, and 0.999, respectively.}

\textbf{Epoch} refers to one forward and one backward pass over the entire dataset. The initial weights of models will be subjected to transformations during the next cycle of the same training dataset. Epoch optimization mainly faces two significant problems: underfitting and overfitting. Using a few epochs will lead to an underfitting of the data during the network's training, which indicates that the network cannot capture the underlying tendency of the data. Increasing the epoch numbers will enable a more optimal solution with better accuracy. However, beyond the optimal number of the epoch, further increases in epoch numbers will lead to overfitting of data, which means that the network is now less accurate as it is capturing noise in the data. Unfortunately, there is no simple solution for choosing the best epoch. Here, we reviewed the 594 selected SLA articles to investigate the choice of epoch numbers in the past twelve years' studies. We find that the epoch numbers are 1000 \citep{nersisson2021dermoscopic, xiao2021boosting}, 500 \citep{bdair2021fedperl, xie2018multi, chen2021nl, deng2020weakly, shan2020automatic, xie2020mutual, wang2021boundary}, 400 \citep{zhang2022dynamic}, 300 \citep{shamsolmoali2022salient, anjum2020deep, he2017skin, li2018dense, gu2022net, li2018skin}, 360 \citep{alfaro2019brief}, 250 \citep{nguyen2022skin, tang2022fusionm4net, mishra2017deep, wu2020automated, chowdary2021automated, mu2021channel}, 200 \citep{tong2021ascu, wang2022skin, feng2022bla, dong2022learning, dong2022tc,arora2021automated, kwasigroch2020neural, dong2021fac, wang2021focal, ammar2018learning, bi2019improving, goyal2019skin, shahin2019deep, jiang2020skin, wei2020attentive}, 192 \citep{van2018visualizing}, 165 \citep{low2020automating}, 150 \citep{ding2021deep, shetty2022skin, ren2022serial, han2022hwa, jiang2019decision, le2022antialiasing}, 120 \citep{xie2021semi}, 110 \citep{saini2021b}, 100 \citep{bayasi2021culprit, popescu2022skin, shan2022automatic, bozkurt2022skin, wan2022mslanet, yilmaz2022mobileskin, sevli2021deep, wang2021multi, shahabi2021performance, khan2021pixels, bissoto2018deep, alahmadi2022multiscale, alahmadi2022semi, aghdam2022attention, bagheri2021skin_2, lee2018wonderm, albahar2019skin, bi2020multi, gessert2020skin, kwasigroch2020self, liu2020automatic, sun2020comparative, wu2020skin, nampalle2020efficient, nathan2020lesion, osadebey2020evaluation, ribeiro2020less, zafar2020skin}, 90 \citep{saini2019detector, ding2021efficient}, 80 \citep{chen2018multi, de2019skin, wang2020donet}, 75 \citep{hu2022net}, 70 \citep{yao2021single}, 60 \citep{sun2021skin, khouloud2022w, aldwgeri2019ensemble, al2022weakly, burdick2017impact}, 50 \citep{tschandl2019domain, ruan2022malunet, dayananda2021skin, miglani2020skin, qin2020gan, tao2021attention, jiang2021residual, jiang2021approximated, bagheri2021skin, zanddizari2021new, wang2021unlabeled}, 40 \citep{zhuang2022cs, kulhalli2019hierarchical, liu2022ncrnet}, 35 \citep{ozturk2020skin, balabantaray2021melanoma}, 30 \citep{liu2021skin, bayasi2021culprit, rahman2021approach, young2019deep}, 24 \citep{rashid2019skin}, 20 \citep{aldhyani2022multi, hoang2022multiclass, khan2021skin, hussain2021recu, zhou2022superpixel}, 15 \citep{canalini2019skin, somfai2022handling}, 13 \citep{saha2020leveraging}, 10 \citep{das2021skin}, and 3 \citep{liu2019skin}. 
\MKH{Those epoch values and their appliance frequencies demonstrate that the fewer epochs ($<50$) are applied to a smaller number of articles, approximately in $16.1\,\%$ of the total articles, whereas the high number of epochs ($>200$) are also employed in fewer articles ($19.5\,\%$). On the other hand, the epochs between 50 and 200 were most commonly employed in $64.4\,\%$ of the articles that confirm the epoch numbers to train the SLA models.}

\subsection{Training Environments}
\label{Training_Environments}
In order to train different ML and DL algorithms and architectures, there are numerous frameworks and libraries that provide a convenient training protocol. These frameworks assemble complex mathematical functions, training algorithms, and statistical modeling tools, allowing their convenient usage without the need to program from scratch. Table~\ref{tab:Frameworks} shows the commonly used DL and/or ML frameworks in the SLA articles that explicitly specified the training protocol(s).
\begin{table}[!ht]
\caption{\MKH{Widespread ML and/or DL frameworks in SLA tasks with their common attributes.}}
\label{tab:Frameworks}
\scriptsize
\begin{tabular}{lllp{7.1cm}}
\Xhline{1pt}
\textbf{Framework (Year)} & \textbf{Written in} & \textbf{Popularity$^\dag$} & \textbf{Corresponding SLA articles} \\ \Xhline{1pt}
 Caffe (2015)             &   C++   &    33K   &   \citep{pour2020transform, huang2019skin, bozorgtabar2017investigating, nasr2017dense, pour2017automated}                     \\ 
 Keras (2015)               &   Python    &  56.6K    &           \citep{zanddizari2021new, nguyen2022skin, balabantaray2021melanoma, shahabi2021performance,samanta2021skin, yilmaz2022mobileskin, ren2022serial, jiang2022seacu, wibowo2021lightweight, hafhouf2020modified, dayananda2021skin, hasan2020dsnet, kamalakannan2020self, low2020automating, saha2020leveraging, shan2020automatic, xie2020mutual, brahmbhatt2019skin, dash2019pslsnet, hasan2019skin, liu2019skin, shan2019improving, soudani2019image, tang2019multi, ammar2018learning, burdick2018rethinking, kolekar2018skin, li2018semi, venkatesh2018deep, xu2018automatic, burdick2017impact, mishra2017deep, qamar2021dense, phan2021skin, gajera2021improving, hu2022net}             \\ 
TensorFlow (2015)                 &   C++, Python, CUDA    &  169K    &    \citep{nersisson2021dermoscopic, nguyen2022skin, kamalakannan2020self, chauhan2021multi, low2020automating, saha2020leveraging,baghersalimi2019dermonet, dash2019pslsnet, de2019skin, goyal2019skin, jiang2019decision, liu2019skin, tang2019multi, jiang2018skin, luo2018fast, nguyen2018isic, ross2018effects, wang2018skin, xu2018automatic, zeng2018multi, mishra2017deep, ramachandram2017skin, jiang2022seacu}                    \\ 

Theano (2008)                &    Python        &  9.6K    &  \citep{yuan2017automatic}                  \\
 
PyTorch (2016)        &   Lua   &  60.1K   &      \citep{wang2021unlabeled, tang2022fusionm4net, ayas2022multiclass, zhuang2022cs, qian2022skin, sun2021skin, liu2021multiscale, ding2021deep, shan2022automatic, yao2021single, xiao2021boosting, wan2022mslanet, zuo2022efficient, tong2021ascu, xiao2021prior, liu2021skin, ding2021efficient, hussain2021recu, tang2021afln, jiang2021residual, wang2021focal, xie2021semi, deng2020weakly, jiang2020skin, qin2020asymmetric, ribeiro2020less, wang2020cascaded, wei2020attentive, wu2020automated, xie2020skin, alfaro2019brief, goyal2019skin, jiang2019decision, liu2019enhanced, ma2019light, shahin2019deep, singh2019fca, tschandl2019domain,  tu2019dense, wang2019automated, wei2019attention, bissoto2018deep, mirikharaji2018deep, qian2018detection, tao2021attention, chen2021nl, chen2021mt, zhou2022superpixel, dong2022tc, dong2022learning, aghdam2022attention, al2022weakly, alahmadi2022semi, alahmadi2022multiscale, zhang2022dynamic, zhang2022dense, han2022hwa, chen2022skin, wu2022fat, liu2022ncrnet, gu2022net}                  \\ \Xhline{1pt}
\multicolumn{4}{l}{$^\dag$These values are the number of stars, taken from their official GitHub accounts [Access date: 07-Nov-2022].}

\end{tabular}
\end{table}
It can be observed from Table~\ref{tab:Frameworks} that Keras, TensorFlow, and PyTorch with Python and/or MATLAB programming languages (most commonly Python) have been most popularly applied to train automated SLA models. Remarkably, most recent articles preferred PyTorch with Python as an SLA development tool. Moreover, those three frameworks are also most popular in other domains according to their popularity. However, applying those DL frameworks depends on many factors, for example, the level of API, computational speed, architecture, and debugging facilities.

\subsection{Model Evaluation}
\label{Evaluation}
Once the SLS and SLC are accomplished, the next step is the assessment of those two methods. The successes of SLS and SLC are typically reported in a confusion matrix that includes statistics about actual and predicted pixels or classes \citep{hasan2021missing}. However, the employed metrics for the SLS and SLC evaluations in the past from 594 selected articles are Accuracy (ACC), Sensitivity (SEN), Precision (PRE), Specificity (SPE), Figure of Merit (FOM), Segmentation Error (SE), False Positive Rate (FPR), F1-score (F1S), Negative Predictive Value (NPV), DSC, Hausdorff Distance (HD), Area Under Curve (AUC), JI, False Negative Rate (FNR), XOR \citep{mirikharaji2022survey}, Correlation Coefficient (CC), Hammoude Distance (HMD), Matthew Correlation Coefficient (MCC), Peak to Signal Ratio (PSNR), Structural Similarity Indices (SSIM), and Balanced Accuracy (BACC). These metrics are summarized in Table~\ref{tab:metrics_evaluation}, along with the articles using them.

{\tiny
\begin{longtable}[c]{l|p{7.2cm}p{7cm}}
\caption{\MKH{A list of SLS and SLC evaluation metrics with their corresponding articles applied in the past (2011 to 2022), examining a total of 594 articles (356 for segmentation and 238 for classification).}}
\label{tab:metrics_evaluation}\\
\Xhline{1pt}
\textbf{Metrics} & \textbf{SLS articles} & \textbf{SLC articles} \\ \Xhline{1pt}
\endfirsthead
\multicolumn{3}{c}%
{{\bfseries Table \thetable\ Continued from previous page}} \\
\Xhline{1pt}
\textbf{Metrics} & \textbf{SLS articles} & \textbf{SLC articles} \\ \Xhline{1pt}
\endhead
\hline
\endfoot
\endlastfoot
ACC                    &   \citep{wu2022fat,barin2022automatic, gu2022net, jiang2022seacu, wang2022skin, hu2022net, csahin2022robust, dai2022ms, chen2022skin, tran2022fully, han2022hwa, zhang2022dense, zhang2022dynamic, alahmadi2022multiscale, alahmadi2022semi, feng2022slt, le2022antialiasing, lee2022progressive, al2022weakly, singh2022empirical, dong2022learning, aghdam2022attention, dong2022tc, fan2022egfnet, khan2022ensemble, khan2022skin, khouloud2022w, ren2022serial, ruan2022malunet, zuo2022efficient, gajera2021improving, liu2021skin, peter2021internet, wibowo2021lightweight, reddy2021handling, yang2021deep, xiao2021prior, chen2021mt, adegun2021probabilistic, yacin2021deep, wang2021knowledge, tong2021ascu, ramya2021segmentation, mu2021channel, kosgiker2021segcaps, hussain2021recu, chauhan2021multi, arora2021automated, ali2021skin, bagheri2021skin, ding2021efficient, chen2021nl, gangwar2021study, kaur2021deep, phan2021skin, redha2021skin, tang2021afln, bagheri2021skin_2, chowdary2021automated, dong2021fac, garg2021skin, jiang2021residual, khan2021skin, qamar2021dense, tao2021attention, saini2021b, wang2021focal, xie2021semi, abbas2013improved, cavalcanti2013macroscopic, nisar2013color, abbas2014combined, al2014automatic, ch2014two, jeniva2015efficient, torkashvand2015automatic, trabelsi2015skin, azehoun2016novel, bi2016automated, jafari2016skin, pennisi2016skin, sagar2016color, yuvaraju2016segmentation, bi2017semi, burdick2017impact, george2017automatic, he2017skin, lynn2017segmentation, martinez2017pigmented, mishra2017deep, nasr2017dense, pour2017automated, qi2017global, ramachandram2017skin, yuan2017automatic, adeyinka2018skin, ahmed2018segmentation, akram2018skin, aljanabi2018skin, ammar2018learning, bi2018improving, burdick2018rethinking, chakkaravarthy2018automatic, chen2018multi, hardie2018skin, he2018dense, jaisakthi2018automated, khan2018implementation, kolekar2018skin, li2018dense, li2018semi, li2018skin, luo2018fast, mirikharaji2018deep, nasir2018improved, navarro2018accurate, olugbara2018segmentation, patino2018automatic, salih2018comparison, ur2018classification, venkatesh2018deep, vesal2018multi, vesal2018skinnet, youssef2018deep, zeng2018multi, abdullah2019deep, adegun2019deep, al2019deep, aljanabi2019various, bisla2019skin, brahmbhatt2019skin, canalini2019skin, cui2019ensemble, dash2019pslsnet, de2019skin, filali2019improved, filali2019multi, goyal2019skin, hasan2019skin, javed2019intelligent, javed2019region, jiang2019decision, khan2019construction, khan2019skin, liu2019skin, ma2019light, rawas2019hcet, saini2019detector, salih2019skin, seeja2019deep, shan2019improving, singh2019fca, soudani2019image, tang2019efficient, tang2019multi, tu2019dense, unver2019skin, wang2019automated, wang2019bi,wei2019attention, xie2020mutual, zhang2019automatic, al2020automatic, anjum2020deep, azad2020attention, bagheri2020two, deng2020weakly, devi2020fuzzy, hafhouf2020modified, hawas2020oce, iranpoor2020skin, jabbari2020segmentation, jayapriya2020hybrid, jiang2020skin, justin2020skin, kamalakannan2020self, kaymak2020skin, khan2020frequency, abhishek2020illumination, lei2020skin, low2020automating, mahbod2020effects, nampalle2020efficient, ozturk2020skin, parida2020transition, pillay2020macroscopic, pour2020transform, qin2020asymmetric, qiu2020inferring, rizzi2020skin, rout2020transition, salih2020skin, santos2020skin, setiawan2020image, shan2020automatic, sivaraj2020detecting, tang2020imscgnet, wang2020cascaded, wang2020donet, wei2020attentive, wu2020automated, xie2020skin, zhu2020asnet, rashid2015novel, sanjar2020improved}           &   \citep{wang2021unlabeled, pereira2021skin_2,moldovanu2021skin, bian2021skin, wang2021multi, shetty2022skin, samia2021skin, reddy2021handling, melbin2021integration, jain2021deep, bansal2021skin, yacin2021deep,samanta2021skin, rahman2021approach, afza2022multiclass, nersisson2021dermoscopic, carvalho2021multimodal, hsu2022hierarchy, balabantaray2021melanoma, xiao2021boosting, bdair2021fedperl, calderon2021bilsk, mukherjee2021transfer, peter2021internet, khouloud2022w, batista2022classification, yilmaz2022mobileskin, hoang2022multiclass, yue2022towards, wan2022mslanet, rasel2022convolutional, khan2022ensemble, deng2022efficient, wei2022dual, thapar2022novel, sarker2022transslc, qian2022skin, nigar2022deep, nakai2022enhanced, aldhyani2022multi, wang2022ssd, samsudin2022skin, hosny2022refined, ayas2022multiclass, wang2022adversarial, shan2022automatic, ngo2022skin, ramlakhan2011mobile, amelard2012extracting, cavalcanti2013macroscopic, she2013lesion, celebi2014automated, jamil2014comparative, patil2014detection, surowka2014optimal, di2015hierarchical, sirakov2015skin, chakravorty2016dermatologist, farooq2016automatic, liao2016skin, majtner2016combining, pomponiu2016deepmole, danpakdee2017classification, devries2017skin, lopez2017skin, lynn2017segmentation, mahdiraji2017skin, mirunalini2017deep, murphree2017transfer, ozkan2017skin, sousa2017araguaia, wahba2017combined, chen2018multi, filali2018study, khan2018implementation, lee2018wonderm, namozov2018adaptive, nasir2018improved, navarro2018webly, pham2018deep, songpan2018improved, thandiackal2018structure, ur2018classification, wahba2018novel, yap2018multimodal, zhang2018skin, abbas2019efficient, agilandeeswari2019skin, al2019deep, albahar2019skin, aldwgeri2019ensemble, aljanabi2019various, bisla2019skin, chatterjee2019integration, chung2019toporesnet, eddine2019skin, filali2019improved, filali2019texture, fisher2019classification, javed2019intelligent, kassani2019depthwise, khan2019construction, khan2019multi, khan2019skin, mahbod2019fusing, seeja2019deep, wang2019mutual, zheng2019relation, afza2020skin, ahmed2020skin, dutta2021skin, akram2020multilevel, almaraz2020melanoma, anjum2020deep, bi2020multi, chaturvedi2020skin, damian2020feature, dhivyaa2020skin, ghalejoogh2020hierarchical, goceri2020comparative, guha2020performance, harangi2020assisted, kamalakannan2020self, kwasigroch2020neural, kwasigroch2020self, mahbod2020transfer, molina2020classification, mporas2020color, nunnari2020study, pereira2020skin, qin2020gan, rahman2020transfer, rodrigues2020new, salian2020skin, wu2020skin, wu2020multi, xie2020mutual, yan2020scalable, yildirim2020pre, yilmaz2020benign}           \\ \hline

SEN        &     \citep{han2022hwa, gu2022net, wang2022skin, wu2022fat, csahin2022robust, dai2022ms, tran2022fully, hu2022net, chen2022skin, zhang2022dense, zhang2022dynamic, alahmadi2022multiscale, alahmadi2022semi, feng2022slt, le2022antialiasing, al2022weakly, singh2022empirical, dong2022learning, aghdam2022attention, ahmed2022new, dong2022tc, fan2022egfnet, khan2022ensemble, khouloud2022w, ren2022serial, ruan2022malunet, zhou2022superpixel, gajera2021improving, liu2021skin, peter2021internet, reddy2021handling, wibowo2021lightweight, xiao2021prior, chen2021mt, adegun2021probabilistic, yacin2021deep, tong2021ascu, ramya2021segmentation, mu2021channel, kosgiker2021segcaps, hussain2021recu, hussain2021recu, das2021skin, chauhan2021multi, arora2021automated, ali2021skin, bagheri2021skin, chen2021nl, marosan2021automated, redha2021skin, tang2021afln, bagheri2021skin_2, chowdary2021automated, dong2021fac, garg2021skin, jiang2021residual, qamar2021dense, tao2021attention, wang2021focal, xie2021semi, li2011estimating, zhou2011gradient, he2012automatic, khakabi2012multi, madooei2012automated, abbas2013improved, cavalcanti2013macroscopic, nisar2013color, wu2013automatic, abbas2014combined, al2014automatic, ch2014two, jeniva2015efficient, rashid2015novel, torkashvand2015automatic, bi2016automated, hassan2016skin, jafari2016skin, pennisi2016skin, yuvaraju2016segmentation, bi2017semi, burdick2017impact, george2017automatic, lynn2017segmentation, mishra2017deep, nasr2017dense, pour2017automated, qi2017global, ramachandram2017skin, yuan2017automatic, adeyinka2018skin, akram2018skin, aljanabi2018skin, ammar2018learning, bi2018improving, burdick2018rethinking, chakkaravarthy2018automatic, hardie2018skin, jaisakthi2018automated, khan2018implementation, kolekar2018skin, li2018dense, li2018semi, luo2018fast, mirikharaji2018deep, nasir2018improved, patino2018automatic, salih2018comparison, ur2018classification, venkatesh2018deep, vesal2018multi, vesal2018skinnet, youssef2018deep, zeng2018multi, adegun2019deep, al2019deep, aljanabi2019various, bisla2019skin, cui2019ensemble, dash2019pslsnet, de2019skin, filali2019improved, filali2019multi, goyal2019skin, huang2019skin, javed2019intelligent, jiang2019decision, khan2019skin, liu2019skin, saini2019detector, salih2019skin, sarker2021slsnet, shahin2019deep, shan2019improving, singh2019fca, soudani2019image, tang2019efficient, tang2019multi, unver2019skin, wei2019attention, zhang2019automatic, azad2020attention, bagheri2020two, deng2020weakly, devi2020fuzzy, hafhouf2020modified, hasan2020dsnet, jabbari2020segmentation, jayapriya2020hybrid, jiang2020skin, khan2020frequency, abhishek2020illumination, lei2020skin, nathan2020lesion, ozturk2020skin, parida2020transition, pillay2020macroscopic, qin2020asymmetric, rizzi2020skin, rout2020transition, saha2020leveraging, setiawan2020image, shan2020automatic, sivaraj2020detecting, wang2020cascaded, wei2020attentive, wu2020automated, xie2020mutual, xie2020skin, madooei2012automated, hardie2018skin, hu2018skin, louhichi2018skin, olugbara2018segmentation, brahmbhatt2019skin, javed2019region, seeja2019deep, anjum2020deep, huang2020skin, wang2020donet, azehoun2016novel, khalid2016segmentation, khan2019construction, santos2020skin, wong2011automatic, pereira2019image, pereira2020dermoscopic, ahmed2018segmentation}         &       
\citep{nersisson2021dermoscopic, calderon2021bilsk, khan2022ensemble, lihacova2022multi, hsu2022hierarchy, batista2022classification, sun2021skin, reddy2021handling, pereira2021skin_2, moldovanu2021skin, liu2021multiscale, ding2021deep,bian2021skin, wang2021multi, tang2022fusionm4net, samsudin2022skin, shetty2022skin, samia2021skin, melbin2021integration, krohling2021smartphone, jain2021deep, bansal2021skin, yacin2021deep, samanta2021skin, rahman2021approach, nersisson2021dermoscopic, carvalho2021multimodal, balabantaray2021melanoma, sevli2021deep, bdair2021fedperl, calderon2021bilsk, mukherjee2021transfer, shen2022low, peter2021internet, wan2022mslanet,rasel2022convolutional, ngo2022skin, bozkurt2022skin, alptekin2022analysis, wei2022dual, thapar2022novel, sarker2022transslc, qian2022skin, aldhyani2022multi, zhuang2022cs, khouloud2022w, hosny2022refined, ayas2022multiclass, wang2022adversarial, hoang2022multiclass, dong2022learning, ramlakhan2011mobile, surowka2014optimal, li2018skin, thandiackal2018structure, eddine2019skin, rashid2019skin, seeja2019deep, ahmed2020skin, akram2020multilevel, anjum2020deep, chaturvedi2020skin, hassan2020skin, jibhakate2020skin, rodrigues2020new, amelard2012extracting, cavalcanti2013macroscopic, she2013lesion, celebi2014automated, chakravorty2016dermatologist, majtner2016combining, pomponiu2016deepmole, al2017classification, filali2017multiscale, lopez2017skin, lynn2017segmentation, mahdiraji2017skin, ozkan2017skin, satheesha2017melanoma, wahba2017combined, filali2018study, khan2018implementation, nasir2018improved, pham2018deep, ur2018classification, wahba2018novel, abbas2019efficient, al2019deep, albahar2019skin, aldwgeri2019ensemble, aljanabi2019various, bisla2019skin, chatterjee2019integration, filali2019improved, filali2019texture, gessert2019skin, javed2019intelligent, kassani2019depthwise, khan2019multi, khan2019skin, zheng2019relation, afza2020skin, dutta2021skin, almaraz2020melanoma, bi2020multi, damian2020feature, dhivyaa2020skin, gessert2020skin, ghalejoogh2020hierarchical, goceri2020comparative, harangi2020assisted, molina2020classification, nunnari2020study, pereira2020skin, qin2020gan, rahman2020transfer, wu2020skin, thomsen2020deep, wu2020multi, xie2020mutual, yan2020scalable, yildirim2020pre, zhuang2020cs}           \\ \hline

PRE  & \citep{han2022hwa, dai2022ms, zhang2022dense, le2022antialiasing, dong2022learning, dong2022tc, khan2022skin, khouloud2022w, zhou2022superpixel, gajera2021improving, reddy2021handling, tong2021ascu, ramya2021segmentation, das2021skin, marosan2021automated, dong2021fac, madooei2012automated, hassan2016skin, akram2018skin, burdick2018rethinking, hardie2018skin, hu2018skin, khan2018implementation, louhichi2018skin, olugbara2018segmentation, brahmbhatt2019skin, huang2019skin, javed2019intelligent, khan2019construction, khan2019skin, seeja2019deep, huang2020skin, setiawan2020image, wang2020donet, nisar2013color, rashid2015novel, anjum2020deep, sivaraj2020detecting, bozorgtabar2016sparse} &  
\citep{moldovanu2021skin, liu2021multiscale, shetty2022skin, samia2021skin, nguyen2022skin, reddy2021handling, krohling2021smartphone, hoang2022multiclass, wang2022adversarial, jain2021deep, bansal2021skin, rahman2021approach, hosny2022refined, nersisson2021dermoscopic, batista2022classification, afza2022multiclass, balabantaray2021melanoma, sevli2021deep, bdair2021fedperl, wang2022ssd, calderon2021bilsk, yilmaz2022mobileskin, rasel2022convolutional, bozkurt2022skin, alptekin2022analysis, wei2022dual, aldhyani2022multi, thapar2022novel, sarker2022transslc, nigar2022deep, nakai2022enhanced, khouloud2022w, tang2022fusionm4net, ngo2022skin, dong2022learning, ramlakhan2011mobile, surowka2014optimal, chakravorty2016dermatologist, majtner2016combining, haofu2017deep, lopez2017skin, mahdiraji2017skin, murphree2017transfer, ozkan2017skin, khan2018implementation, li2018skin, liao2018deep, yap2018multimodal, zhang2018skin, eddine2019skin, javed2019intelligent, khan2019multi, khan2019skin, rashid2019skin, seeja2019deep, wang2019mutual, zheng2019relation, ahmed2020skin, akram2020multilevel, almaraz2020melanoma, bi2020multi, chaturvedi2020skin, hassan2020skin, jibhakate2020skin, molina2020classification, qin2020gan, rahman2020transfer, rodrigues2020new, pham2018deep, aldwgeri2019ensemble, aljanabi2019various, anjum2020deep, harangi2020assisted, thomsen2020deep, yan2020scalable}    \\ \hline

SPE                     &     \citep{han2022hwa, gu2022net, wang2022skin, wu2022fat, csahin2022robust, tran2022fully, chen2022skin, hu2022net, zhang2022dense, zhang2022dynamic, alahmadi2022multiscale, alahmadi2022semi, feng2022slt, al2022weakly, singh2022empirical, dong2022tc, aghdam2022attention, ahmed2022new, fan2022egfnet, khan2022ensemble, khan2022skin, khouloud2022w, ruan2022malunet, gajera2021improving, liu2021skin, peter2021internet, reddy2021handling, wibowo2021lightweight, xiao2021prior, chen2021mt, adegun2021probabilistic, yacin2021deep, tong2021ascu, mu2021channel, kosgiker2021segcaps, hussain2021recu, hussain2021recu, arora2021automated, ali2021skin, bagheri2021skin, chen2021nl, redha2021skin, tang2021afln, bagheri2021skin_2, chowdary2021automated, dong2021fac, garg2021skin, jiang2021residual, qamar2021dense, tao2021attention, wang2021focal, li2011estimating, zhou2011gradient, he2012automatic, khakabi2012multi, madooei2012automated, abbas2013improved, cavalcanti2013macroscopic, nisar2013color, wu2013automatic, abbas2014combined, al2014automatic, ch2014two, jeniva2015efficient, rashid2015novel, torkashvand2015automatic, azehoun2016novel, bi2016automated, hassan2016skin, jafari2016skin, pennisi2016skin, yuvaraju2016segmentation, bi2017semi, burdick2017impact, george2017automatic, martinez2017pigmented, mishra2017deep, nasr2017dense, pour2017automated, qi2017global, ramachandram2017skin, yuan2017automatic, adeyinka2018skin, akram2018skin, aljanabi2018skin, ammar2018learning, bi2018improving, chakkaravarthy2018automatic, hardie2018skin, jaisakthi2018automated, khan2018implementation, kolekar2018skin, li2018dense, li2018semi, luo2018fast, mirikharaji2018deep, nasir2018improved, patino2018automatic, salih2018comparison, ur2018classification, venkatesh2018deep, vesal2018multi, vesal2018skinnet, youssef2018deep, zeng2018multi, adegun2019deep, al2019deep, aljanabi2019various, bisla2019skin, cui2019ensemble, dash2019pslsnet, de2019skin, filali2019improved, filali2019multi, goyal2019skin, huang2019skin, javed2019intelligent, jiang2019decision, khan2019skin, liu2019skin, lynn2017segmentation, saini2019detector, salih2019skin, sarker2021slsnet, shahin2019deep, shan2019improving, singh2019fca, soudani2019image, tang2019efficient, tang2019multi, unver2019skin, wei2019attention, xie2020mutual, zhang2019automatic, azad2020attention, bagheri2020two, devi2020fuzzy, hafhouf2020modified, hasan2020dsnet, jabbari2020segmentation, jayapriya2020hybrid, jiang2020skin, khan2020frequency, abhishek2020illumination, lei2020skin, nathan2020lesion, ozturk2020skin, parida2020transition, pillay2020macroscopic, qin2020asymmetric, rizzi2020skin, rout2020transition, saha2020leveraging, setiawan2020image, shan2020automatic, sivaraj2020detecting, wang2020cascaded, wei2020attentive, wu2020automated, xie2020skin}         &     \citep{khan2022ensemble, lihacova2022multi, hsu2022hierarchy, sun2021skin,pereira2021skin_2, bian2021skin, zhuang2022cs,shen2022low, wang2021multi,samia2021skin, reddy2021handling, samsudin2022skin, melbin2021integration, yacin2021deep, hoang2022multiclass, ayas2022multiclass, peter2021internet, qian2022skin, samanta2021skin, hosny2022refined, carvalho2021multimodal, mukherjee2021transfer, wan2022mslanet, rasel2022convolutional, wei2022dual, thapar2022novel, khouloud2022w, wang2022adversarial, tang2022fusionm4net, ngo2022skin, amelard2012extracting, cavalcanti2013macroscopic, she2013lesion, celebi2014automated, surowka2014optimal, chakravorty2016dermatologist, majtner2016combining, pomponiu2016deepmole, al2017classification, filali2017multiscale, lynn2017segmentation, mahdiraji2017skin, ozkan2017skin, satheesha2017melanoma, wahba2017combined, filali2018study, khan2018implementation, nasir2018improved, pham2018deep, thandiackal2018structure, ur2018classification, wahba2018novel, abbas2019efficient, al2019deep, albahar2019skin, aldwgeri2019ensemble, aljanabi2019various, bisla2019skin, chatterjee2019integration, filali2019improved, filali2019texture, gessert2019skin, javed2019intelligent, kassani2019depthwise, khan2019multi, khan2019skin, zheng2019relation, afza2020skin, dutta2021skin, almaraz2020melanoma, bi2020multi, damian2020feature, dhivyaa2020skin, gessert2020skin, ghalejoogh2020hierarchical, goceri2020comparative, harangi2020assisted, molina2020classification, nunnari2020study, pereira2020skin, qin2020gan, rahman2020transfer, wu2020skin, thomsen2020deep, wu2020multi, xie2020mutual, yan2020scalable, yildirim2020pre, zhuang2020cs}             \\ \hline

FOM                     &     \citep{li2011estimating}         &               \\ \hline
        
SE                     &    \citep{marosan2021automated, wong2011automatic, masood2014integrating, nasr2017dense, martinez2017pigmented, hardie2018skin, iranpoor2020skin, pereira2019image, pereira2020dermoscopic, liu2019enhanced, khalid2016segmentation}          &                \\ \hline

FPR                    &     \citep{khan2022skin, reddy2021handling, wong2011automatic, azehoun2016novel, khalid2016segmentation, akram2018skin, khan2018implementation, khan2019construction, khan2019skin, pereira2019image, huang2013robust, pereira2020dermoscopic, sivaraj2020detecting}         &      \citep{wang2021unlabeled, reddy2021handling, nersisson2021dermoscopic, calderon2021bilsk, surowka2014optimal, khan2018implementation, khan2019multi, khan2019skin, afza2020skin, akram2020multilevel}            \\ \hline
        
F1S                   &      \citep{alahmadi2022multiscale, le2022antialiasing, shamsolmoali2022salient, dong2022learning, gajera2021improving, reddy2021handling, tong2021ascu, kaur2021deep, filali2021graph, li2021digital, madooei2012automated, anjum2020deep, azad2020attention, pennisi2016skin, patino2018automatic, youssef2018deep, huang2019skin, javed2019region, nasir2018improved, liu2019enhanced, seeja2019deep, kamalakannan2020self, sivaraj2020detecting}        &      \citep{liu2021multiscale, ding2021deep, shetty2022skin,reddy2021handling, jain2021deep, bansal2021skin, rahman2021approach, sevli2021deep, hoang2022multiclass, bdair2021fedperl, calderon2021bilsk, bozkurt2022skin, yue2022towards, batista2022classification, yilmaz2022mobileskin,rasel2022convolutional, alptekin2022analysis, thapar2022novel, sarker2022transslc, ngo2022skin, nigar2022deep, nakai2022enhanced, aldhyani2022multi, nguyen2022skin, dong2022learning, surowka2014optimal, chakravorty2016dermatologist, ozkan2017skin, li2018skin, nasir2018improved, thandiackal2018structure, eddine2019skin, gessert2019skin, rashid2019skin, seeja2019deep, anjum2020deep, chaturvedi2020skin, goceri2020comparative, hassan2020skin, jibhakate2020skin, kamalakannan2020self, nunnari2020study, rahman2020transfer, rodrigues2020new, salian2020skin, wahba2017combined, wahba2018novel, almaraz2020melanoma}            \\ \hline

NPV                    &     \citep{reddy2021handling, nisar2013color, rashid2015novel, akram2018skin, sivaraj2020detecting}         &      \citep{mukherjee2021transfer, aldwgeri2019ensemble, aljanabi2019various, thomsen2020deep}            \\ \hline

DSC                    &    \citep{han2022hwa, gu2022net, jiang2022seacu, wang2022skin, wu2022fat, csahin2022robust, chen2022skin, tran2022fully, hu2022net, zhang2022dense, zhang2022dynamic, alahmadi2022multiscale, alahmadi2022semi, feng2022slt, feng2022bla, le2022antialiasing, al2022weakly, singh2022empirical, dong2022learning, aghdam2022attention, ahmed2022new, dong2022tc, fan2022egfnet, khan2022ensemble, khouloud2022w, rehman2022machine, ren2022serial, ruan2022malunet, zhao2022self, zhou2022superpixel, zuo2022efficient, liu2021skin, wibowo2021lightweight, yang2021deep, xiao2021prior, dayananda2021skin, chen2021mt, adegun2021probabilistic, wang2021knowledge, ramya2021segmentation, kosgiker2021segcaps, hussain2021recu, hussain2021recu, das2021skin, chauhan2021multi, arora2021automated, abhishek2021matthews, bagheri2021skin, chen2021nl, ding2021efficient, gangwar2021study, marosan2021automated, phan2021skin, tang2021afln, bagheri2021skin_2, chowdary2021automated, dong2021fac, garg2021skin, jiang2021residual, qamar2021dense, saini2021b, tao2021attention, wang2021focal, xie2021semi, masood2014integrating, he2017skin, lin2017skin, mishra2017deep, pour2017automated, qi2017global, yuan2017automatic, bi2017semi, bozorgtabar2017investigating, bozorgtabar2017skin, adeyinka2018skin, aljanabi2018skin, ammar2018learning, chen2018multi, he2018dense, jaisakthi2018automated, kolekar2018skin, li2018dense, li2018skin, mirikharaji2018deep, navarro2018accurate, olugbara2018segmentation, venkatesh2018deep, vesal2018multi, vesal2018skinnet, zeng2018multi, bi2018improving, hu2018skin, louhichi2018skin, riaz2018active, abdullah2019deep, adegun2019deep, al2019deep, alfaro2019brief, ali2019supervised, baghersalimi2019dermonet, bi2019improving, brahmbhatt2019skin, cui2019ensemble, dash2019pslsnet, de2019skin, goyal2019skin, jiang2019decision, liu2019skin, lynn2017segmentation, ma2019light, ninh2019skin, ooi2019interactive, saini2019detector, salih2019skin, sarker2021slsnet, sengupta2019segmentation, shahin2019deep, shan2019improving, singh2019fca, soudani2019image, tang2019efficient, tang2019multi, thanh2019adaptive, thanh2019skin, tschandl2019domain, tu2019dense, unver2019skin, wang2019automated, wang2019bi, wei2019attention, wu2019skin, zhang2019automatic, al2020automatic, bagheri2020two, deng2020weakly, hafhouf2020modified, jayapriya2020hybrid, jiang2020skin, justin2020skin, kaymak2020skin, khan2020frequency, abhishek2020illumination, lei2020skin, nathan2020lesion, ozturk2020skin, parida2020transition, pillay2020macroscopic, qin2020asymmetric, qiu2020inferring, ramella2020automatic, rizzi2020skin, rout2020transition, saha2020leveraging, salih2020skin, sanjar2020improved, santos2020skin, sayed2020novel, setiawan2020image, shan2020automatic, tang2020imscgnet, thanh2020adaptive, wang2020cascaded, wang2020donet, wei2020attentive, wu2020automated, xie2020mutual, xie2020skin, zafar2020skin, zhu2020asnet, agarwal2017automated, nasr2017dense}          &      \citep{farooq2016automatic, lynn2017segmentation, al2019deep, tschandl2019domain, chatterjee2019integration, moldovanu2021skin, melbin2021integration, balabantaray2021melanoma}            \\ \hline

HD                     &     \citep{azehoun2016novel, pereira2020dermoscopic, zhou2022superpixel}         &                 \\ \hline
        
AUC                   &     \citep{dong2022learning, khan2022skin, kosgiker2021segcaps, filali2021graph, das2021skin, bozorgtabar2016sparse, burdick2017impact, burdick2018rethinking, javed2019intelligent, khan2019skin, mahbod2020effects, hasan2020dsnet}         &       \citep{wang2021unlabeled, tschandl2019expert, sun2021skin, moldovanu2021skin, bian2021skin, yao2021single, wang2021multi, rahman2021approach, shen2022low, mahbod2021investigating, wan2022mslanet, deng2022efficient, alptekin2022analysis, wei2022dual, sarker2022transslc, wang2022ssd, nguyen2022skin, yue2022towards, wang2022adversarial, tang2022fusionm4net, shan2022automatic, dong2022learning, chakravorty2016dermatologist, devries2017skin, jia2017skin, murphree2017transfer, pham2018deep, thandiackal2018structure, wahba2018novel, xie2018multi, yap2018multimodal, zhang2018skin, albahar2019skin, javed2019intelligent, khan2019multi, khan2019skin, mahbod2019fusing, mahbod2019skin, serte2019gabor, serte2019wavelet, wang2019mutual, zheng2019relation, afza2020skin, almaraz2020melanoma, bagchi2020learning, gessert2020skin, harangi2020assisted, kwasigroch2020neural, kwasigroch2020self, liu2020automatic, mahbod2020effects, miglani2020skin, rahman2020transfer, salian2020skin, wu2020skin, wu2020multi, xie2020mutual, yan2020scalable, bi2020multi, yildirim2020pre}           \\ \hline
        
JI                   &      \citep{barin2022automatic, han2022hwa, gu2022net, jiang2022seacu, wang2022skin, wu2022fat, csahin2022robust, dai2022ms, hu2022net, chen2022skin, tran2022fully, stofa2022u, zhang2022dense, alahmadi2022multiscale, zhang2022dynamic, feng2022slt, feng2022bla, le2022antialiasing, lee2022progressive, al2022weakly, dong2022learning, ahmed2022new, dong2022tc, fan2022egfnet, khan2022ensemble, rehman2022machine, ren2022serial, ruan2022malunet, zhou2022superpixel, zuo2022efficient, liu2021skin, wibowo2021lightweight, xiao2021prior, yang2021deep, dayananda2021skin, chen2021mt, wang2021knowledge, tong2021ascu, mu2021channel, kosgiker2021segcaps, hussain2021recu, hussain2021recu, chauhan2021multi, arora2021automated, abhishek2021matthews, bagheri2021skin, chen2021nl, ding2021efficient, gangwar2021study, kaur2021deep, le2021modified, phan2021skin, tang2021afln, bagheri2021skin_2, chowdary2021automated, dong2021fac, jiang2021residual, li2021digital, qamar2021dense, saini2021b, tao2021attention, wang2021focal, xie2021semi, majtner2016improving, adeyinka2018skin, aljanabi2018skin, chen2018multi, guth2018skin, he2018dense, jaisakthi2018automated, jiang2018skin, kolekar2018skin, li2018dense, li2018semi, mirikharaji2018deep, navarro2018accurate, nguyen2018isic, rekha2018log, ross2018effects, salih2018skin, venkatesh2018deep, vesal2018multi, vesal2018skinnet, wang2018skin, zeng2018multi, abdullah2019deep, adegun2019deep, al2019deep, alfaro2019brief, ali2019supervised, cui2019ensemble, dash2019pslsnet, de2019skin, goyal2019skin, hasan2019comparative, jiang2019decision, lameski2019skin, liu2019skin, lynn2017segmentation, ma2019light, saini2019detector, salih2019skin, sarker2021slsnet, shahin2019deep, shan2019improving, soudani2019image, tang2019efficient, tang2019multi, thanh2019adaptive, thanh2019skin, tschandl2019domain, tu2019dense, unver2019skin, wang2019automated, wang2019bi, wei2019attention, wu2019skin, xie2020mutual, zhang2019automatic, al2020automatic, bagheri2020two, deng2020weakly, hawas2020oce, jayapriya2020hybrid, justin2020skin, abhishek2020illumination, lei2020skin, low2020automating, nathan2020lesion, ozturk2020skin, parida2020transition, pereira2020dermoscopic, qiu2020inferring, rizzi2020skin, rout2020transition, saha2020leveraging, salih2020skin, sayed2020novel, setiawan2020image, shan2020automatic, tang2020imscgnet, thanh2020adaptive, wang2020cascaded, wang2020donet, wei2020attentive, xie2020skin, zafar2020skin, zhu2020asnet, bozorgtabar2016sparse, bozorgtabar2017investigating, bozorgtabar2017skin, baghersalimi2019dermonet, bi2019improving, ninh2019skin, azad2020attention, hafhouf2020modified, jiang2020skin, kaymak2020skin, pillay2020macroscopic, qin2020asymmetric, wu2020automated, agarwal2017automated, bi2018improving, guth2018skin, qian2018detection, wu2020automated, singh2019fca, hasan2020dsnet, nampalle2020efficient, sanjar2020improved, masood2014integrating}        &      \citep{farooq2016automatic, lynn2017segmentation, chen2018multi, al2019deep, khan2019construction, tschandl2019domain, chatterjee2019integration}            \\ \hline

FNR                    &    \citep{reddy2021handling, khalid2016segmentation, akram2018skin, khan2018implementation, javed2019intelligent, khan2019construction, sivaraj2020detecting, ahmed2018segmentation}          &     \citep{reddy2021handling, khan2018implementation, javed2019intelligent, khan2019construction, afza2020skin, akram2020multilevel}             \\ \hline
        
XOR                    &     \citep{filali2021graph, li2011estimating, schaefer2011colour, he2012automatic, abbas2014combined, hassan2016skin, kasmi2016biologically, hu2018skin, riaz2018active, huang2019skin, ooi2019interactive, yang2019sampling}         &           \\ \hline

CC                   &     \citep{agarwal2017automated, gupta2017adaptive, sayed2020novel}         &            \\ \hline

HMD                   &      \citep{hu2018skin, riaz2018active, ooi2019interactive}        &            \\ \hline

MCC                 &      \citep{reddy2021handling, jiang2020skin, xiao2021prior, setiawan2020image, sivaraj2020detecting, tao2021attention}        &    \citep{almaraz2020melanoma, goceri2020comparative, ding2021deep, thapar2022novel, reddy2021handling, rahman2021approach, calderon2021bilsk}          \\ \hline

PSNR                  &      \citep{sayed2020novel}        &           \\ \hline
        
SSIM                     &     \citep{sayed2020novel}         &            \\ \hline

BACC              &      \citep{ali2021skin, bi2016automated, redha2021skin}        &    \citep{ozkan2017skin, rahman2020transfer, dos2018robust, kitada2018skin, almaraz2020melanoma, qin2020gan, surowka2014optimal, pereira2021skin_2, yao2021single, krohling2021smartphone, somfai2022handling, wang2022ssd, shen2022low, nguyen2022skin, foahom2022end, ngo2022skin}          \\ \Xhline{1pt}
 
\end{longtable}}

Qualitative evaluations are also reported in many articles in addition to the quantitative SLS evaluation \citep{hasan2020dsnet, hasan2021dermo, hasan2021dermoexpert, setiawan2020image, parida2020transition, osadebey2020evaluation, qin2020asymmetric, hafhouf2020modified, liu2020multi, saha2020leveraging, rout2020transition, sayed2020novel, ramella2020automatic, tang2020imscgnet, devi2020fuzzy, 
jiang2020skin, lei2020skin, mahbod2020effects, xie2020skin, baghersalimi2019dermonet, bhakta2019tsalli, filali2019multi, ribeiro2019handling, saini2019detector, singh2019fca, tang2019multi, wang2019bi, guth2018skin, jiang2018skin, luo2018fast, mirikharaji2018deep, olugbara2018segmentation, youssef2018deep, zeng2018multi, bozorgtabar2017investigating, bozorgtabar2017skin, george2017automatic, mishra2017deep, nasr2017dense, qi2017global, jafari2016skin, khalid2016segmentation, pennisi2016skin, khattak2015maximum, hamd2013skin, he2012automatic, khakabi2012multi}. Overlaying the segmented masks onto the ground truth masks is one such way to perform a qualitative evaluation. The intersected region(s) indicates true positive results, while non-overlapping regions correspond to false-positive and false-negative results. However, in some cases, the multi-class evaluation metrics are required for the SLS and SLC tasks, which are the expansions of the binary metrics in Table~\ref{tab:metrics_evaluation}. These metrics are averaged to achieve a multi-class metric across all the classes in two possible ways: micro and macro. The former averaging method (micro) considers class imbalance, calculating the metrics globally by counting the number of times each class was correctly and incorrectly predicted. In contrast, the latter approach (macro) does not assume class imbalance, measuring each class's metrics independently and attaining their unweighted mean. Sometimes, investigators employ k-fold cross-validation to verify the robustness of the trained model. In such a case, the obtained metrics from each fold are expressed as an average value with a standard deviation \citep{hasan2020automatic}, which can be expressed as Eq.~\ref{metric}. If the standard deviation has a higher value, it symbolizes higher inter-fold variation, consequently poor robustness, and vice-versa. 
\begin{equation}
\label{metric}
    M = \frac{1}{K} \times \sum_{i=1}^K M_i \pm \sqrt \frac{\sum_{i=1}^K \big( M_i -\mu \big)^2}{K},
\end{equation}
where $M_i\in \mathbb{R}$, $i\forall K$, denotes an estimated metric with a mean value of $\mu$ and $K$ is the fold numbers. Additionally, the authors in \citep{ribeiro2020less, wong2011automatic, almaraz2020melanoma} exploit the statistical validation test, i.e., Analysis of Variance (ANOVA), to evaluate the SLS and SLC techniques for revealing the significance of the experimentations with a suitable $p$-value. Moreover, Total Computation Time (TCT) is occasionally employed to judge the SLS and SLC methods for determining the time the algorithms need. Short computation times allow SLS and SLC methods to become real-time applications in CAD systems. However, among the reviewed articles, a few papers used TCT in their evaluation strategy, which are: \citep{ivanovici2012color, pennisi2016skin, agarwal2017automated, gupta2017adaptive, akram2018skin, he2018dense, khan2018implementation, khan2019skin, ma2019light, low2020automating} for SLS; and \citep{khan2018implementation, javed2019intelligent, khan2019skin, afza2020skin, akram2020multilevel, afza2022multiclass, khan2021skin} for SLC.

\begin{figure}[!ht]
  \centering
\includegraphics[width=15cm, height= 3cm]{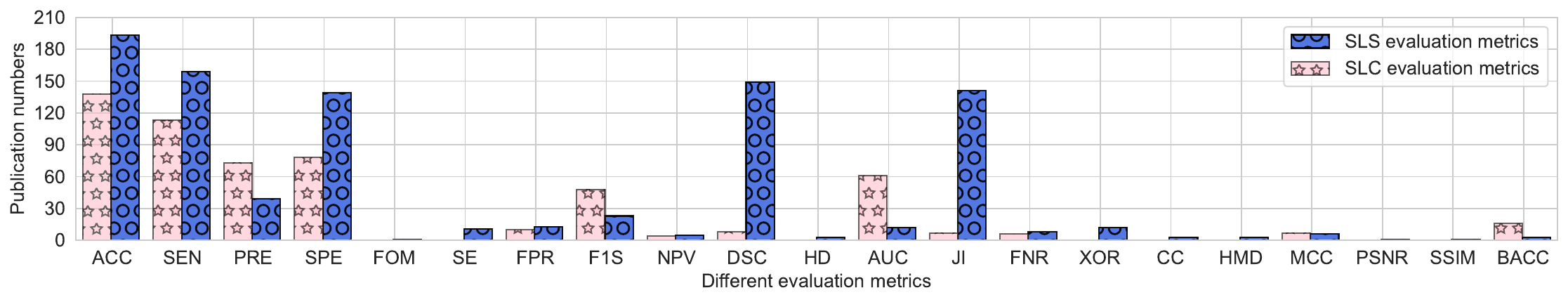}
\caption{\MKH{The number of articles employing different evaluation metrics, conferring the commonly applied metrics.}}
\label{fig:SLC_SLS_Eval}
\end{figure}

Fig.~\ref{fig:SLC_SLS_Eval} summarizes the frequency with that various metrics are used in the 594 SLA articles we reviewed.
\MKH{Accuracy, sensitivity, specificity, dice similarity coefficient, and Jaccard index are observed to be the 5 most frequently used metrics, being employed in $54.2\,\%$, $44.7\,\%$, $39.0\,\%$, $41.9\,\%$, and $39.6\,\%$ of the total articles. Therefore, they can be treated as the \textbf{\textit{High-frequency}} SLS evaluation metrics. In contrast, accuracy, sensitivity, precision, specificity, and AUC are the most-5 regularly applied metrics for the SLC evaluation respectively practiced in $58.0\,\%$, $47.5\,\%$, $30.7\,\%$, $32.8\,\%$, and $25.6\,\%$ articles, leading to the \textbf{\textit{High-frequency}} SLC evaluation metrics. The other metrics are much less frequently used than the top 5 in both SLS and SLC tasks. Thus, we have considered them as the \textbf{\textit{Low-frequency}} metrics rather than \textbf{\textit{Medium-frequency}}.} 
Again, although the dice similarity coefficient and Jaccard index are frequently used in SLS evaluation, they are only sometimes employed for SLC evaluation. Some metrics like accuracy, sensitivity, specificity, precision, F1-score, balance accuracy, and false-positive and false-negative rates are employed for the SLS and SLC evaluations in approximately the same number of articles. Moreover, although many articles applied accuracy as measured by the SLS and SLC metrics, they need to consider the class imbalance measurement. If there are many more samples in one particular class than others, the trained models tend to learn that class more accurately at the expense of accuracy for other classes, leading to a general increase in false false-positive or false-negative rates. If there is no penalty for class imbalance, conventional metrics like accuracy can still have high values despite a poorly performing model. Some metrics, such as ROC and AUC estimation \citep{hasan2020dsnet, hasan2021dermo, hasan2021dermoexpert, al2018skin}, however, address this imbalance. Nevertheless, the barely $3.4\,\%$ for SLS and $25.6\,\%$ for SLC articles acknowledged this issue by determining the AUC values for their assessments.

\section{\MKH{Explainability and Clinical Impact of SLA Methods}}
\label{Explainability_of_SLA_Methods}
\MKH{Explainable Artificial Intelligence (XAI) aims to unbox how AI systems make decisions, investigate the criteria and models that go into making decisions, and look for ways to describe such criteria and models in more detail \citep{yang2022unbox, arrieta2020explainable}. \citet{barata2021explainable} used visual attention mechanisms to direct the model to the most discriminative regions and attributes of the lesion at each decision level. They presented a heatmap as a mode of SLA explanation. A Local Interpretable Model-agnostic Explanation (LIME), a post hoc method, was used in \citep{nigar2022deep} that generated a few cues like the mole’s size, bleeding, and shape to help the dermatologist make a final decision. Class Activation Map (CAM) was applied in some articles \citep{chowdhury2021exploring, li2020fusing, xie2020mutual, yan2019melanoma, yang2018classification, zhang2019attention}, where the explanation was presented using a heatmap and/or histogram. A modification of CAM called Gradient CAM (GradCAM) \citep{hasan2022challenges}, which weights the feature maps based on gradients, was also used in some SLA articles \citep{nunnari2021overlap, rieger2020interpretations, wang2021interpretability, zunair2020melanoma, young2019deep}. GradCAM used a heatmap and/or histogram for the SLA explanation. The other XAI methods in the past SLA literature are t-SNE \citep{sadeghi2020using, barata2021improving}, content-based image retrieval \citep{tschandl2019diagnostic, sadeghi2018users, barata2021improving}, Shapley Additive Explanations (SHAP) \citep{wang2021interpretability, young2019deep}, Kernel-SHAP \citep{young2019deep}, and feature activation maps \citep{wei2020automatic}.} 

\MKH{In terms of clinical applications of SLA CAD, this is not reviewed here as our focus is on the SLA techniques, but a detailed review can be found in \citep{hauser2022explainable}. However, it is noteworthy that there is evidence of CNN’s ability to outperform human experts. \citet{haggenmuller2021skin} evaluated CNN’s SLA task performance and compared it to a systematic evaluation by clinical human experts. The performance of an AI-based SLA algorithm trained exclusively by open-source images was compared to that of a large number of dermatologists covering all levels within the clinical hierarchy by \citet{brinker2019deep}. The authors demonstrated that AI-based SLA outperformed 136 of 157 dermatologists for the melanoma image classification task. \citet{tschandl2019expert} demonstrated that the performance of AI-based SLA is better than human raters for the skin cancer detection task. A case study on skin cancer diagnosis in the UK’s General Practitioners (GPs) was carried out by \citet{micocci2021gps}, and the author showed that AI is accurate; there is a positive effect on GPs’ performance and confidence, indicating the possibility of reducing referrals for benign lesions.}

\MKH{Currently, several AI-based SLA technologies are approved by regulatory bodies for use clinically; for example, MelaFind and Nevisense were approved by the US FDA, and an AI-powered digital tool for diagnosing skin cancers (Skin Analytics) was approved by the UK MHRA as a Class IIA medical device.}

\section{Observations and Recommendations}
\label{Observations_and_Recommendations}
This section highlights our key findings from a critical inspection of the selected 594 articles (356 for segmentation and 238 for classification). Here, we provide suggestions for future SLA tasks, including input data (dataset utilization, preprocessing, augmentation, and imbalance problem solving), approach configuration (techniques, architectures, module frameworks, and loss functions), training protocol (hyperparameter settings), and evaluation criteria (metrics). These suggestions are intended to inform future SLA work.

\begin{itemize}
    \item Currently, ISIC datasets have been most commonly applied to the SLS and SLC tasks after 2016, turning them into a representative dataset for SLA (see Fig.~\ref{fig:data_used_per_year}). Our review also reveals that there is a necessity for larger, more diverse, and more representative skin lesion datasets (with more inter-class diversity) to train DL models to make them more accurate and robust. The most popular ISIC Archive contains over 69,900 publicly available skin lesion images, but many cases lack proper ground truth. One way to improve this is to crowdsource annotations from investigators who are experts in this field and then run an ensemble based on the annotations that come up most often.

    \item Our review suggests that there need to be specific clarifications to the questions: is preprocessing essential, and which preprocessing approach is best for SLA tasks? To answer the first question, two considerations have been uncovered in the past SLA articles: If manual feature engineering with ML models or other computer vision-based models is employed, preprocessing is strongly suggested; conversely, if DL-based automated models are applied, preprocessing is ill-advised. To answer the second question, the proper selection of the preprocessing method depends on the algorithm types. The utilization frequency history provided in this review can help future investigators in their preprocessing design (see details in section~\ref{Preprocessing}, especially in Fig.~\ref{fig:preprocessing}).
    
    \item The applications of the augmentations need to be recommended only for the DL-based computerized SLA techniques, as they were seldom involved in the other ML (or computer vision)-based SLA techniques in the past (see Fig~\ref{fig:Augmentaion}). The typical baseline augmentation strategies can be found in section~\ref{Data_scarcity_and_imbalance_problem}, leading to help for augmentation preference for SLA. Although augmentations are involved in the minority class to increase its sample numbers, the class weighting has been most commonly suggested to resolve the class imbalance problem (see details in section~\ref{Data_scarcity_and_imbalance_problem}).
    
    \item The insight overlooked in the past SLS articles reveals that DL-based procedures, especially CNN models to learn a lesion's spatial information, are the recent trends, especially after 2016 (see Fig.~\ref{fig:year_wise_paper}, Fig.~\ref{fig:sls_yearwise_method} (a) and (b)). Therefore, CNN-based encoder-decoder networks are the most prioritized procedures for the SLS task. Again, the investigation of the encoder structure reveals that the ResNet, VGG, and DenseNet family networks are the most widely applied encoder architectures and are the most popular trends for the SLSs. In contrast, U-Net-like structures are the preferred decoder mechanism in the CNN-based SLS job. Additionally, segmented lesion masks are generally improved by employing several post-processing techniques (see Table~\ref{tab:Postprocessing}), where the various morphological operations are the most applied methods in the last twelve years (see back-and-forth in section~\ref{Segmentation_Techniques}, particularly in Table~\ref{tab:Postprocessing}).

    \item Although this survey article discloses the most commonly employed lesion's attributes (see Fig.~\ref{fig:Feature_Used}) and ML-based classifiers (see Fig.~\ref{fig:Pie_slc_ML_methods}) for the manual feature-based SLC strategies, the DL-based automated approaches have been the last few years' trends (see Fig.~\ref{fig:Pie_slc_DL_methods}). Further investigation reveals the most frequent CNN architectures like ResNet, VGG, DenseNet, InceptionNet, and AlexNet for lesion's attribute learning, which are discussed in section~\ref{DL_SLC}. Using lightweight pre-trained architectures that are appropriate for the smaller training set is also observed and recommended, as they have been widely applied in the last twelve years with fewer training examples. According to previous research, engaging in ensemble learning and/or combining manual and automated CNN's lesion features is advantageous for capturing the discernment lesion characteristics, thereby improving the SLC outcome(s).
    
    \item The 594 SLA literature also enabled researchers to determine the best experimental settings that are most typically used for both SLS and SLC tasks. Although it is challenging to decide the values of the proper hyperparameters, recent history could support resolving it. For illustration, Fig.~\ref{fig:BS_Comparison} and Fig.~\ref{fig:LR_Comparison} would aid in understanding the last twelve years' history of batch size and LR employment, respectively, revealing that their low and/or very high values are rarely used. Cross-entropy (binary or multi-class) is the most suggested loss function for the SLS and SLC tasks. However, many authors have enforced extra care in the SLS's loss function, considering the DSC (or JI) function as in Eq.~\ref{eq:dsc} and Eq.~\ref{eq:ji} (see details reflection in Eq.~\ref{eq:loss}). This review would help to determine good optimizers and epoch numbers; for example, adaptive and SGD optimizers and mid-range epoch numbers like 100 to 200 have been most commonly employed in the past. The suitable values of the optimizers' parameters could be chosen by overlooking Table~\ref{tab:control_params}. Lastly, the trends of ML and/or DL frameworks for developing the SLA pipeline can be seen in section~\ref{Training_Environments}.
    
    \item There is no universally appropriate evaluation criterion \citep{mirikharaji2022survey}, as different benchmarks grasp different characteristics of an SLS and/or SLC algorithm’s performance on a provided dataset. Our analysis in Fig.~\ref{fig:SLC_SLS_Eval} informs future investigators on which metrics have been most commonly used in the recent past to help investigators decide the best metrics to use. A direct quantitative assessment system using the metrics (see Table~\ref{tab:metrics_evaluation}), qualitative evaluation, and reflection of TCT are the three most standard approaches to estimating the SLA method(s). In order to explain the performance of the SLA technique(s), all three of these evaluation techniques could be involved, as in \citep{ivanovici2012color, pennisi2016skin, agarwal2017automated, gupta2017adaptive, akram2018skin, he2018dense, khan2018implementation, khan2019skin, ma2019light, low2020automating} for SLS and \citep{khan2018implementation, javed2019intelligent, khan2019skin, afza2020skin, akram2020multilevel, afza2022multiclass, khan2021skin} for SLC. Unfortunately, it has been infrequently used in the past. This is one of the shortcomings of the current practice observed in the SLA articles, and the researchers would consider it in future experimentation.
    
    \item \MKH{Although an enormous number of SLA articles have been published in the past, only a few studies have developed prototype applications, and none of the 594 SLA articles has considered developing user-friendly clinical applications in practical settings. Our review could not find the answer in the 594 articles to how the technologies have been integrated into clinical settings. The researchers should concentrate on real-time clinical applications in conjunction with SLA technique improvement, providing a case study of AI methods' successes.}

    \item \MKH{Despite some papers studying XAI strategies to unlock the black box of the decision-making SLA pipeline, only a few of these studies \citep{sadeghi2020using, sadeghi2018users, tschandl2020human} have examined the influence on diagnosis accuracy and dermatologist acceptance. This human factor would be considered in future SLA research. Optimizing the reasoning method for best human usage and improving interpretability beyond visualization approaches are issues that demand more study in this SLA area. Future research should demonstrate how to include XAI in a comprehensive optimization process in order to increase the model's efficacy while reducing its complexity. In addition, defining a model to be explainable based on its capacity to instill confidence may not satisfy the condition of model explainability \citep{arrieta2020explainable}. Trustworthiness is the assurance that a model will respond as intended when presented with a given problem; nonetheless, this must be added to future SLA literature \citep{lekadir2021future}.}
    
\end{itemize}

\section{Conclusion}
\label{Conclusion}
This survey summarizes the 594 SLA articles published over the course of the past year, with an emphasis on dataset utilization, preprocessing, augmentations, solving imbalance problems, SLS \& SLC techniques, training tactics (frameworks and hyperparameter settings), and evaluation benchmarks (metrics). The survey analysis conducted on SLA reveals that ISIC is the most commonly applied dataset that best exemplifies this area. When using ML models with manual feature engineering, preprocessing was typically used; however, when using automatically generated DL models, it had yet to be generally applied in the past twelve years. DL-based computational SLA techniques are becoming the usual trend in the skin lesion arena. This is due to the fact that they perform better than the other SLA methods, require less parameter adjusting, and are end-to-end. In order to evaluate SLA methods, this survey suggests that in addition to qualitative and quantitative evaluations, SLA's TCT would be applied to assess the method(s) from diverse aspects. Our review may be helpful to future investigators in designing their SLA approach. Potential researchers should translate suitable SLA methodologies into real-time applications considering clinical settings, as there is room for such translation.

\section*{Acknowledgement}
Guang Yang was supported in part by the BHF (TG/18/5/34111, PG/16/78/32402), the ERC IMI (101005122), the H2020 (952172), the MRC (MC/PC/21013), the Royal Society (IEC/NS-FC/211235), the NVIDIA Academic Hardware Grant Program, the SABER project supported by Boehringer Ingelheim
Ltd, NIHR Imperial Biomedical Research Centre (RDA01), and the UKRI Future Leaders Fellowship (MR/V023799/1).

\bibliographystyle{model1-num-names}

\bibliography{sample}
\end{document}